\documentclass[%
 reprint,
 amsmath,amssymb,
 aps,nofootinbib,
]{revtex4-1}

\usepackage{graphicx}
\usepackage{dcolumn}
\usepackage{aasjournal} 
\usepackage{bm}
\usepackage{xcolor}

\begin{document}


\title{Sample Variance for Supernovae Distance Measurements and the Hubble tension}

\author{Zhongxu Zhai}
  \email{zhongxu.zhai@uwaterloo.ca}
\author{Will J. Percival}%
  \altaffiliation[Also at ]{Perimeter Institute for Theoretical Physics, 31 Caroline St. North, Waterloo, ON N2L 2Y5, Canada}

\affiliation{%
Waterloo Center for Astrophysics, University of Waterloo, Waterloo, ON N2L 3G1, Canada \\
Department of Physics and Astronomy, University of Waterloo, Waterloo, ON N2L 3G1, Canada
}%

\date{\today}

\begin{abstract}
Recent local measurements of the Hubble constant made using supernovae have delivered a value that differs by $\sim$5$\sigma$ (statistical error) from predictions using the Cosmic Microwave Background (CMB), or using Baryon Acoustic Oscillations (BAO) and Big-Bang Nucleosynthesis (BBN) constraints, which are themselves consistent. The effective volume covered by the supernovae is small compared to the other probes, and it is therefore interesting to consider whether sample variance (often also called cosmic variance) is a significant contributor to the offset. We consider four ways of calculating the sample variance: (i) perturbation theory applied to the luminosity distance, which is the most common method considered in the literature; (ii) perturbation of cosmological parameters, as is commonly used to alleviate super-sample covariance in sets of N-body simulations; (iii) a new method based on the variance between perturbed spherical top-hat regions; (iv) using numerical N-body simulations. All give consistent results showing that, for the Pantheon supernova sample, sample variance can only lead to fluctuations in $H_0$ of order $\pm1$ km s$^{-1}$Mpc$^{-1}$ or less. While this is not in itself a new result, the agreement between the methods used adds to its robustness. Furthermore, it is instructive to see how the different methods fit together. We also investigate the internal variance of the $H_{0}$ measurement using SH0ES and Pantheon data. By searching for an offset between measurements in opposite hemispheres, we find that the direction coincident with the CMB dipole has a higher $H_{0}$ measurement than the opposite hemisphere by roughly 4 km s$^{-1}$Mpc$^{-1}$. We compare this with a large number of simulations and find that the size of this asymmetry is statistically likely, but the preference of direction may indicate that further calibration is needed.

\end{abstract}

\maketitle


\section{Introduction} 

Recent measurements of the Hubble constant \cite{localH0-2016,localH0-2021} using a local distant ladder combining observations of Cepheids and supernovae (SNe) have given us the constraint $H_{0}=73.04\pm1.04$\,km s$^{-1}$Mpc$^{-1}$. This is in strong tension with the constraint from Planck \cite{Planck2018} Cosmic Microwave Background observations, which gives $H_{0}=67.37\pm0.54$\,km s$^{-1}$Mpc$^{-1}$ for a flat $\Lambda$CDM model, and recent observations of the Baryon Acoustic Oscillations (BAO) standard ruler \citep{eBOSS-cosmo} combined with the Big Bang Nucleosynthesis (BBN) observations \citep{BBNobs} required to standardize the ruler, which give $H_{0}=67.35\pm0.97$\,km s$^{-1}$Mpc$^{-1}$. Using the error bars provided, the tension is at the 5$\sigma$ level. Many potential deviations from the $\Lambda$CDM model have been discussed as a solution to this problem \cite{Hubble-hunters,H0-tension-review}. Alternative solutions include an unknown systematic problem with one of more data or that the error bars are underestimated. It is also possible that the solution will require a number of separate contributions.

Each SN observation probes a region of space-time that covers the line of sight (LOS) from the observer to the SN. There will be inhomogeneities along the LOS that will alter the Hubble Constant measured using the redshift and luminosity distance, compared with that averaged over a larger patch of space-time. When considered over all of the supernovae within current samples, recent studies have shown that this error cannot explain the 5$\sigma$ tension recently observed alone \cite{Marra2013,Wojtak2014,Romano2016,Wu2017}.

The sample variance in the Hubble parameter between different patches of the Universe is driven by changes in the matter density. Density fluctuations cause the local universe to behave differently, dependent on the amplitude of the fluctuation, which is related to its scale with smaller fluctuations having more scatter in amplitude than larger fluctuations. We consider the methods previously used to determine the sample variance including considering perturbations to the luminosity distance \cite{Sasaki1987}, and calculated using simulations \cite{Wojtak2014,Wu2017}. We introduce two further ways of measuring sample variance, borrowing ideas from work on super-sample covariance\cite{Sirko2005}, and from homogeneous spherical models\cite{Percival2005}. Other than using simulations, the methods can all be considered to be part of perturbation theory, and differ in what we are perturbing: the luminosity distance, cosmological parameters or the curvature of patches in the Universe. 

We consider two simulation based methods: one using all halos within a particular radius, which is closely matched to our analytic approaches, and one based on simulating the Pantheon \cite{Pantheon} distribution of SN. Apart from the latter approach where we directly use the SN sample, all of the methods require us to define a volume in order to find the distribution of fluctuations in the overdensity. We present a new method to estimate this for Pantheon based on defining a zone of influence for each SN observation. With the same definition of this volume, the perturbative results all give similar results, matching that from N-body simulations where we use all halos within the same volume.

As well as considering larger samples from which the Pantheon SN sample is assumed to be a typical draw, we also use internal methods to compare the distribution of measurements across the Pantheon sample. We split the sample into opposite hemispheres, optionally including the Cepheid calibration in this split. After subsampling, we measure $H_{0}$ from each hemisphere and compare the variation in values recovered to that from simulations. 

This paper is organized as follows, Section~II presents the methods and results estimating $H_{0}$ variations due to a local inhomogeneity. Section~III presents the estimate of the volume covered by the Pantheon SNe Ia sample and compares analytic results for this sample to the variance of $H_{0}$ based on simulations. Section~IV presents our reanalysis of SH0ES and Pantheon data cut into various subsamples. Section~V includes our discussion and conclusions. 

\section{Sample Variance for $H_0$}

The starting point for analysing sample variance is fluctuations in density in the Universe. The variance of these scale-dependent fluctuations at early times and on large scales can be estimated by integrating the linear power spectrum $P(k)$ multiplied by a window function 
\begin{equation}  \label{eq:PkconvW}
  \sigma_R^2=\frac{1}{2\pi^2}\int_0^{\infty}P(k)
      \tilde{W}^2(k;R)k^2\,dk\,,
\end{equation}
where it is common to assume a top-hat (in real-space) filter
\begin{equation}
  \tilde{W}(k;R)=3\frac{\sin(kR)-(kR)\cos(kR)}{(kR)^3}\,.
\end{equation}
For $R=8$\,h$^{-1}$Mpc, we recover the standard definition of $\sigma_8$, often used to normalise the power spectrum. We now consider three methods for translating from $\delta$ to give the local value of $H_0$ measured in a patch of the Universe of a given size.

\subsection{Perturbing the luminosity distance} 

The sample variance in local distance-ladder based measurements of $H_0$ can be determined by considering the effect of changes in $\delta$ on the luminosity distance directly \cite{Sasaki1987,Barausse2005,Bonvin2006,Hui2006}. Traditionally, the derivation starts by considering fluctuations to the Angular Diameter distance, $D_A=D_{A,b}[1-k(z)]$, where the convergence $k(z)$ has many terms, including a term related to magnification. The dominant term is a shift related to the peculiar velocity of the source, 
\begin{equation}
  k_v=\left[1-\frac{1}{a_e\chi_e H_e}\right]v_e\cdot n+\frac{1}{a_e\chi_e H_e }v_o\cdot n\,,
\end{equation}
where $\chi$ is the comoving distance, and $v$ the peculiar velocity relative to the background model. A subscript $e$ denotes a quantity evaluated at the point of emission of the photons, and the unit vector $n$ is in the direction of propagation from the emitter to the observer. This is calculated for the same $\Delta z$ in perturbed and unperturbed frames. After a number of approximations (e.g. \cite{Romano2016}), we find that
\begin{equation}
k_v=\frac{1}{3}f\delta\left(aH\chi-1\right)\,,
\end{equation}
where $f$ is the logarithmic derivative of the linear growth rate $f=d\log D/d\log a$. For local fluctuations, $aH\chi\approx z$, so that the variation in luminosity distance can be directly related to a change in $H_0$
\begin{equation}
    \frac{\Delta H_{0}}{H_{0}} = -\frac{1}{3}f\delta\,.
\end{equation}
Thus we can translate directly from a distribution of $\delta_0$ to a distribution in the locally measured value of $H_0$.

\subsection{Perturbing the cosmological parameters}

So called super-sample covariance (SSC) is commonly considered in the field of N-body simulations \cite{Sirko2005} and for attributing errors to clustering measurements made from galaxy clustering in small surveys \cite{Takada2013}. Simply put, on large scales, SSC refers to the changes in cosmological quantities that occur patch-to-patch due to variations in the large-scale density. For galaxy clustering, there is also a link from the large-scale modes driving SSC to small-scale non-linear modes \cite{Takada2013}, but that does not affect us here. One way of considering how SSC works is to think about running a large number of small N-body simulations to understand the patch-to-patch variance where each patch is the same size as the box. The simplest way to run the simulations would be to fix the cosmological parameters at the background values for all boxes and set the average overdensity within each simulation box to zero. However, such a set of simulations would not include all of the variance between patches of the Universe represented by the boxes. Large-scale modes affect the "DC-level" density, and one should really sample the properties of each box (or patch) from a distribution of parameters reflecting the range of densities driven by large-scale modes on sizes larger than the box \cite{Frenk88,Sirko2005}. This was recently applied to give covariance matrices for galaxy surveys including SSC \cite{Howlett2017}.

We can use the same formalism to estimate the sample variance for an analysis of SNe, where we wish to understand the impact of fluctuations larger than the volume covered by the SNe. Within a background cosmology, we can consider the situation where we have a large-scale value of $H_0$ (analogous to a large simulation), but different patches in the Universe (analogous to small simulations) each have a local value that changes because of the DC-mode density. \citet{Sirko2005} showed that the parameters used for each simulation (or patch) should be modified to allow for SSC via:
\begin{equation}\label{eq:ssc}
\begin{aligned}
a_{\mathrm{patch}} &= a\biggl(1-\frac{D(a)\delta_{b,0}}{3D(1)}\biggl), \\
H_{0,\mathrm{patch}} &= H_{0}(1+\phi)^{-1}, \\
\Omega_{m,0,\mathrm{patch}} &= \Omega_{m,0}(1+\phi)^{2}, \\
\Omega_{\Lambda,0,\mathrm{patch}} &= \Omega_{\Lambda,0}(1+\phi)^{2}, \\
\Omega_{k,0,\mathrm{patch}} &= 1-(1+\phi)^{2}(\Omega_{m,0}+\Omega_{\Lambda,0}),
\end{aligned}
\end{equation}
where
\begin{equation}
\phi = \frac{5\Omega_{m,0}}{6}\frac{\delta_{b,0}}{D(1)},
\end{equation}
$\delta_{b,0}$ is the background mode at redshift 0, $D$ is the linear growth factor, $a$, $H_{0}$, $\Omega_{m,0}$, $\Omega_{\Lambda,0}$, $\Omega_{k,0}$ define the output scale factor and cosmology of the ensemble, and $a_{\mathrm{patch}}$, $H_{0,\mathrm{patch}}$, $\Omega_{m,0,\mathrm{patch}}$, $\Omega_{\Lambda,0,\mathrm{patch}}$, $\Omega_{k,0,\mathrm{patch}}$ are the parameters given to each realisation. The size of the SSC component depends on the over-density of the patch considered, with large patches naturally leading to smaller variations in overdensity and hence a smaller sample variance contribution to the errors. 

\subsection{Perturbed spherical top-hat regions}

Rather than consider the effect of variations in overdensity perturbing the cosmological parameters directly, we now develop a new method using the spherical top-hat model to understand variations in $H_0$ between different patches of the Universe. To do this, we borrow heavily from the methodology developed to estimate the critical density for collapse, and follow the notation and much of the general derivation presented in \cite{Percival2005}. This follows from derivations for an Einstein-de Sitter cosmology \citep{gunn72}, for open cosmologies \citet{lacey93} and for flat $\Lambda$ cosmologies \citep{eke96}.

Following the standard top-hat model, we consider two equal mass spheres: one following the background with radius $a$, and a sphere perturbed by a homogeneous change in overdensity of radius $a_p$. To leading order, we can define the overdensity as $a_p=a(1-\delta/3)$. We assume that the dark energy component is negligible at early times so we can write the Friedman equation for both $a$ and $a_p$ (for simplicity we use a subscript $X$ for quantities that differ for the two spheres so we do not have to duplicate similar equations),
\begin{equation}
  \left(\frac{da_X}{d(H_0t)}\right)^2=\frac{\Omega_M}{a_X}+\epsilon_X\,,
  \label{eq:friedmann1}
\end{equation}
where the curvature term $\epsilon_p$ is allowed to take any real value for the perturbation, while for the background, $\epsilon=\Omega_K\equiv(1-\Omega_M-\Omega_\Lambda)$. A series
solution for $a_X$ in the limit $H_0t\to0$ is given
by $a_X=\alpha(H_0t)^{2/3}+\beta_X(H_0t)^{4/3}+O[(H_0t)^{6/3}]$, where
\begin{equation}
  \alpha = \left(\frac{9\Omega_M}{4}\right)^{1/3}, \,\,\,
  \beta_X = \frac{3\epsilon_X}{20}\left(\frac{12}{\Omega_M}\right)^{1/3}\,.
\end{equation}
From the definition of $\delta$, we have that
\begin{equation}
  \lim_{H_0t\to0}\delta(H_0t)=
    \frac{3}{5}\left(\frac{3}{2\Omega_M}\right)^{2/3}
    \left[(\Omega_K-\epsilon_p\right](H_0t)^{2/3}\,.
\end{equation}
This links the limiting density at early times to the subsequent curvature of that patch of space-time. We can link this to the linearly extrapolated overdensity at present day using linear growth
\begin{equation}
  \frac{\delta_0}{D_0}=
    \lim_{H_0t\to0}\left[\frac{\delta(t)}{D(t)}\right] 
    =\frac{3}{5\Omega_M}\left[\Omega_K-\epsilon_p\right]\,.
\end{equation}
Thus given $\delta_0$ we can find the curvature $\epsilon_p$ for that patch. In order to determine the value of $H_0$ (or other cosmological parameters) in that patch of space-time at present-day, we can solve the Friedmann equation
\begin{equation}  \label{eq:fried}
  \left(\frac{da_p}{d(H_0t)}\right)^2=\frac{\Omega_M}{a_p}+\epsilon_p
    + \Omega_\Lambda a_p^2\,.
\end{equation}
We define present-day as matching the cosmic age for the background model. Note that with this definition, the value of $H_0$ assumed for the patch simply fixes the normalisation of $a_p=1$, and is not the value of $H_0$ at the present-day time. To find this, we numerically find the value of $a_p$ that gives an age of the universe matching that of the background by integrating Eq.~\ref{eq:fried}. We then use the same equation to measure the value of the Hubble parameter $\dot{a}_p/a_p$ at this time.

\begin{figure}
\begin{center}
\includegraphics[width=8.5cm]{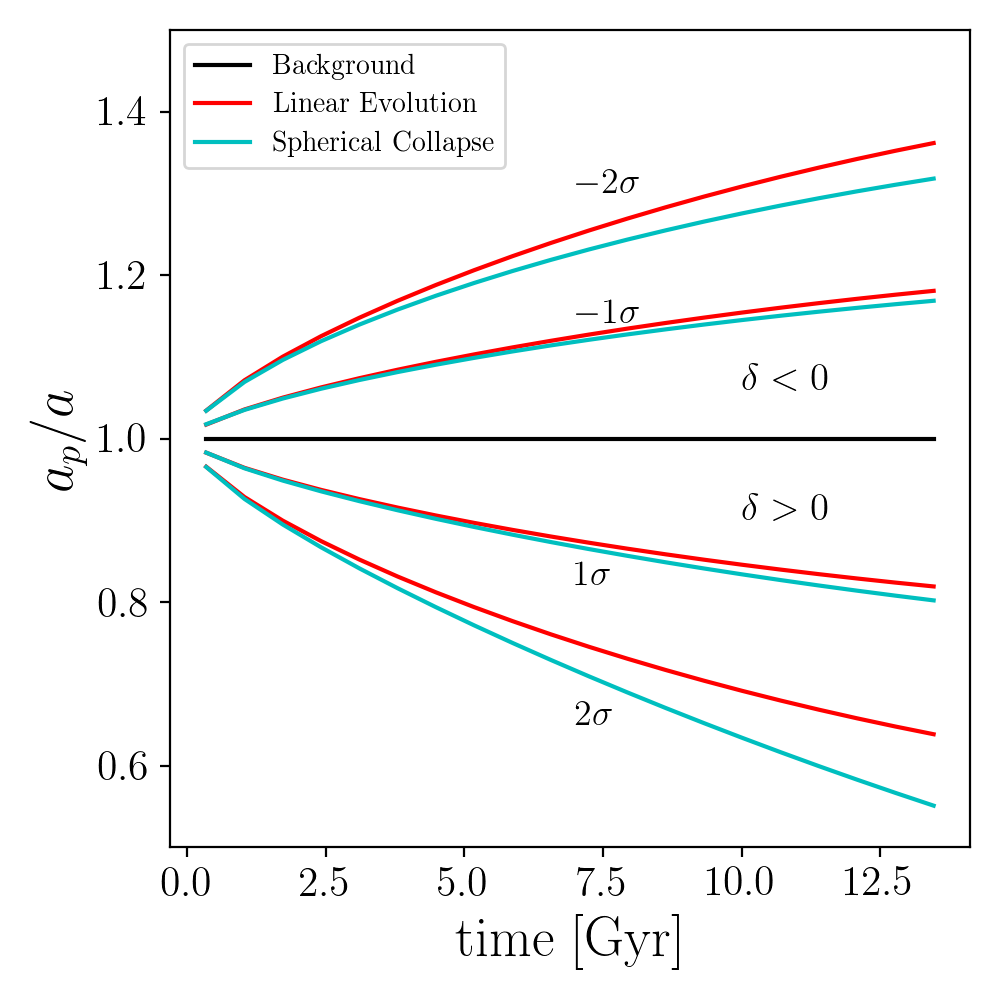}
\caption{Illustration for the scale factor of a local sphere of the expected fluctuations for a sphere of radius 20Mpc, with over/under density as a function of time. The linear model (red) and top-hat model (blue) are plotted as a ratio with respect to the background universe (black). The variance for $\delta$ is computed using Eq.~(\ref{eq:PkconvW}), and we plot curves for both 1 and 2$\sigma$ fluctuations as labeled.}
\label{fig:collapse}
\end{center}
\end{figure}

The process is illustrated in Fig.~\ref{fig:collapse}, which shows the evolution of the scale factor of the patch $a_p$ relative to the background universe $a$. Due to inhomogeneity, which leads to a change in curvature, the patch experiences a different evolution compared with the background. To get the curvature, we need to link the early behaviour of the patch where the linear theory is valid to that of the background. Once we have the curvature we can evolve the patch forwards using the Friedmann equation for the patch until we match the present-day time of the patch to the background universe. At this time, the scale factor of the patch $a_p$ is different from unity. The observer inside the patch can redefine the scale factor just like the background universe, and this requires a scaling of the cosmological parameters including $H_{0}$. Thus we can link a fluctuation in $\delta$ to a fluctuation in $H_0$. While the previous two methods used perturbation theory linking the value of $\delta_0$ to changes in the recovered cosmological parameters, the spherical top-hat allows for the full non-linear evolution of the patch, albeit within the context of a homogeneous spherical region.

\subsection{Numerical simulations}

Within a cosmological N-body simulation, all distances and velocities are measured with respect to the background model. Provided that the peculiar velocities are included, the luminosity distance and redshift will be the same as if we had followed the evolution of the patch using the patch or background cosmologies. To see why we need to include peculiar velocities, consider a patch within a simulation with a perturbed initial density (as in the SSC discussion above): the peculiar velocities with respect to the background of objects at the edges of a patch simply tell us the difference between the expansion rate of that patch when considered within the background or within the cosmological model appropriate to that patch. The change in distances similarly tell us the change in the size of the patch. Thus, in order to measure local variations in $H_{0}$ for the simulation we simply need to estimate the variance in $H_{0}$ measurements using a local distance ladder \cite{Wojtak2014, Odderskov2014, Odderskov2017, Wu2017}. 

For this purpose, we rely on a large-scale N-body simulation from UNIT \footnote{\url{http://www.unitsims.org/}} \cite{Chuang2019} and the identified halo catalog. We start with an observer residing in a randomly chosen dark matter halo of $M\sim10^{12-15}h^{-1}M_{\odot}$ in the simulation box, taken to mimic the location of the Milky-Way galaxy. The variance caused by peculiar velocity on the $H_{0}$ measurement of the observer can be estimated using the method from \cite{Wu2017}. In particular, the distance ladder method requires calibration of the absolute magnitude $M_{B}$ of SNe Ia,  and the cosmic expansion parameter $a_{B}$, defined as in Eq.~(\ref{eq:local_H0}). Observations of Cepheids can determine the value of $M_{B}$, and $a_{B}$ can be constrained from the Hubble diagram which is affected by the peculiar velocity of the dark matter halos. Its variance can be estimated via \cite{Wu2017}
\begin{equation}\label{eq:Delta_ab}
    \Delta a_{B} = \frac{1}{N}\sum_{i=1}^{N}\frac{1}{\ln{10}}\frac{v_{i}}{r_{i}H_{0}},
\end{equation}
where $v_{i}$ is the peculiar velocity along the line-of-sight, $r_{i}$ is the distance to the SNe Ia (or host halos), the summation is through all the halos within distance $R$ to the observer. Then we can convert this uncertainty to $\Delta H_{0}$ by randomly choosing a large number of observers.

The previous analytical methods are based on the local inhomogeneity, computed from the variance of the entire sub-volume of the top-hat window function. To provide a fair comparison using simulation, we include all the halos around the observer within some distance. In Section~\ref{sec:Pantheon-sim} we consider a more direct way of matching the Pantheon sample geometry using a simulation-based method, where we use the positions of the SNe Ia and match them to the neighboring halos and inherit the peculiar velocity of these halos. This will match the measured $H_{0}$ and thus the variance calculated here provided that we use the correct effective volume.

\subsection{Comparison of models}

\begin{figure}
\begin{center}
\includegraphics[width=8.5cm]{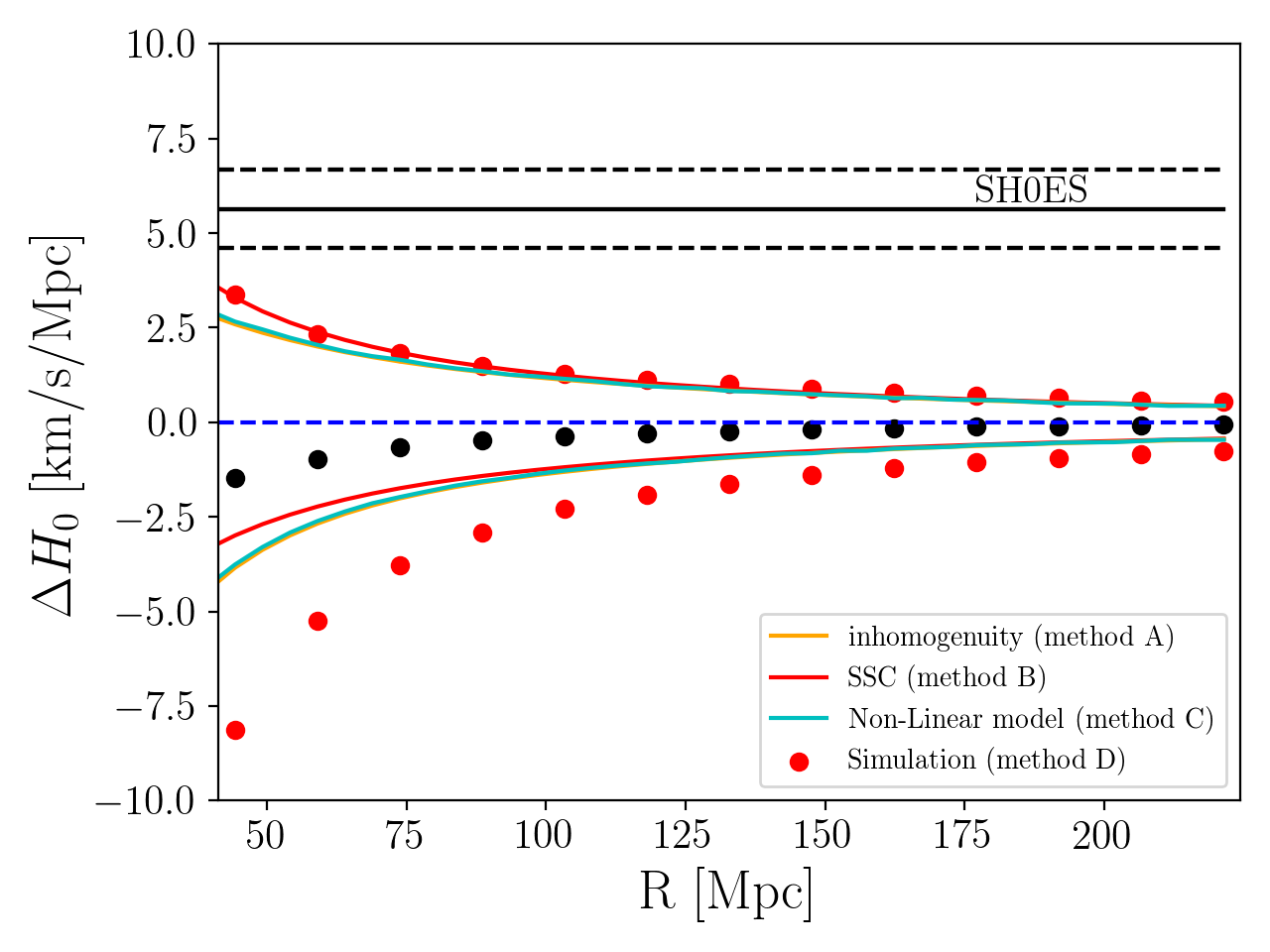}
\caption{Uncertainty of $H_{0}$ measurement caused by local inhomogeneities as a function of scale, comparing analytical and simulation based methods. For analytical methods, the scale $R$ is used to compute over-density from Eq.~(\ref{eq:PkconvW}) assuming a Planck 2016 \cite{Planck_2016} cosmology to be consistent with the simulation. For the simulation-based method, $R$ is the maximum scale to include halos around the observer. For analytical methods, the lines correspond to $+\delta$ and $-\delta$, while for simulation-based method, the dots correspond to 16\% (red), 50\% (black) and 84\% (red) intervals by randomly choosing thousands of observers in the simulation.}
\label{fig:DeltaH0_density}
\end{center}
\end{figure}

In Fig.~\ref{fig:DeltaH0_density}, we show the change in $H_{0}$ estimated using the different methods described above for $1\sigma$ shifts in $\delta$ as a function of scale. For the analytical methods, the scale $R$ is used to compute the change in the overdensity (Eq.~\ref{eq:PkconvW}), while for the simulation-based analysis, $R$ is the maximum distance to include halos. In this work, we apply the Planck 2016 cosmology \cite{Planck_2016} throughout the computation, to be consistent with the UNIT simulation. The result shows that the different analytical methods give consistent estimates of $\Delta H_{0}$ over a wide range of scales. The simulation based method is the most discrepant, giving an estimate that is in agreement at large scales, but shows some deviations at small scales. At small scale, the average $H_{0}$ prefers lower values as expected since the halos are formed in preferentially over-density regions with large scale infall velocity \cite{Wojtak2014, Wu2017}. The distribution of over(under)-density becomes more skewed towards smaller scales due to non-linear evolution of the density field, and this causes wider distribution for positive $\delta$ and a preference for lower values of $H_{0}$. Note that in this analysis, the observer is located in the CMB rest frame \cite{Wojtak2014}, and thus the average measurement of $H_{0}$ is close to the background value. If the observer is in included in its own frame, the impact from its peculiar velocity will shift the overall distribution, but the level of uncertainty is not affected significantly, see the comparison from \cite{Wojtak2014}. None of the methods predicts fluctuations in $H_0$ that could explain the current tension between observations unless the local measurement is made within a very small volume.

\section{Sample variance for Pantheon}  \label{sec:Pantheon-sim}

In order to use the perturbation theory based estimates of sample variance, we need an estimate of the average overdensity in the patch covered by the SNe. I.e. to determine the radius $R$ to use in Eq.~\ref{eq:PkconvW}, we need to know the effective volume of the region of the Universe probed by distance measurements from all of the SNe in a particular sample. In general, this is far smaller than might be thought given the maximum SN redshift. For the Pantheon sample \citep{Pantheon} containing $\sim$1000 SN, there are many more SN at low redshift and the likelihood calculation includes each SN approximately equally. If we assume that each SN $i$ contributes equally, and the LOS traces the density within a region of influence $R_i({\bf x})$ with volume $V_i$, then the set of SN traces a weighted region $w({\bf x}) = \sum_i [R_i({\bf x})/V_i]$, where the $V_i$ normalises the region to provide equal weighting for each SN. We can define an effective volume for the sample $V_{\rm eff} = [\int w({\bf x})]^2 / \int w^2({\bf x})$ assuming that each small volume element contributes equally to the sample variance. We can then estimate $R$ as the radius of a sphere with the same volume.

If we assume that the region of influence of every LOS to a SN is a sphere centred on the mid-point of the LOS and touching the SN and us, then for the full Pantheon sample, using the formula above, the effective volume corresponds to a sphere of radius $\sim270$Mpc, significantly smaller than the total volume covered by the sample. Limiting to SN with $z<0.15$ corresponding to a comoving distance of $\sim610$Mpc, the SNe sample within this range only probes a volume of radius $\sim120$Mpc. Having the region of influence defined in this way is motivated by the spherical top-hat model where we consider spherical patches of the Universe as mini-Universes, each behaving according its internal density. If instead we were to assume that the volume of influence occupies a smaller volume around the LOS to each SN, then the effective volume would be smaller. We consider $\sim170$Mpc to be a conservative estimate, which matches that adopted in previous analyses: \cite{Wu2017} noted that the number distribution of SNe used for $H_{0}$ measurement peaks at $z\sim0.04$, a scale of $\sim170$Mpc. Considering all the halos within this distance of the observer, we estimate from Fig.~\ref{fig:DeltaH0_density} that the uncertainty of $H_{0}$ caused by peculiar velocity is about 1\%, consistent with earlier studies \cite{Wojtak2014} and insufficient to explain the Hubble tension.

Within a time-slice of an N-body simulation, we can directly incorporate the spatial distribution of the Pantheon data, without having to define the volume separately. The ideal framework for such an investigation would be a light-cone simulation, which provide the correct age for structures given their distance from us. Creating these requires a careful extraction of particles \cite{L-PICOLA}, or interpolation of merger-tree halos in terms of their position and velocity with fine time steps \cite{Merson_2013}. The single time-slice simulation adopted here is conservative as it assumes that all structure has evolved to present day, leading to larger fluctuations than for a lightcone.  We consider the Pantheon sample and, for each SN with $0.023<z<0.15$, we use the Pantheon redshift to determine the distance using the background cosmological model. This redshift range is chosen to match the primary fit in the distance ladder analysis \cite{localH0-2016,localH0-2021}. For this, we ignore that there is a component of the redshift from the observed peculiar velocity which will change the sample slightly, but not affect our results. We then assign each SN to the nearest dark matter halo. This results a subsample of the haloes compared with the previous method.  We repeat this process more than $10^{4}$ times to get a distribution of $H_{0}$ values for different observers and rotations of the Pantheon sample. This then gives an estimate of the local sample variance of $H_{0}$ due to inhomogeneities along the LOS. Although our calculation differs from previous investigations \cite{Wu2017} in terms of simulation volume, mass resolution, halo mass cut and other details, we find that the impact from peculiar velocity on $H_{0}$ measurement is at a similar level of $\sim0.4~\text{km}~\text{s}^{-1}\text{Mpc}^{-1}$. This result is comparable to but slightly lower than the prediction using analytical models assuming a scale of $170$ Mpc (Fig.~\ref{fig:DeltaH0_density}), since the lower redshift cut ($z>0.023$) removes nearby halos that can dominate the variance (Eq. \ref{eq:Delta_ab}). 

\section{Internal measurements of $H_0$ variations}

In this section, we investigate the sample variance for $H_{0}$ estimated for the latest distance ladder measurements by considering differences obtained when splitting the sample. This serves as an internal examination of variance within the Cepheid and SNe data. As we are interested in spatial variations, rather than remove individual objects as in a Jacknife approach, we instead consider removing angular regions. 

\subsection{(An)isotropic $H_{0}$ measurement}\label{sec:H0_spatial}

We investigate the variance of $H_{0}$ measurement based on the method from \cite{localH0-2009, localH0-2016, localH0-2021}. With observations of both standardizable Cepheids and SNe Ia, we can construct a three-rung distance ladder up to redshift where the cosmic expansion is dominant. Then we fit the relations characterizing the luminosity and distance of these objects through a likelihood analysis. The result can include the fiducial luminosity of Cepheids, SNe Ia and $H_{0}$. In practice, we measure Hubble constant via
\begin{equation}\label{eq:local_H0}
    \log{H_{0}} = 0.2M_{B}^{0}+a_{B}+5,
\end{equation}
where $M_{B}^{0}$ is the fiducial luminosity of SN Ia, and $a_{B}$ is the parameter describing luminosity distance and redshift \cite{localH0-2009}. Therefore $H_{0}$ can be fully determined with these two parameters. 

The first two rungs of the distance ladder constrain the absolute magnitude of SN Ia, while the third rung determines the intercept $a_{B}$ of the redshift-distance relation. The ladder parameters $M_{B}^{0}$ and $a_{B}$ can be constrained separately, as in the method of \cite{localH0-2016} (hereafter R16) or simultaneously \cite{localH0-2021}. In the following, we apply the R16 approach and have tested that our main results are not affected significantly by different methods. The equations for the calculation are well described in a compact matrix form as Section 2 of \cite{localH0-2021} and removing the corresponding columns and rows for SNe Ia can return to a R16 style analysis easily.   

For the first two rungs of the distance ladder, we use the newly released SH0ES data \cite{localH0-2021}, including the 37 Cepheids hosts and their SNe Ia. The external constraints and anchors are also from \cite{localH0-2021} (see table 4 of \cite{localH0-2021} for instance). For the SNe Ia in the Hubble flow, we use the Pantheon supernovae data \cite{Pantheon}\footnote{The Pantheon plus data was not fully publicly available when this work was started.} within the redshift range of $0.023<z<0.15$ to be consistent with \cite{localH0-2021}. This data has 40 fewer objects compared with the latest Pantheon plus compilation \cite{PantheonPlus} but this shouldn't impact our main result significantly. With the combined SH0ES Cepheids and Pantheon SNe Ia data, we find H$_{0}=72.74\pm1.08$ km s$^{-1}$Mpc$^{-1}$, a 0.3$\sigma$ offset compared with \cite{localH0-2021}.

In this work, we perform a simple resampling analysis by splitting the data based on their angular positions. We first use the healpy \cite{healpix, Zonca2019}\footnote{\url{http://healpix.sourceforge.net}} code to pixelize the sky with parameter NSIDE=4, resulting in 192 equal-sized pixels that are uniformly distributed on the sky. For each pixel, we define the center as the new North pole and select objects (Cepheids and/or SNe Ia) within an angular separation smaller than 90 degrees. This forms a subset that only distribute in one hemisphere, and the rest of the data produces the other hemisphere. Throughout the analysis, we only apply this sub-sampling to the host galaxies in the second rung (galaxies that have both Cepheids and SNe Ia) and SNe Ia in the third rung. The external constraints and Cepheids in the anchors are not split based on their angular positions for all analyses. 

\begin{figure*}
\begin{center}
\includegraphics[width=17cm]{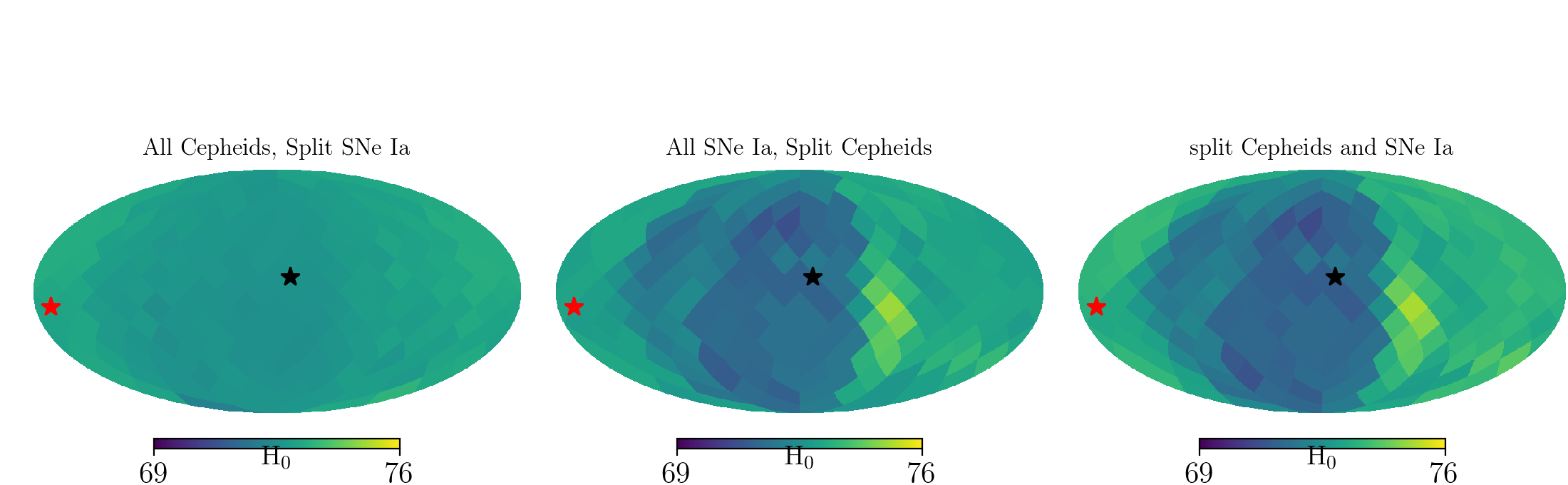}
\includegraphics[width=17cm]{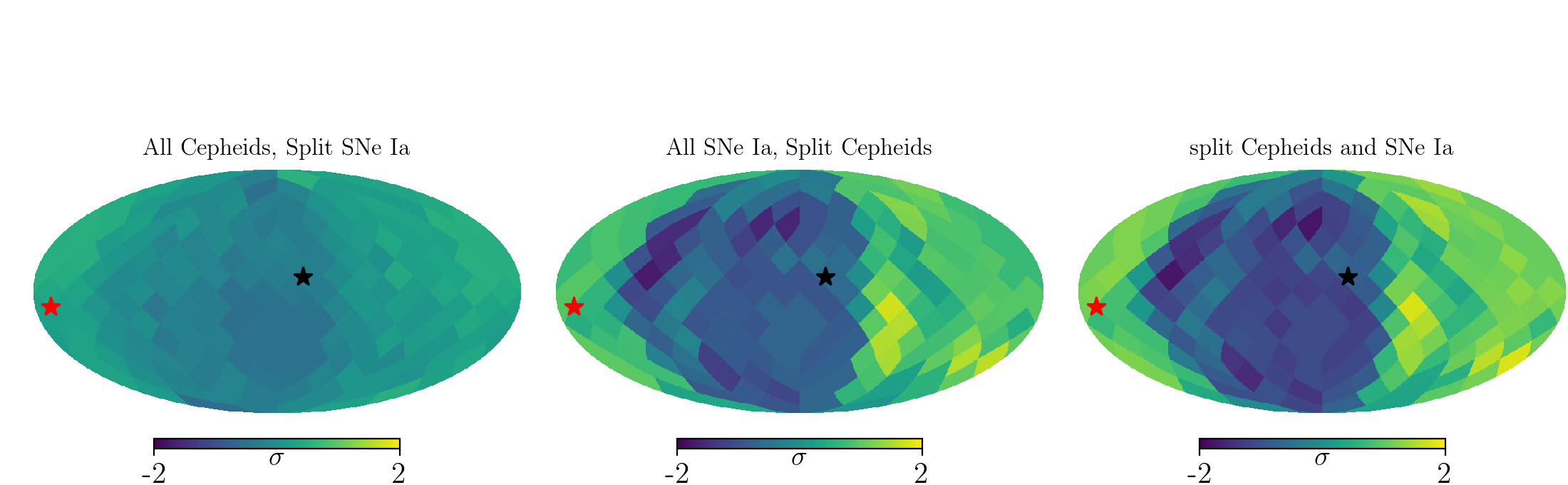}
\caption{$\mathbf{Top:}$ Measurement of $H_{0}$ using SH0ES Cepheids and Pantheon SNe Ia as a function of angular coordinates. The value of each pixel represents the result using a subset of data within an angular separation smaller than 90$^{\circ}$ around the center of the pixel. $\mathbf{Bottom:}$ Offset of $H_{0}$ between two hemispheres, normalized by the uncertainty, i.e. Eq.~(\ref{eq:sigma}). $\mathbf{Left:}$ Only SNe Ia are split into hemispheres; $\mathbf{Middle}:$ Only Cepheid fields are split into hemispheres; $\mathbf{Right:}$ Both Cepheid fields and SNe Ia are split into hemispheres.}
\label{fig:H0map}
\end{center}
\end{figure*}

In the top row of Fig.~\ref{fig:H0map}, we present measurements of $H_{0}$ where we consider three variants of the split performed: (1) use all 37 Cepheid fields but split SNe Ia (left), (2) use all SNe Ia but split Cepheids (middle), (3) split Cepheid fields and SNe Ia simultaneously. Note that the value of each pixel represents the measurement over the entire hemisphere, so measurements from nearby pixels are correlated. This leads to a smooth pattern for the measurements, since the neighboring pixels have significant overlaps of their hemispheres and thus the variations change gradually from pixel to pixel. The results show a few features: first, the SNe Ia sample has a much weaker variation (72.08 to 73.54 km s$^{-1}$Mpc$^{-1}$) than the Cepheid fields (70.72 to 74.93 km s$^{-1}$Mpc$^{-1}$). This is explained by the relative sample sizes (there are more Cepheids than SNe Ia, but the spatial variation is based on Cepheid hosts instead of Cepheids). Second, both data sets seem to indicate similar direction preference. When we split both Cepheid fields and SNe Ia simultaneously, the pattern is enhanced slightly. For comparison, we also plot the CMB dipole direction (168$^\circ$ for right ascension and -7$^\circ$ for declination \cite{Planck_CMB_dipole}, red star) and its opposite (black star). The observation implies that roughly, the hemisphere along the CMB dipole direction gives a higher measurement of $H_{0}$ than the opposite direction. Depending on particular direction, the measurement varies from 70.61 to 75.08 km s$^{-1}$Mpc$^{-1}$. This amount of variation was also found in \cite{Kenworthy_2022} when only the first two rungs of the distance ladder were used to measure $H_{0}$, and larger than just simply halving the data set (Section \ref{sec:H0error}). 

In order to evaluate the significance of the variation, we define a metric that can normalize the difference of $H_{0}$ in two hemispheres by their uncertainties
\begin{equation}\label{eq:sigma}
    \sigma = \frac{H_{0,A}-H_{0,B}}{\sqrt{\sigma_{H_{0,A}}^{2}+\sigma_{H_{0,B}}^{2}}},
\end{equation}
where the subscripts ``A" and ``B" denote the two opposite hemispheres respectively. This allows for anisotropic distribution of both Cepheids and SNe, which give rise to varying errors for the $H_0$ measurements from different hemispheres. We plot the variance weighted measurements in the bottom row of Fig.~\ref{fig:H0map}. The overall pattern is consistent with the top row. When both SNe Ia and Cepheid fields are split into hemispheres (lower right panel), we find that variation is less than $2\sigma$ for all the directions that we consider. The maximum asymmetry is around $1.78\sigma$. We note that this type of asymmetry is also found in other studies, for instance \cite{Krishnan_2021, Krishnan_2022,  Luongo_2022} based on SNe Ia and QSO data. The amplitude of the asymmetry can vary, but multiple analyses show a higher $H_{0}$ value in the CMB dipole direction. This may come from the effect of peculiar velocities at low redshift and a thorough examination of systematics may be of importance. We now consider the statistical significance of this offset.

\subsection{Statistical significance of the asymmetry}

\begin{figure*}
\begin{center}
\includegraphics[width=17cm]{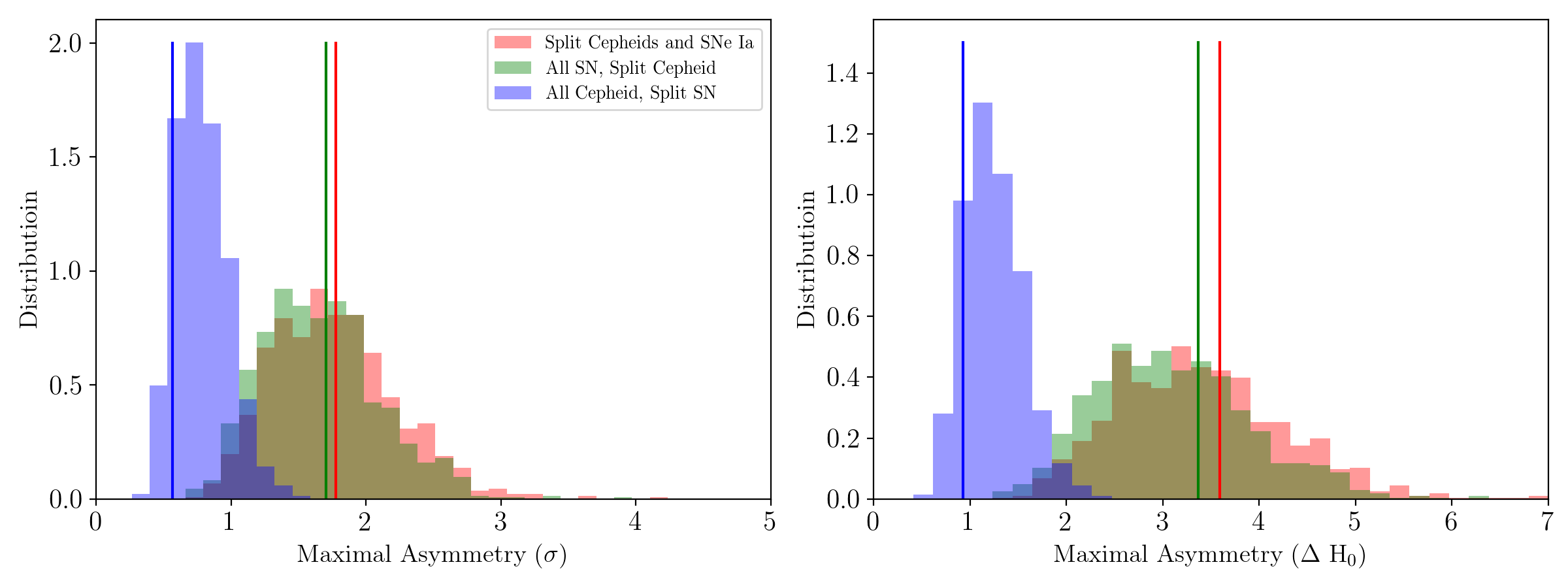}
\caption{Distribution of the maximal hemispherical asymmetry of $H_{0}$ measurement from one thousand simulations, using current observations from SH0ES and Pantheon. The vertical lines represent the value from real data. $\mathbf{Left:}$ Results normalized by uncertainty (Eq.~\ref{eq:sigma}); $\mathbf{Right:}$ Raw measurement in unit of km s$^{-1}$Mpc$^{-1}$. }
\label{fig:maximum_distribution}
\end{center}
\end{figure*}

In order to access the significance of the observed 1.78$\sigma$ difference for the maximal asymmetry, we generate simulations for both Cepheids and SNe Ia following a variant of the method described in Section~\ref{sec:Pantheon-sim} and repeat the above analysis. In particular, we choose the best-fit parameters using the entire SH0ES Cepheids and Pantheon data to get the theoretical predictions of the magnitude for Cepheids and/or SNe Ia, then we add noise generated from the observed covariance matrix as the simulated data vector. Note that in this process, the SNe Ia in the Cepheid hosts are also included in the simulation, and the covariance matrix for the Pantheon sample include contributions from both statistical error and systematics. In addition, the angular coordinates of the Cepheid fields and SNe Ia are randomly distributed on the sky. For each simulated data set, we perform an anisotropic measurement as in the previous section and find the maximal difference of $H_{0}$ and $\sigma$ (Eq.~\ref{eq:sigma}). The distribution from one thousand simulations is presented in Fig.~\ref{fig:maximum_distribution} considering three scenarios: split SNe Ia only (blue), split Cepheid fields only (green) and split both (red). The result from SH0ES and Pantheon is the vertical lines with the same color. For comparison, the two panels show the difference of $H_{0}$ with (left) and without (right) normalization.

The result clearly shows that the observed asymmetry is statistically likely, for both SNe Ia and Cepheid fields. A more quantitative evaluation such as $p-$value can be easily computed as the fraction of the simulation with more extremal asymmetry. In our analysis, it is above 0.3, indicating that the asymmetry with this amplitude is consistent with statistical fluctuations. This is in somewhat tension with other results based on QSO and SNe Ia \cite{Krishnan_2022} that find that the variation is more significant. 

\subsection{Dependence of $\Delta H_{0}$ on the number of Cepheid fields and SNe Ia} \label{sec:H0error}

\begin{figure}
\begin{center}
\includegraphics[width=9cm]{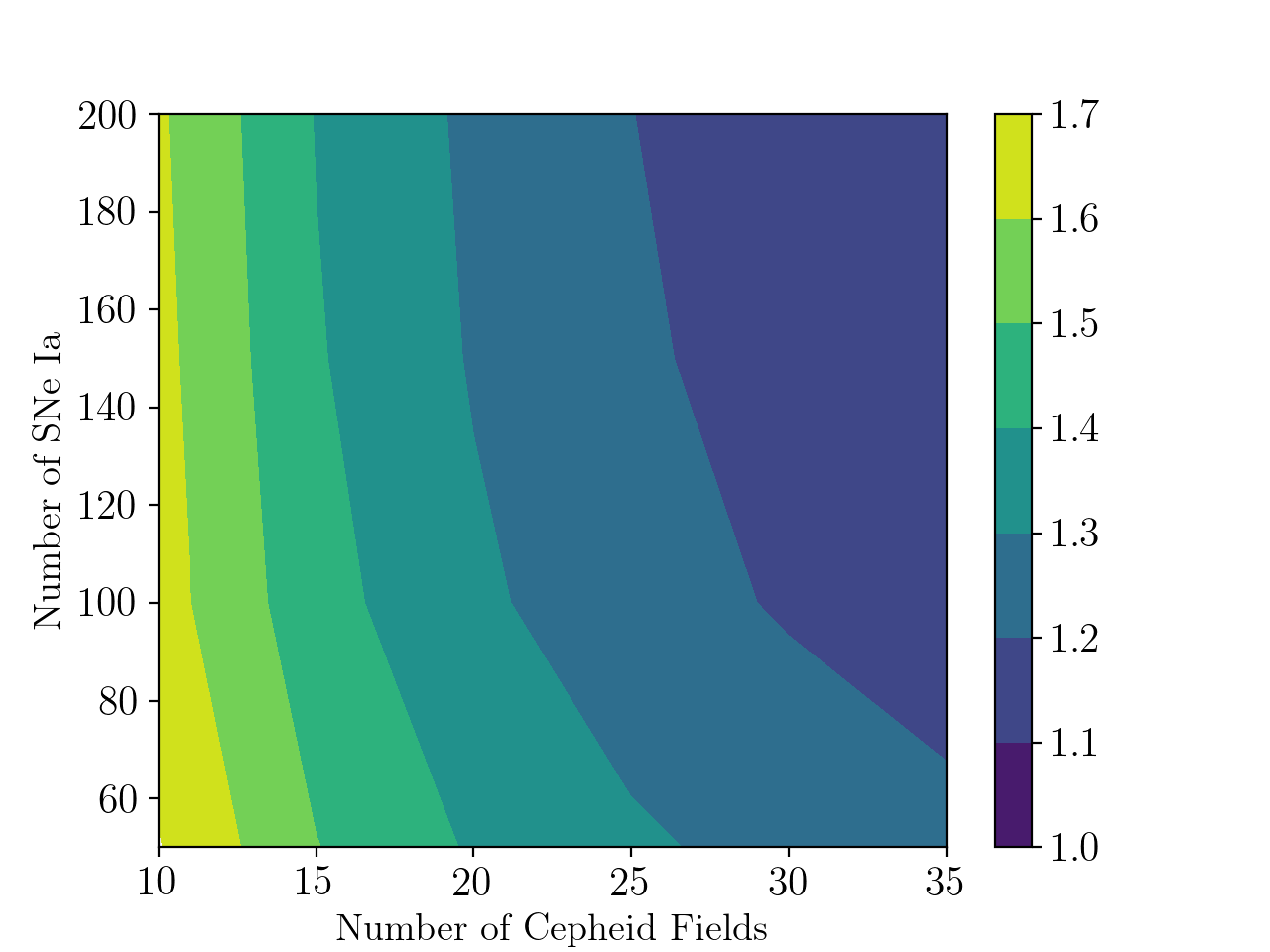}
\caption{Contour plot of the uncertainty of $H_{0}$ using distance ladder, as a function of the number of Cepheid fields and SNe Ia. The value represents the average of multiple subsamplings based on current data. }
\label{fig:errorMap}
\end{center}
\end{figure}

Our previous analysis reveals a variation of $\Delta H_{0}\sim4$ km s$^{-1}$Mpc$^{-1}$ between opposite hemispheres. We quantify this offset using current data in this section.

The number of Cepheid fields is of critical importance in the local measurement of $H_{0}$, with the accuracy improving from 4.8\% (or 3.6 km s$^{-1}$Mpc$^{-1}$) with 6 Cepheid hosts \cite{localH0-2009}, to 2.4\% (or 1.74 km s$^{-1}$Mpc$^{-1}$) with 19 hosts \cite{localH0-2016} and 1.4\% (or 1.04 km s$^{-1}$Mpc$^{-1}$) with 37 hosts \cite{localH0-2021}. The size of the current SH0ES data set enables an exploration of the scaling relation between $\Delta H_{0}$ and the number of data points. In order to do so, we perform a jackknife-like sub-sampling method by randomly choosing a number of Cepheid fields and SNe Ia and remeasure $\Delta H_{0}$ for each realization. Since the number of Cepheid in one host can vary significantly, for instance M101 has 259 fits, while N0105 only has 5 fits, we repeat this process 500 times with different random seeds and take the average to represent the uncertainty of $H_{0}$. Fig.~\ref{fig:errorMap} displays the contour plot as a function of the number of Cepheid fields and SNe Ia respectively. 

The result clearly shows the monotonic increase of accuracy with more and more Cepheid fields, especially when the number of Cepheid fields is low. Compared with Cepheids, the number of SNe Ia contributes less in the determination of $H_{0}$. However, both objects become saturated at certain threshold, i.e. the accuracy of $H_{0}$ increases more and more slowly with higher number of Cepheids and SNe Ia, which is not surprising. On the other hand, the average error of $H_{0}$ is around 1.3-1.4 km s$^{-1}$Mpc$^{-1}$ with $\sim20$ Cepheid fields and $\sim120$ SNe Ia. Assuming the independence of each Cepheid field and SNe Ia, the $\sim4$ km s$^{-1}$Mpc$^{-1}$ difference found by maximizing the measurement in hemispheres in previous section is statistically likely.

\section{Discussion and Summary}

We have considered how local inhomogeneities affect measurements of $H_{0}$. We have compared the ``standard" method which considers fluctuations in the luminosity distance with methods that consider the parameters of different perturbed patches in the Universe. Specifically we present a new application of a method previously developed to correct for super-sample covariance in numerical N-body simulations, and present a new method based on spherical top-hat regions similar to that used to measure the critical density for collapse.

We compare these analytical methods with a simulation-based approach and find that they give similar estimates of $\Delta H_{0}$ over a wide range of scales. A full relativistic description of a locally perturbed background within a cosmological model is provided by the Lemaitre-Tolman-Bondi (LTB) model. This model provides a more accurate framework for local perturbations, but does not predict the distribution of expected perturbation sizes. It simply provides a model to link the perturbed spherical top-hat regions considered here with the background. Postulating that we live in a LTB Universe requires too large a local underdensity to explain Dark Energy \cite{LTB-DE1,LTB-DE2}, but it has also been used to understand sample variance for $H_{0}$ measurements \cite{LTB-H0,LTB-H02}. The key issue for such models is the same that we address here and requires similar techniques: statistically, how likely is it that we see a large enough fluctuation to address the Hubble tension? This is the question we have tried to address, and thus we consider our analysis to be applicable to LTB - based analyses as well as the more general ideas we have used to understand the sample variance of local $H_{0}$ measurements.

The size of the patch in our analysis is of critical importance since it determines the amplitude of the density contrast and thus $\Delta H_{0}$. One may simply choose the maximum redshift of SNe to define spherical mini-universe. However, the Cepheids and SNe Ia are not randomly sampling the underlying density field: their distribution is sparse and highly non-uniform in both angular and radial directions which significantly reduces the probed volume. On the other hand, one could directly estimate $\delta$ using the simulated density field along the LOS to each SN and simply take the average. We find that this method usually gives a higher variance for $\delta$ than a sphere of $\sim100 h^{-1}$ Mpc. This indicates that the scale of $\sim~120h^{-1}$Mpc used in previous studies \cite{Wu2017} to estimate the density contrast may be a conservative choice. The measured $\Delta H_{0}$ may just represent a lower limit and the actual variance due to this small volume can be larger and even accommodate the observed $H_{0}$ tension. However, this may require a better and robust estimate of the volume that Cepheids and SNe Ia data have sampled, which can be non-trivial. 

In addition to sample variance, we note that there are other effects that also contribute to the $\Delta H_{0}$ estimate. One is the gravitational redshift resulted from the difference of gravitational potential between the observer and SNe. Although the amplitude of this effect is rather small, $\sim10^{-5}$, earlier studies such as \cite{Wojtak2015} show that ignoring this effect can lead to a 1\% change in the constraints on cosmological parameters. We model this effect in our simulation-based analysis by approximately assuming that the gravitational redshift can change the total velocity by a factor of $(1+z_{g})$, where $z_{g}$ is determined by the difference of gravitational potential between observer and SNe, which depends on the density contrast. Assuming that the typical amplitude of $\delta$ is a few percent, this effect is negligible in estimating $\Delta H_{0}$.

Another effect comes from the redshift uncertainty of the SN/host galaxies. The data from BOSS survey shows that this uncertainty grows with redshift \cite{Bolton2012} and the amplitude is at the level of a few tens of km/s. Although the SNe sample is at lower redshift and this redshift uncertainty is not fully relevant to the host galaxies of the SNe, we can artificially model this effect in the $\Delta H_{0}$ estimate using simulation. We add an independent velocity component into the peculiar velocity by random draw from a Gaussian distribution with dispersion $\sim50$km/s. Note that this dispersion is higher than the measurement using repeat observations from \cite{Bolton2012}. The result shows that this additional term also has a negligible impact on $\Delta H_{0}$. This is not surprising since the Hubble flow and peculiar velocity dominate the variance. 

As a complementary analysis to understanding the sample variance due to local inhomogeneity using models, we have also investigated the uncertainty of $H_{0}$ measurement using internal methods applied to real data. We split the data of Cepheid and SNe Ia based on their angular coordinates and investigate the spatial variation of $H_{0}$. By changing the angular direction to define hemispheres, we find that the maximal difference of $H_{0}$ between two opposing hemispheres is around 4 km s$^{-1}$Mpc$^{-1}$, larger than the typical uncertainty quoted using half of the data. We further examine the significance by running a large amount of simulations and obtain a distribution of this maximal asymmetry. The result shows that the amplitude of the signal from real data is consistent with statistical fluctuations. However, it is interesting that the direction of this maximal asymmetry is close to the CMB dipole direction, similar to the results from literature \cite{Krishnan_2021, Krishnan_2022,  Luongo_2022} using data other than Cepheids, which may indicate that the calibration of the data can be improved to further tighten the constraint on $H_{0}$. 

\begin{acknowledgments}

We thank the SH0ES team for making their data publicly available. ZZ thanks Niayesh Afshordi for helpful discussions. Research at Perimeter Institute is supported in part by the Government of Canada through the Department of Innovation, Science and Economic Development Canada and by the Province of Ontario through the Ministry of Colleges and Universities.

This research was enabled in part by support provided by Compute Ontario (computeontario.ca) and the Digital Research Alliance of Canada (alliancecan.ca).

\end{acknowledgments}

\appendix

\bibliography{SN-SSC}

\providecommand{\noopsort}[1]{}\providecommand{\singleletter}[1]{#1}%
\begin{thebibliography}{45}%
\makeatletter
\providecommand \@ifxundefined [1]{%
 \@ifx{#1\undefined}
}%
\providecommand \@ifnum [1]{%
 \ifnum #1\expandafter \@firstoftwo
 \else \expandafter \@secondoftwo
 \fi
}%
\providecommand \@ifx [1]{%
 \ifx #1\expandafter \@firstoftwo
 \else \expandafter \@secondoftwo
 \fi
}%
\providecommand \natexlab [1]{#1}%
\providecommand \enquote  [1]{``#1''}%
\providecommand \bibnamefont  [1]{#1}%
\providecommand \bibfnamefont [1]{#1}%
\providecommand \citenamefont [1]{#1}%
\providecommand \href@noop [0]{\@secondoftwo}%
\providecommand \href [0]{\begingroup \@sanitize@url \@href}%
\providecommand \@href[1]{\@@startlink{#1}\@@href}%
\providecommand \@@href[1]{\endgroup#1\@@endlink}%
\providecommand \@sanitize@url [0]{\catcode `\\12\catcode `\$12\catcode
  `\&12\catcode `\#12\catcode `\^12\catcode `\_12\catcode `\%12\relax}%
\providecommand \@@startlink[1]{}%
\providecommand \@@endlink[0]{}%
\providecommand \url  [0]{\begingroup\@sanitize@url \@url }%
\providecommand \@url [1]{\endgroup\@href {#1}{\urlprefix }}%
\providecommand \urlprefix  [0]{URL }%
\providecommand \Eprint [0]{\href }%
\providecommand \doibase [0]{http://dx.doi.org/}%
\providecommand \selectlanguage [0]{\@gobble}%
\providecommand \bibinfo  [0]{\@secondoftwo}%
\providecommand \bibfield  [0]{\@secondoftwo}%
\providecommand \translation [1]{[#1]}%
\providecommand \BibitemOpen [0]{}%
\providecommand \bibitemStop [0]{}%
\providecommand \bibitemNoStop [0]{.\EOS\space}%
\providecommand \EOS [0]{\spacefactor3000\relax}%
\providecommand \BibitemShut  [1]{\csname bibitem#1\endcsname}%
\let\auto@bib@innerbib\@empty
\bibitem [{\citenamefont {{Riess}}\ \emph {et~al.}(2016)\citenamefont
  {{Riess}}, \citenamefont {{Macri}}, \citenamefont {{Hoffmann}}, \citenamefont
  {{Scolnic}}, \citenamefont {{Casertano}}, \citenamefont {{Filippenko}},
  \citenamefont {{Tucker}}, \citenamefont {{Reid}}, \citenamefont {{Jones}},
  \citenamefont {{Silverman}}, \citenamefont {{Chornock}}, \citenamefont
  {{Challis}}, \citenamefont {{Yuan}}, \citenamefont {{Brown}},\ and\
  \citenamefont {{Foley}}}]{localH0-2016}%
  \BibitemOpen
  \bibfield  {author} {\bibinfo {author} {\bibfnamefont {A.~G.}\ \bibnamefont
  {{Riess}}}, \bibinfo {author} {\bibfnamefont {L.~M.}\ \bibnamefont
  {{Macri}}}, \bibinfo {author} {\bibfnamefont {S.~L.}\ \bibnamefont
  {{Hoffmann}}}, \bibinfo {author} {\bibfnamefont {D.}~\bibnamefont
  {{Scolnic}}}, \bibinfo {author} {\bibfnamefont {S.}~\bibnamefont
  {{Casertano}}}, \bibinfo {author} {\bibfnamefont {A.~V.}\ \bibnamefont
  {{Filippenko}}}, \bibinfo {author} {\bibfnamefont {B.~E.}\ \bibnamefont
  {{Tucker}}}, \bibinfo {author} {\bibfnamefont {M.~J.}\ \bibnamefont
  {{Reid}}}, \bibinfo {author} {\bibfnamefont {D.~O.}\ \bibnamefont {{Jones}}},
  \bibinfo {author} {\bibfnamefont {J.~M.}\ \bibnamefont {{Silverman}}},
  \bibinfo {author} {\bibfnamefont {R.}~\bibnamefont {{Chornock}}}, \bibinfo
  {author} {\bibfnamefont {P.}~\bibnamefont {{Challis}}}, \bibinfo {author}
  {\bibfnamefont {W.}~\bibnamefont {{Yuan}}}, \bibinfo {author} {\bibfnamefont
  {P.~J.}\ \bibnamefont {{Brown}}}, \ and\ \bibinfo {author} {\bibfnamefont
  {R.~J.}\ \bibnamefont {{Foley}}},\ }\href {\doibase
  10.3847/0004-637X/826/1/56} {\bibfield  {journal} {\bibinfo  {journal}
  {\apj}\ }\textbf {\bibinfo {volume} {826}},\ \bibinfo {eid} {56} (\bibinfo
  {year} {2016})},\ \Eprint {http://arxiv.org/abs/1604.01424} {arXiv:1604.01424
  [astro-ph.CO]} \BibitemShut {NoStop}%
\bibitem [{\citenamefont {{Riess}}\ \emph {et~al.}(2021)\citenamefont
  {{Riess}}, \citenamefont {{Yuan}}, \citenamefont {{Macri}}, \citenamefont
  {{Scolnic}}, \citenamefont {{Brout}}, \citenamefont {{Casertano}},
  \citenamefont {{Jones}}, \citenamefont {{Murakami}}, \citenamefont
  {{Breuval}}, \citenamefont {{Brink}}, \citenamefont {{Filippenko}},
  \citenamefont {{Hoffmann}}, \citenamefont {{Jha}}, \citenamefont
  {{Kenworthy}}, \citenamefont {{Mackenty}}, \citenamefont {{Stahl}},\ and\
  \citenamefont {{Zheng}}}]{localH0-2021}%
  \BibitemOpen
  \bibfield  {author} {\bibinfo {author} {\bibfnamefont {A.~G.}\ \bibnamefont
  {{Riess}}}, \bibinfo {author} {\bibfnamefont {W.}~\bibnamefont {{Yuan}}},
  \bibinfo {author} {\bibfnamefont {L.~M.}\ \bibnamefont {{Macri}}}, \bibinfo
  {author} {\bibfnamefont {D.}~\bibnamefont {{Scolnic}}}, \bibinfo {author}
  {\bibfnamefont {D.}~\bibnamefont {{Brout}}}, \bibinfo {author} {\bibfnamefont
  {S.}~\bibnamefont {{Casertano}}}, \bibinfo {author} {\bibfnamefont {D.~O.}\
  \bibnamefont {{Jones}}}, \bibinfo {author} {\bibfnamefont {Y.}~\bibnamefont
  {{Murakami}}}, \bibinfo {author} {\bibfnamefont {L.}~\bibnamefont
  {{Breuval}}}, \bibinfo {author} {\bibfnamefont {T.~G.}\ \bibnamefont
  {{Brink}}}, \bibinfo {author} {\bibfnamefont {A.~V.}\ \bibnamefont
  {{Filippenko}}}, \bibinfo {author} {\bibfnamefont {S.}~\bibnamefont
  {{Hoffmann}}}, \bibinfo {author} {\bibfnamefont {S.~W.}\ \bibnamefont
  {{Jha}}}, \bibinfo {author} {\bibfnamefont {W.~D.}\ \bibnamefont
  {{Kenworthy}}}, \bibinfo {author} {\bibfnamefont {J.}~\bibnamefont
  {{Mackenty}}}, \bibinfo {author} {\bibfnamefont {B.~E.}\ \bibnamefont
  {{Stahl}}}, \ and\ \bibinfo {author} {\bibfnamefont {W.}~\bibnamefont
  {{Zheng}}},\ }\href@noop {} {\bibfield  {journal} {\bibinfo  {journal} {arXiv
  e-prints}\ ,\ \bibinfo {eid} {arXiv:2112.04510}} (\bibinfo {year} {2021})},\
  \Eprint {http://arxiv.org/abs/2112.04510} {arXiv:2112.04510 [astro-ph.CO]}
  \BibitemShut {NoStop}%
\bibitem [{\citenamefont {{Planck Collaboration}}\ \emph
  {et~al.}(2020{\natexlab{a}})\citenamefont {{Planck Collaboration}},
  \citenamefont {{Aghanim}}, \citenamefont {{Akrami}}, \citenamefont
  {{Ashdown}}, \citenamefont {{Aumont}}, \citenamefont {{Baccigalupi}},
  \citenamefont {{Ballardini}}, \citenamefont {{Banday}}, \citenamefont
  {{Barreiro}}, \citenamefont {{Bartolo}}, \citenamefont {{Basak}},
  \citenamefont {{Battye}}, \citenamefont {{Benabed}}, \citenamefont
  {{Bernard}}, \citenamefont {{Bersanelli}}, \citenamefont {{Bielewicz}},
  \citenamefont {{Bock}}, \citenamefont {{Bond}}, \citenamefont {{Borrill}},
  \citenamefont {{Bouchet}}, \citenamefont {{Boulanger}}, \citenamefont
  {{Bucher}}, \citenamefont {{Burigana}}, \citenamefont {{Butler}},
  \citenamefont {{Calabrese}}, \citenamefont {{Cardoso}}, \citenamefont
  {{Carron}}, \citenamefont {{Challinor}}, \citenamefont {{Chiang}},
  \citenamefont {{Chluba}}, \citenamefont {{Colombo}}, \citenamefont
  {{Combet}}, \citenamefont {{Contreras}}, \citenamefont {{Crill}},
  \citenamefont {{Cuttaia}}, \citenamefont {{de Bernardis}}, \citenamefont {{de
  Zotti}}, \citenamefont {{Delabrouille}}, \citenamefont {{Delouis}},
  \citenamefont {{Di Valentino}}, \citenamefont {{Diego}}, \citenamefont
  {{Dor{\'e}}}, \citenamefont {{Douspis}}, \citenamefont {{Ducout}},
  \citenamefont {{Dupac}}, \citenamefont {{Dusini}}, \citenamefont
  {{Efstathiou}}, \citenamefont {{Elsner}}, \citenamefont {{En{\ss}lin}},
  \citenamefont {{Eriksen}}, \citenamefont {{Fantaye}}, \citenamefont
  {{Farhang}}, \citenamefont {{Fergusson}}, \citenamefont {{Fernandez-Cobos}},
  \citenamefont {{Finelli}}, \citenamefont {{Forastieri}}, \citenamefont
  {{Frailis}}, \citenamefont {{Fraisse}}, \citenamefont {{Franceschi}},
  \citenamefont {{Frolov}}, \citenamefont {{Galeotta}}, \citenamefont
  {{Galli}}, \citenamefont {{Ganga}}, \citenamefont {{G{\'e}nova-Santos}},
  \citenamefont {{Gerbino}}, \citenamefont {{Ghosh}}, \citenamefont
  {{Gonz{\'a}lez-Nuevo}}, \citenamefont {{G{\'o}rski}}, \citenamefont
  {{Gratton}}, \citenamefont {{Gruppuso}}, \citenamefont {{Gudmundsson}},
  \citenamefont {{Hamann}}, \citenamefont {{Handley}}, \citenamefont
  {{Hansen}}, \citenamefont {{Herranz}}, \citenamefont {{Hildebrandt}},
  \citenamefont {{Hivon}}, \citenamefont {{Huang}}, \citenamefont {{Jaffe}},
  \citenamefont {{Jones}}, \citenamefont {{Karakci}}, \citenamefont
  {{Keih{\"a}nen}}, \citenamefont {{Keskitalo}}, \citenamefont {{Kiiveri}},
  \citenamefont {{Kim}}, \citenamefont {{Kisner}}, \citenamefont {{Knox}},
  \citenamefont {{Krachmalnicoff}}, \citenamefont {{Kunz}}, \citenamefont
  {{Kurki-Suonio}}, \citenamefont {{Lagache}}, \citenamefont {{Lamarre}},
  \citenamefont {{Lasenby}}, \citenamefont {{Lattanzi}}, \citenamefont
  {{Lawrence}}, \citenamefont {{Le Jeune}}, \citenamefont {{Lemos}},
  \citenamefont {{Lesgourgues}}, \citenamefont {{Levrier}}, \citenamefont
  {{Lewis}}, \citenamefont {{Liguori}}, \citenamefont {{Lilje}}, \citenamefont
  {{Lilley}}, \citenamefont {{Lindholm}}, \citenamefont {{L{\'o}pez-Caniego}},
  \citenamefont {{Lubin}}, \citenamefont {{Ma}}, \citenamefont
  {{Mac{\'\i}as-P{\'e}rez}}, \citenamefont {{Maggio}}, \citenamefont {{Maino}},
  \citenamefont {{Mandolesi}}, \citenamefont {{Mangilli}}, \citenamefont
  {{Marcos-Caballero}}, \citenamefont {{Maris}}, \citenamefont {{Martin}},
  \citenamefont {{Martinelli}}, \citenamefont {{Mart{\'\i}nez-Gonz{\'a}lez}},
  \citenamefont {{Matarrese}}, \citenamefont {{Mauri}}, \citenamefont
  {{McEwen}}, \citenamefont {{Meinhold}}, \citenamefont {{Melchiorri}},
  \citenamefont {{Mennella}}, \citenamefont {{Migliaccio}}, \citenamefont
  {{Millea}}, \citenamefont {{Mitra}}, \citenamefont {{Miville-Desch{\^e}nes}},
  \citenamefont {{Molinari}}, \citenamefont {{Montier}}, \citenamefont
  {{Morgante}}, \citenamefont {{Moss}}, \citenamefont {{Natoli}}, \citenamefont
  {{N{\o}rgaard-Nielsen}}, \citenamefont {{Pagano}}, \citenamefont
  {{Paoletti}}, \citenamefont {{Partridge}}, \citenamefont {{Patanchon}},
  \citenamefont {{Peiris}}, \citenamefont {{Perrotta}}, \citenamefont
  {{Pettorino}}, \citenamefont {{Piacentini}}, \citenamefont {{Polastri}},
  \citenamefont {{Polenta}}, \citenamefont {{Puget}}, \citenamefont {{Rachen}},
  \citenamefont {{Reinecke}}, \citenamefont {{Remazeilles}}, \citenamefont
  {{Renzi}}, \citenamefont {{Rocha}}, \citenamefont {{Rosset}}, \citenamefont
  {{Roudier}}, \citenamefont {{Rubi{\~n}o-Mart{\'\i}n}}, \citenamefont
  {{Ruiz-Granados}}, \citenamefont {{Salvati}}, \citenamefont {{Sandri}},
  \citenamefont {{Savelainen}}, \citenamefont {{Scott}}, \citenamefont
  {{Shellard}}, \citenamefont {{Sirignano}}, \citenamefont {{Sirri}},
  \citenamefont {{Spencer}}, \citenamefont {{Sunyaev}}, \citenamefont
  {{Suur-Uski}}, \citenamefont {{Tauber}}, \citenamefont {{Tavagnacco}},
  \citenamefont {{Tenti}}, \citenamefont {{Toffolatti}}, \citenamefont
  {{Tomasi}}, \citenamefont {{Trombetti}}, \citenamefont {{Valenziano}},
  \citenamefont {{Valiviita}}, \citenamefont {{Van Tent}}, \citenamefont
  {{Vibert}}, \citenamefont {{Vielva}}, \citenamefont {{Villa}}, \citenamefont
  {{Vittorio}}, \citenamefont {{Wandelt}}, \citenamefont {{Wehus}},
  \citenamefont {{White}}, \citenamefont {{White}}, \citenamefont {{Zacchei}},\
  and\ \citenamefont {{Zonca}}}]{Planck2018}%
  \BibitemOpen
  \bibfield  {author} {\bibinfo {author} {\bibnamefont {{Planck
  Collaboration}}}, \bibinfo {author} {\bibfnamefont {N.}~\bibnamefont
  {{Aghanim}}}, \bibinfo {author} {\bibfnamefont {Y.}~\bibnamefont {{Akrami}}},
  \bibinfo {author} {\bibfnamefont {M.}~\bibnamefont {{Ashdown}}}, \bibinfo
  {author} {\bibfnamefont {J.}~\bibnamefont {{Aumont}}}, \bibinfo {author}
  {\bibfnamefont {C.}~\bibnamefont {{Baccigalupi}}}, \bibinfo {author}
  {\bibfnamefont {M.}~\bibnamefont {{Ballardini}}}, \bibinfo {author}
  {\bibfnamefont {A.~J.}\ \bibnamefont {{Banday}}}, \bibinfo {author}
  {\bibfnamefont {R.~B.}\ \bibnamefont {{Barreiro}}}, \bibinfo {author}
  {\bibfnamefont {N.}~\bibnamefont {{Bartolo}}}, \bibinfo {author}
  {\bibfnamefont {S.}~\bibnamefont {{Basak}}}, \bibinfo {author} {\bibfnamefont
  {R.}~\bibnamefont {{Battye}}}, \bibinfo {author} {\bibfnamefont
  {K.}~\bibnamefont {{Benabed}}}, \bibinfo {author} {\bibfnamefont {J.~P.}\
  \bibnamefont {{Bernard}}}, \bibinfo {author} {\bibfnamefont {M.}~\bibnamefont
  {{Bersanelli}}}, \bibinfo {author} {\bibfnamefont {P.}~\bibnamefont
  {{Bielewicz}}}, \bibinfo {author} {\bibfnamefont {J.~J.}\ \bibnamefont
  {{Bock}}}, \bibinfo {author} {\bibfnamefont {J.~R.}\ \bibnamefont {{Bond}}},
  \bibinfo {author} {\bibfnamefont {J.}~\bibnamefont {{Borrill}}}, \bibinfo
  {author} {\bibfnamefont {F.~R.}\ \bibnamefont {{Bouchet}}}, \bibinfo {author}
  {\bibfnamefont {F.}~\bibnamefont {{Boulanger}}}, \bibinfo {author}
  {\bibfnamefont {M.}~\bibnamefont {{Bucher}}}, \bibinfo {author}
  {\bibfnamefont {C.}~\bibnamefont {{Burigana}}}, \bibinfo {author}
  {\bibfnamefont {R.~C.}\ \bibnamefont {{Butler}}}, \bibinfo {author}
  {\bibfnamefont {E.}~\bibnamefont {{Calabrese}}}, \bibinfo {author}
  {\bibfnamefont {J.~F.}\ \bibnamefont {{Cardoso}}}, \bibinfo {author}
  {\bibfnamefont {J.}~\bibnamefont {{Carron}}}, \bibinfo {author}
  {\bibfnamefont {A.}~\bibnamefont {{Challinor}}}, \bibinfo {author}
  {\bibfnamefont {H.~C.}\ \bibnamefont {{Chiang}}}, \bibinfo {author}
  {\bibfnamefont {J.}~\bibnamefont {{Chluba}}}, \bibinfo {author}
  {\bibfnamefont {L.~P.~L.}\ \bibnamefont {{Colombo}}}, \bibinfo {author}
  {\bibfnamefont {C.}~\bibnamefont {{Combet}}}, \bibinfo {author}
  {\bibfnamefont {D.}~\bibnamefont {{Contreras}}}, \bibinfo {author}
  {\bibfnamefont {B.~P.}\ \bibnamefont {{Crill}}}, \bibinfo {author}
  {\bibfnamefont {F.}~\bibnamefont {{Cuttaia}}}, \bibinfo {author}
  {\bibfnamefont {P.}~\bibnamefont {{de Bernardis}}}, \bibinfo {author}
  {\bibfnamefont {G.}~\bibnamefont {{de Zotti}}}, \bibinfo {author}
  {\bibfnamefont {J.}~\bibnamefont {{Delabrouille}}}, \bibinfo {author}
  {\bibfnamefont {J.~M.}\ \bibnamefont {{Delouis}}}, \bibinfo {author}
  {\bibfnamefont {E.}~\bibnamefont {{Di Valentino}}}, \bibinfo {author}
  {\bibfnamefont {J.~M.}\ \bibnamefont {{Diego}}}, \bibinfo {author}
  {\bibfnamefont {O.}~\bibnamefont {{Dor{\'e}}}}, \bibinfo {author}
  {\bibfnamefont {M.}~\bibnamefont {{Douspis}}}, \bibinfo {author}
  {\bibfnamefont {A.}~\bibnamefont {{Ducout}}}, \bibinfo {author}
  {\bibfnamefont {X.}~\bibnamefont {{Dupac}}}, \bibinfo {author} {\bibfnamefont
  {S.}~\bibnamefont {{Dusini}}}, \bibinfo {author} {\bibfnamefont
  {G.}~\bibnamefont {{Efstathiou}}}, \bibinfo {author} {\bibfnamefont
  {F.}~\bibnamefont {{Elsner}}}, \bibinfo {author} {\bibfnamefont {T.~A.}\
  \bibnamefont {{En{\ss}lin}}}, \bibinfo {author} {\bibfnamefont {H.~K.}\
  \bibnamefont {{Eriksen}}}, \bibinfo {author} {\bibfnamefont {Y.}~\bibnamefont
  {{Fantaye}}}, \bibinfo {author} {\bibfnamefont {M.}~\bibnamefont
  {{Farhang}}}, \bibinfo {author} {\bibfnamefont {J.}~\bibnamefont
  {{Fergusson}}}, \bibinfo {author} {\bibfnamefont {R.}~\bibnamefont
  {{Fernandez-Cobos}}}, \bibinfo {author} {\bibfnamefont {F.}~\bibnamefont
  {{Finelli}}}, \bibinfo {author} {\bibfnamefont {F.}~\bibnamefont
  {{Forastieri}}}, \bibinfo {author} {\bibfnamefont {M.}~\bibnamefont
  {{Frailis}}}, \bibinfo {author} {\bibfnamefont {A.~A.}\ \bibnamefont
  {{Fraisse}}}, \bibinfo {author} {\bibfnamefont {E.}~\bibnamefont
  {{Franceschi}}}, \bibinfo {author} {\bibfnamefont {A.}~\bibnamefont
  {{Frolov}}}, \bibinfo {author} {\bibfnamefont {S.}~\bibnamefont
  {{Galeotta}}}, \bibinfo {author} {\bibfnamefont {S.}~\bibnamefont {{Galli}}},
  \bibinfo {author} {\bibfnamefont {K.}~\bibnamefont {{Ganga}}}, \bibinfo
  {author} {\bibfnamefont {R.~T.}\ \bibnamefont {{G{\'e}nova-Santos}}},
  \bibinfo {author} {\bibfnamefont {M.}~\bibnamefont {{Gerbino}}}, \bibinfo
  {author} {\bibfnamefont {T.}~\bibnamefont {{Ghosh}}}, \bibinfo {author}
  {\bibfnamefont {J.}~\bibnamefont {{Gonz{\'a}lez-Nuevo}}}, \bibinfo {author}
  {\bibfnamefont {K.~M.}\ \bibnamefont {{G{\'o}rski}}}, \bibinfo {author}
  {\bibfnamefont {S.}~\bibnamefont {{Gratton}}}, \bibinfo {author}
  {\bibfnamefont {A.}~\bibnamefont {{Gruppuso}}}, \bibinfo {author}
  {\bibfnamefont {J.~E.}\ \bibnamefont {{Gudmundsson}}}, \bibinfo {author}
  {\bibfnamefont {J.}~\bibnamefont {{Hamann}}}, \bibinfo {author}
  {\bibfnamefont {W.}~\bibnamefont {{Handley}}}, \bibinfo {author}
  {\bibfnamefont {F.~K.}\ \bibnamefont {{Hansen}}}, \bibinfo {author}
  {\bibfnamefont {D.}~\bibnamefont {{Herranz}}}, \bibinfo {author}
  {\bibfnamefont {S.~R.}\ \bibnamefont {{Hildebrandt}}}, \bibinfo {author}
  {\bibfnamefont {E.}~\bibnamefont {{Hivon}}}, \bibinfo {author} {\bibfnamefont
  {Z.}~\bibnamefont {{Huang}}}, \bibinfo {author} {\bibfnamefont {A.~H.}\
  \bibnamefont {{Jaffe}}}, \bibinfo {author} {\bibfnamefont {W.~C.}\
  \bibnamefont {{Jones}}}, \bibinfo {author} {\bibfnamefont {A.}~\bibnamefont
  {{Karakci}}}, \bibinfo {author} {\bibfnamefont {E.}~\bibnamefont
  {{Keih{\"a}nen}}}, \bibinfo {author} {\bibfnamefont {R.}~\bibnamefont
  {{Keskitalo}}}, \bibinfo {author} {\bibfnamefont {K.}~\bibnamefont
  {{Kiiveri}}}, \bibinfo {author} {\bibfnamefont {J.}~\bibnamefont {{Kim}}},
  \bibinfo {author} {\bibfnamefont {T.~S.}\ \bibnamefont {{Kisner}}}, \bibinfo
  {author} {\bibfnamefont {L.}~\bibnamefont {{Knox}}}, \bibinfo {author}
  {\bibfnamefont {N.}~\bibnamefont {{Krachmalnicoff}}}, \bibinfo {author}
  {\bibfnamefont {M.}~\bibnamefont {{Kunz}}}, \bibinfo {author} {\bibfnamefont
  {H.}~\bibnamefont {{Kurki-Suonio}}}, \bibinfo {author} {\bibfnamefont
  {G.}~\bibnamefont {{Lagache}}}, \bibinfo {author} {\bibfnamefont {J.~M.}\
  \bibnamefont {{Lamarre}}}, \bibinfo {author} {\bibfnamefont {A.}~\bibnamefont
  {{Lasenby}}}, \bibinfo {author} {\bibfnamefont {M.}~\bibnamefont
  {{Lattanzi}}}, \bibinfo {author} {\bibfnamefont {C.~R.}\ \bibnamefont
  {{Lawrence}}}, \bibinfo {author} {\bibfnamefont {M.}~\bibnamefont {{Le
  Jeune}}}, \bibinfo {author} {\bibfnamefont {P.}~\bibnamefont {{Lemos}}},
  \bibinfo {author} {\bibfnamefont {J.}~\bibnamefont {{Lesgourgues}}}, \bibinfo
  {author} {\bibfnamefont {F.}~\bibnamefont {{Levrier}}}, \bibinfo {author}
  {\bibfnamefont {A.}~\bibnamefont {{Lewis}}}, \bibinfo {author} {\bibfnamefont
  {M.}~\bibnamefont {{Liguori}}}, \bibinfo {author} {\bibfnamefont {P.~B.}\
  \bibnamefont {{Lilje}}}, \bibinfo {author} {\bibfnamefont {M.}~\bibnamefont
  {{Lilley}}}, \bibinfo {author} {\bibfnamefont {V.}~\bibnamefont
  {{Lindholm}}}, \bibinfo {author} {\bibfnamefont {M.}~\bibnamefont
  {{L{\'o}pez-Caniego}}}, \bibinfo {author} {\bibfnamefont {P.~M.}\
  \bibnamefont {{Lubin}}}, \bibinfo {author} {\bibfnamefont {Y.~Z.}\
  \bibnamefont {{Ma}}}, \bibinfo {author} {\bibfnamefont {J.~F.}\ \bibnamefont
  {{Mac{\'\i}as-P{\'e}rez}}}, \bibinfo {author} {\bibfnamefont
  {G.}~\bibnamefont {{Maggio}}}, \bibinfo {author} {\bibfnamefont
  {D.}~\bibnamefont {{Maino}}}, \bibinfo {author} {\bibfnamefont
  {N.}~\bibnamefont {{Mandolesi}}}, \bibinfo {author} {\bibfnamefont
  {A.}~\bibnamefont {{Mangilli}}}, \bibinfo {author} {\bibfnamefont
  {A.}~\bibnamefont {{Marcos-Caballero}}}, \bibinfo {author} {\bibfnamefont
  {M.}~\bibnamefont {{Maris}}}, \bibinfo {author} {\bibfnamefont {P.~G.}\
  \bibnamefont {{Martin}}}, \bibinfo {author} {\bibfnamefont {M.}~\bibnamefont
  {{Martinelli}}}, \bibinfo {author} {\bibfnamefont {E.}~\bibnamefont
  {{Mart{\'\i}nez-Gonz{\'a}lez}}}, \bibinfo {author} {\bibfnamefont
  {S.}~\bibnamefont {{Matarrese}}}, \bibinfo {author} {\bibfnamefont
  {N.}~\bibnamefont {{Mauri}}}, \bibinfo {author} {\bibfnamefont {J.~D.}\
  \bibnamefont {{McEwen}}}, \bibinfo {author} {\bibfnamefont {P.~R.}\
  \bibnamefont {{Meinhold}}}, \bibinfo {author} {\bibfnamefont
  {A.}~\bibnamefont {{Melchiorri}}}, \bibinfo {author} {\bibfnamefont
  {A.}~\bibnamefont {{Mennella}}}, \bibinfo {author} {\bibfnamefont
  {M.}~\bibnamefont {{Migliaccio}}}, \bibinfo {author} {\bibfnamefont
  {M.}~\bibnamefont {{Millea}}}, \bibinfo {author} {\bibfnamefont
  {S.}~\bibnamefont {{Mitra}}}, \bibinfo {author} {\bibfnamefont {M.~A.}\
  \bibnamefont {{Miville-Desch{\^e}nes}}}, \bibinfo {author} {\bibfnamefont
  {D.}~\bibnamefont {{Molinari}}}, \bibinfo {author} {\bibfnamefont
  {L.}~\bibnamefont {{Montier}}}, \bibinfo {author} {\bibfnamefont
  {G.}~\bibnamefont {{Morgante}}}, \bibinfo {author} {\bibfnamefont
  {A.}~\bibnamefont {{Moss}}}, \bibinfo {author} {\bibfnamefont
  {P.}~\bibnamefont {{Natoli}}}, \bibinfo {author} {\bibfnamefont {H.~U.}\
  \bibnamefont {{N{\o}rgaard-Nielsen}}}, \bibinfo {author} {\bibfnamefont
  {L.}~\bibnamefont {{Pagano}}}, \bibinfo {author} {\bibfnamefont
  {D.}~\bibnamefont {{Paoletti}}}, \bibinfo {author} {\bibfnamefont
  {B.}~\bibnamefont {{Partridge}}}, \bibinfo {author} {\bibfnamefont
  {G.}~\bibnamefont {{Patanchon}}}, \bibinfo {author} {\bibfnamefont {H.~V.}\
  \bibnamefont {{Peiris}}}, \bibinfo {author} {\bibfnamefont {F.}~\bibnamefont
  {{Perrotta}}}, \bibinfo {author} {\bibfnamefont {V.}~\bibnamefont
  {{Pettorino}}}, \bibinfo {author} {\bibfnamefont {F.}~\bibnamefont
  {{Piacentini}}}, \bibinfo {author} {\bibfnamefont {L.}~\bibnamefont
  {{Polastri}}}, \bibinfo {author} {\bibfnamefont {G.}~\bibnamefont
  {{Polenta}}}, \bibinfo {author} {\bibfnamefont {J.~L.}\ \bibnamefont
  {{Puget}}}, \bibinfo {author} {\bibfnamefont {J.~P.}\ \bibnamefont
  {{Rachen}}}, \bibinfo {author} {\bibfnamefont {M.}~\bibnamefont
  {{Reinecke}}}, \bibinfo {author} {\bibfnamefont {M.}~\bibnamefont
  {{Remazeilles}}}, \bibinfo {author} {\bibfnamefont {A.}~\bibnamefont
  {{Renzi}}}, \bibinfo {author} {\bibfnamefont {G.}~\bibnamefont {{Rocha}}},
  \bibinfo {author} {\bibfnamefont {C.}~\bibnamefont {{Rosset}}}, \bibinfo
  {author} {\bibfnamefont {G.}~\bibnamefont {{Roudier}}}, \bibinfo {author}
  {\bibfnamefont {J.~A.}\ \bibnamefont {{Rubi{\~n}o-Mart{\'\i}n}}}, \bibinfo
  {author} {\bibfnamefont {B.}~\bibnamefont {{Ruiz-Granados}}}, \bibinfo
  {author} {\bibfnamefont {L.}~\bibnamefont {{Salvati}}}, \bibinfo {author}
  {\bibfnamefont {M.}~\bibnamefont {{Sandri}}}, \bibinfo {author}
  {\bibfnamefont {M.}~\bibnamefont {{Savelainen}}}, \bibinfo {author}
  {\bibfnamefont {D.}~\bibnamefont {{Scott}}}, \bibinfo {author} {\bibfnamefont
  {E.~P.~S.}\ \bibnamefont {{Shellard}}}, \bibinfo {author} {\bibfnamefont
  {C.}~\bibnamefont {{Sirignano}}}, \bibinfo {author} {\bibfnamefont
  {G.}~\bibnamefont {{Sirri}}}, \bibinfo {author} {\bibfnamefont {L.~D.}\
  \bibnamefont {{Spencer}}}, \bibinfo {author} {\bibfnamefont {R.}~\bibnamefont
  {{Sunyaev}}}, \bibinfo {author} {\bibfnamefont {A.~S.}\ \bibnamefont
  {{Suur-Uski}}}, \bibinfo {author} {\bibfnamefont {J.~A.}\ \bibnamefont
  {{Tauber}}}, \bibinfo {author} {\bibfnamefont {D.}~\bibnamefont
  {{Tavagnacco}}}, \bibinfo {author} {\bibfnamefont {M.}~\bibnamefont
  {{Tenti}}}, \bibinfo {author} {\bibfnamefont {L.}~\bibnamefont
  {{Toffolatti}}}, \bibinfo {author} {\bibfnamefont {M.}~\bibnamefont
  {{Tomasi}}}, \bibinfo {author} {\bibfnamefont {T.}~\bibnamefont
  {{Trombetti}}}, \bibinfo {author} {\bibfnamefont {L.}~\bibnamefont
  {{Valenziano}}}, \bibinfo {author} {\bibfnamefont {J.}~\bibnamefont
  {{Valiviita}}}, \bibinfo {author} {\bibfnamefont {B.}~\bibnamefont {{Van
  Tent}}}, \bibinfo {author} {\bibfnamefont {L.}~\bibnamefont {{Vibert}}},
  \bibinfo {author} {\bibfnamefont {P.}~\bibnamefont {{Vielva}}}, \bibinfo
  {author} {\bibfnamefont {F.}~\bibnamefont {{Villa}}}, \bibinfo {author}
  {\bibfnamefont {N.}~\bibnamefont {{Vittorio}}}, \bibinfo {author}
  {\bibfnamefont {B.~D.}\ \bibnamefont {{Wandelt}}}, \bibinfo {author}
  {\bibfnamefont {I.~K.}\ \bibnamefont {{Wehus}}}, \bibinfo {author}
  {\bibfnamefont {M.}~\bibnamefont {{White}}}, \bibinfo {author} {\bibfnamefont
  {S.~D.~M.}\ \bibnamefont {{White}}}, \bibinfo {author} {\bibfnamefont
  {A.}~\bibnamefont {{Zacchei}}}, \ and\ \bibinfo {author} {\bibfnamefont
  {A.}~\bibnamefont {{Zonca}}},\ }\href {\doibase 10.1051/0004-6361/201833910}
  {\bibfield  {journal} {\bibinfo  {journal} {\aap}\ }\textbf {\bibinfo
  {volume} {641}},\ \bibinfo {eid} {A6} (\bibinfo {year}
  {2020}{\natexlab{a}})},\ \Eprint {http://arxiv.org/abs/1807.06209}
  {arXiv:1807.06209 [astro-ph.CO]} \BibitemShut {NoStop}%
\bibitem [{\citenamefont {{Alam}}\ \emph {et~al.}(2021)\citenamefont {{Alam}},
  \citenamefont {{Aubert}}, \citenamefont {{Avila}}, \citenamefont {{Balland}},
  \citenamefont {{Bautista}}, \citenamefont {{Bershady}}, \citenamefont
  {{Bizyaev}}, \citenamefont {{Blanton}}, \citenamefont {{Bolton}},
  \citenamefont {{Bovy}}, \citenamefont {{Brinkmann}}, \citenamefont
  {{Brownstein}}, \citenamefont {{Burtin}}, \citenamefont {{Chabanier}},
  \citenamefont {{Chapman}}, \citenamefont {{Choi}}, \citenamefont {{Chuang}},
  \citenamefont {{Comparat}}, \citenamefont {{Cousinou}}, \citenamefont
  {{Cuceu}}, \citenamefont {{Dawson}}, \citenamefont {{de la Torre}},
  \citenamefont {{de Mattia}}, \citenamefont {{Agathe}}, \citenamefont {{des
  Bourboux}}, \citenamefont {{Escoffier}}, \citenamefont {{Etourneau}},
  \citenamefont {{Farr}}, \citenamefont {{Font-Ribera}}, \citenamefont
  {{Frinchaboy}}, \citenamefont {{Fromenteau}}, \citenamefont
  {{Gil-Mar{\'\i}n}}, \citenamefont {{Le Goff}}, \citenamefont
  {{Gonzalez-Morales}}, \citenamefont {{Gonzalez-Perez}}, \citenamefont
  {{Grabowski}}, \citenamefont {{Guy}}, \citenamefont {{Hawken}}, \citenamefont
  {{Hou}}, \citenamefont {{Kong}}, \citenamefont {{Parker}}, \citenamefont
  {{Klaene}}, \citenamefont {{Kneib}}, \citenamefont {{Lin}}, \citenamefont
  {{Long}}, \citenamefont {{Lyke}}, \citenamefont {{de la Macorra}},
  \citenamefont {{Martini}}, \citenamefont {{Masters}}, \citenamefont
  {{Mohammad}}, \citenamefont {{Moon}}, \citenamefont {{Mueller}},
  \citenamefont {{Mu{\~n}oz-Guti{\'e}rrez}}, \citenamefont {{Myers}},
  \citenamefont {{Nadathur}}, \citenamefont {{Neveux}}, \citenamefont
  {{Newman}}, \citenamefont {{Noterdaeme}}, \citenamefont {{Oravetz}},
  \citenamefont {{Oravetz}}, \citenamefont {{Palanque-Delabrouille}},
  \citenamefont {{Pan}}, \citenamefont {{Paviot}}, \citenamefont {{Percival}},
  \citenamefont {{P{\'e}rez-R{\`a}fols}}, \citenamefont {{Petitjean}},
  \citenamefont {{Pieri}}, \citenamefont {{Prakash}}, \citenamefont
  {{Raichoor}}, \citenamefont {{Ravoux}}, \citenamefont {{Rezaie}},
  \citenamefont {{Rich}}, \citenamefont {{Ross}}, \citenamefont {{Rossi}},
  \citenamefont {{Ruggeri}}, \citenamefont {{Ruhlmann-Kleider}}, \citenamefont
  {{S{\'a}nchez}}, \citenamefont {{S{\'a}nchez}}, \citenamefont
  {{S{\'a}nchez-Gallego}}, \citenamefont {{Sayres}}, \citenamefont
  {{Schneider}}, \citenamefont {{Seo}}, \citenamefont {{Shafieloo}},
  \citenamefont {{Slosar}}, \citenamefont {{Smith}}, \citenamefont {{Stermer}},
  \citenamefont {{Tamone}}, \citenamefont {{Tinker}}, \citenamefont
  {{Tojeiro}}, \citenamefont {{Vargas-Maga{\~n}a}}, \citenamefont {{Variu}},
  \citenamefont {{Wang}}, \citenamefont {{Weaver}}, \citenamefont {{Weijmans}},
  \citenamefont {{Y{\`e}che}}, \citenamefont {{Zarrouk}}, \citenamefont
  {{Zhao}}, \citenamefont {{Zhao}},\ and\ \citenamefont
  {{Zheng}}}]{eBOSS-cosmo}%
  \BibitemOpen
  \bibfield  {author} {\bibinfo {author} {\bibfnamefont {S.}~\bibnamefont
  {{Alam}}}, \bibinfo {author} {\bibfnamefont {M.}~\bibnamefont {{Aubert}}},
  \bibinfo {author} {\bibfnamefont {S.}~\bibnamefont {{Avila}}}, \bibinfo
  {author} {\bibfnamefont {C.}~\bibnamefont {{Balland}}}, \bibinfo {author}
  {\bibfnamefont {J.~E.}\ \bibnamefont {{Bautista}}}, \bibinfo {author}
  {\bibfnamefont {M.~A.}\ \bibnamefont {{Bershady}}}, \bibinfo {author}
  {\bibfnamefont {D.}~\bibnamefont {{Bizyaev}}}, \bibinfo {author}
  {\bibfnamefont {M.~R.}\ \bibnamefont {{Blanton}}}, \bibinfo {author}
  {\bibfnamefont {A.~S.}\ \bibnamefont {{Bolton}}}, \bibinfo {author}
  {\bibfnamefont {J.}~\bibnamefont {{Bovy}}}, \bibinfo {author} {\bibfnamefont
  {J.}~\bibnamefont {{Brinkmann}}}, \bibinfo {author} {\bibfnamefont {J.~R.}\
  \bibnamefont {{Brownstein}}}, \bibinfo {author} {\bibfnamefont
  {E.}~\bibnamefont {{Burtin}}}, \bibinfo {author} {\bibfnamefont
  {S.}~\bibnamefont {{Chabanier}}}, \bibinfo {author} {\bibfnamefont {M.~J.}\
  \bibnamefont {{Chapman}}}, \bibinfo {author} {\bibfnamefont {P.~D.}\
  \bibnamefont {{Choi}}}, \bibinfo {author} {\bibfnamefont {C.-H.}\
  \bibnamefont {{Chuang}}}, \bibinfo {author} {\bibfnamefont {J.}~\bibnamefont
  {{Comparat}}}, \bibinfo {author} {\bibfnamefont {M.-C.}\ \bibnamefont
  {{Cousinou}}}, \bibinfo {author} {\bibfnamefont {A.}~\bibnamefont {{Cuceu}}},
  \bibinfo {author} {\bibfnamefont {K.~S.}\ \bibnamefont {{Dawson}}}, \bibinfo
  {author} {\bibfnamefont {S.}~\bibnamefont {{de la Torre}}}, \bibinfo {author}
  {\bibfnamefont {A.}~\bibnamefont {{de Mattia}}}, \bibinfo {author}
  {\bibfnamefont {V.~d.~S.}\ \bibnamefont {{Agathe}}}, \bibinfo {author}
  {\bibfnamefont {H.~d.~M.}\ \bibnamefont {{des Bourboux}}}, \bibinfo {author}
  {\bibfnamefont {S.}~\bibnamefont {{Escoffier}}}, \bibinfo {author}
  {\bibfnamefont {T.}~\bibnamefont {{Etourneau}}}, \bibinfo {author}
  {\bibfnamefont {J.}~\bibnamefont {{Farr}}}, \bibinfo {author} {\bibfnamefont
  {A.}~\bibnamefont {{Font-Ribera}}}, \bibinfo {author} {\bibfnamefont {P.~M.}\
  \bibnamefont {{Frinchaboy}}}, \bibinfo {author} {\bibfnamefont
  {S.}~\bibnamefont {{Fromenteau}}}, \bibinfo {author} {\bibfnamefont
  {H.}~\bibnamefont {{Gil-Mar{\'\i}n}}}, \bibinfo {author} {\bibfnamefont
  {J.-M.}\ \bibnamefont {{Le Goff}}}, \bibinfo {author} {\bibfnamefont {A.~X.}\
  \bibnamefont {{Gonzalez-Morales}}}, \bibinfo {author} {\bibfnamefont
  {V.}~\bibnamefont {{Gonzalez-Perez}}}, \bibinfo {author} {\bibfnamefont
  {K.}~\bibnamefont {{Grabowski}}}, \bibinfo {author} {\bibfnamefont
  {J.}~\bibnamefont {{Guy}}}, \bibinfo {author} {\bibfnamefont {A.~J.}\
  \bibnamefont {{Hawken}}}, \bibinfo {author} {\bibfnamefont {J.}~\bibnamefont
  {{Hou}}}, \bibinfo {author} {\bibfnamefont {H.}~\bibnamefont {{Kong}}},
  \bibinfo {author} {\bibfnamefont {J.}~\bibnamefont {{Parker}}}, \bibinfo
  {author} {\bibfnamefont {M.}~\bibnamefont {{Klaene}}}, \bibinfo {author}
  {\bibfnamefont {J.-P.}\ \bibnamefont {{Kneib}}}, \bibinfo {author}
  {\bibfnamefont {S.}~\bibnamefont {{Lin}}}, \bibinfo {author} {\bibfnamefont
  {D.}~\bibnamefont {{Long}}}, \bibinfo {author} {\bibfnamefont {B.~W.}\
  \bibnamefont {{Lyke}}}, \bibinfo {author} {\bibfnamefont {A.}~\bibnamefont
  {{de la Macorra}}}, \bibinfo {author} {\bibfnamefont {P.}~\bibnamefont
  {{Martini}}}, \bibinfo {author} {\bibfnamefont {K.}~\bibnamefont
  {{Masters}}}, \bibinfo {author} {\bibfnamefont {F.~G.}\ \bibnamefont
  {{Mohammad}}}, \bibinfo {author} {\bibfnamefont {J.}~\bibnamefont {{Moon}}},
  \bibinfo {author} {\bibfnamefont {E.-M.}\ \bibnamefont {{Mueller}}}, \bibinfo
  {author} {\bibfnamefont {A.}~\bibnamefont {{Mu{\~n}oz-Guti{\'e}rrez}}},
  \bibinfo {author} {\bibfnamefont {A.~D.}\ \bibnamefont {{Myers}}}, \bibinfo
  {author} {\bibfnamefont {S.}~\bibnamefont {{Nadathur}}}, \bibinfo {author}
  {\bibfnamefont {R.}~\bibnamefont {{Neveux}}}, \bibinfo {author}
  {\bibfnamefont {J.~A.}\ \bibnamefont {{Newman}}}, \bibinfo {author}
  {\bibfnamefont {P.}~\bibnamefont {{Noterdaeme}}}, \bibinfo {author}
  {\bibfnamefont {A.}~\bibnamefont {{Oravetz}}}, \bibinfo {author}
  {\bibfnamefont {D.}~\bibnamefont {{Oravetz}}}, \bibinfo {author}
  {\bibfnamefont {N.}~\bibnamefont {{Palanque-Delabrouille}}}, \bibinfo
  {author} {\bibfnamefont {K.}~\bibnamefont {{Pan}}}, \bibinfo {author}
  {\bibfnamefont {R.}~\bibnamefont {{Paviot}}}, \bibinfo {author}
  {\bibfnamefont {W.~J.}\ \bibnamefont {{Percival}}}, \bibinfo {author}
  {\bibfnamefont {I.}~\bibnamefont {{P{\'e}rez-R{\`a}fols}}}, \bibinfo {author}
  {\bibfnamefont {P.}~\bibnamefont {{Petitjean}}}, \bibinfo {author}
  {\bibfnamefont {M.~M.}\ \bibnamefont {{Pieri}}}, \bibinfo {author}
  {\bibfnamefont {A.}~\bibnamefont {{Prakash}}}, \bibinfo {author}
  {\bibfnamefont {A.}~\bibnamefont {{Raichoor}}}, \bibinfo {author}
  {\bibfnamefont {C.}~\bibnamefont {{Ravoux}}}, \bibinfo {author}
  {\bibfnamefont {M.}~\bibnamefont {{Rezaie}}}, \bibinfo {author}
  {\bibfnamefont {J.}~\bibnamefont {{Rich}}}, \bibinfo {author} {\bibfnamefont
  {A.~J.}\ \bibnamefont {{Ross}}}, \bibinfo {author} {\bibfnamefont
  {G.}~\bibnamefont {{Rossi}}}, \bibinfo {author} {\bibfnamefont
  {R.}~\bibnamefont {{Ruggeri}}}, \bibinfo {author} {\bibfnamefont
  {V.}~\bibnamefont {{Ruhlmann-Kleider}}}, \bibinfo {author} {\bibfnamefont
  {A.~G.}\ \bibnamefont {{S{\'a}nchez}}}, \bibinfo {author} {\bibfnamefont
  {F.~J.}\ \bibnamefont {{S{\'a}nchez}}}, \bibinfo {author} {\bibfnamefont
  {J.~R.}\ \bibnamefont {{S{\'a}nchez-Gallego}}}, \bibinfo {author}
  {\bibfnamefont {C.}~\bibnamefont {{Sayres}}}, \bibinfo {author}
  {\bibfnamefont {D.~P.}\ \bibnamefont {{Schneider}}}, \bibinfo {author}
  {\bibfnamefont {H.-J.}\ \bibnamefont {{Seo}}}, \bibinfo {author}
  {\bibfnamefont {A.}~\bibnamefont {{Shafieloo}}}, \bibinfo {author}
  {\bibfnamefont {A.}~\bibnamefont {{Slosar}}}, \bibinfo {author}
  {\bibfnamefont {A.}~\bibnamefont {{Smith}}}, \bibinfo {author} {\bibfnamefont
  {J.}~\bibnamefont {{Stermer}}}, \bibinfo {author} {\bibfnamefont
  {A.}~\bibnamefont {{Tamone}}}, \bibinfo {author} {\bibfnamefont {J.~L.}\
  \bibnamefont {{Tinker}}}, \bibinfo {author} {\bibfnamefont {R.}~\bibnamefont
  {{Tojeiro}}}, \bibinfo {author} {\bibfnamefont {M.}~\bibnamefont
  {{Vargas-Maga{\~n}a}}}, \bibinfo {author} {\bibfnamefont {A.}~\bibnamefont
  {{Variu}}}, \bibinfo {author} {\bibfnamefont {Y.}~\bibnamefont {{Wang}}},
  \bibinfo {author} {\bibfnamefont {B.~A.}\ \bibnamefont {{Weaver}}}, \bibinfo
  {author} {\bibfnamefont {A.-M.}\ \bibnamefont {{Weijmans}}}, \bibinfo
  {author} {\bibfnamefont {C.}~\bibnamefont {{Y{\`e}che}}}, \bibinfo {author}
  {\bibfnamefont {P.}~\bibnamefont {{Zarrouk}}}, \bibinfo {author}
  {\bibfnamefont {C.}~\bibnamefont {{Zhao}}}, \bibinfo {author} {\bibfnamefont
  {G.-B.}\ \bibnamefont {{Zhao}}}, \ and\ \bibinfo {author} {\bibfnamefont
  {Z.}~\bibnamefont {{Zheng}}},\ }\href {\doibase 10.1103/PhysRevD.103.083533}
  {\bibfield  {journal} {\bibinfo  {journal} {\prd}\ }\textbf {\bibinfo
  {volume} {103}},\ \bibinfo {eid} {083533} (\bibinfo {year} {2021})},\ \Eprint
  {http://arxiv.org/abs/2007.08991} {arXiv:2007.08991 [astro-ph.CO]}
  \BibitemShut {NoStop}%
\bibitem [{\citenamefont {{Cooke}}\ \emph {et~al.}(2018)\citenamefont
  {{Cooke}}, \citenamefont {{Pettini}},\ and\ \citenamefont
  {{Steidel}}}]{BBNobs}%
  \BibitemOpen
  \bibfield  {author} {\bibinfo {author} {\bibfnamefont {R.~J.}\ \bibnamefont
  {{Cooke}}}, \bibinfo {author} {\bibfnamefont {M.}~\bibnamefont {{Pettini}}},
  \ and\ \bibinfo {author} {\bibfnamefont {C.~C.}\ \bibnamefont {{Steidel}}},\
  }\href {\doibase 10.3847/1538-4357/aaab53} {\bibfield  {journal} {\bibinfo
  {journal} {\apj}\ }\textbf {\bibinfo {volume} {855}},\ \bibinfo {eid} {102}
  (\bibinfo {year} {2018})},\ \Eprint {http://arxiv.org/abs/1710.11129}
  {arXiv:1710.11129 [astro-ph.CO]} \BibitemShut {NoStop}%
\bibitem [{\citenamefont {{Knox}}\ and\ \citenamefont
  {{Millea}}(2020)}]{Hubble-hunters}%
  \BibitemOpen
  \bibfield  {author} {\bibinfo {author} {\bibfnamefont {L.}~\bibnamefont
  {{Knox}}}\ and\ \bibinfo {author} {\bibfnamefont {M.}~\bibnamefont
  {{Millea}}},\ }\href {\doibase 10.1103/PhysRevD.101.043533} {\bibfield
  {journal} {\bibinfo  {journal} {\prd}\ }\textbf {\bibinfo {volume} {101}},\
  \bibinfo {eid} {043533} (\bibinfo {year} {2020})},\ \Eprint
  {http://arxiv.org/abs/1908.03663} {arXiv:1908.03663 [astro-ph.CO]}
  \BibitemShut {NoStop}%
\bibitem [{\citenamefont {{Di Valentino}}\ \emph {et~al.}(2021)\citenamefont
  {{Di Valentino}}, \citenamefont {{Mena}}, \citenamefont {{Pan}},
  \citenamefont {{Visinelli}}, \citenamefont {{Yang}}, \citenamefont
  {{Melchiorri}}, \citenamefont {{Mota}}, \citenamefont {{Riess}},\ and\
  \citenamefont {{Silk}}}]{H0-tension-review}%
  \BibitemOpen
  \bibfield  {author} {\bibinfo {author} {\bibfnamefont {E.}~\bibnamefont {{Di
  Valentino}}}, \bibinfo {author} {\bibfnamefont {O.}~\bibnamefont {{Mena}}},
  \bibinfo {author} {\bibfnamefont {S.}~\bibnamefont {{Pan}}}, \bibinfo
  {author} {\bibfnamefont {L.}~\bibnamefont {{Visinelli}}}, \bibinfo {author}
  {\bibfnamefont {W.}~\bibnamefont {{Yang}}}, \bibinfo {author} {\bibfnamefont
  {A.}~\bibnamefont {{Melchiorri}}}, \bibinfo {author} {\bibfnamefont {D.~F.}\
  \bibnamefont {{Mota}}}, \bibinfo {author} {\bibfnamefont {A.~G.}\
  \bibnamefont {{Riess}}}, \ and\ \bibinfo {author} {\bibfnamefont
  {J.}~\bibnamefont {{Silk}}},\ }\href {\doibase 10.1088/1361-6382/ac086d}
  {\bibfield  {journal} {\bibinfo  {journal} {Classical and Quantum Gravity}\
  }\textbf {\bibinfo {volume} {38}},\ \bibinfo {eid} {153001} (\bibinfo {year}
  {2021})},\ \Eprint {http://arxiv.org/abs/2103.01183} {arXiv:2103.01183
  [astro-ph.CO]} \BibitemShut {NoStop}%
\bibitem [{\citenamefont {{Marra}}\ \emph {et~al.}(2013)\citenamefont
  {{Marra}}, \citenamefont {{Amendola}}, \citenamefont {{Sawicki}},\ and\
  \citenamefont {{Valkenburg}}}]{Marra2013}%
  \BibitemOpen
  \bibfield  {author} {\bibinfo {author} {\bibfnamefont {V.}~\bibnamefont
  {{Marra}}}, \bibinfo {author} {\bibfnamefont {L.}~\bibnamefont {{Amendola}}},
  \bibinfo {author} {\bibfnamefont {I.}~\bibnamefont {{Sawicki}}}, \ and\
  \bibinfo {author} {\bibfnamefont {W.}~\bibnamefont {{Valkenburg}}},\ }\href
  {\doibase 10.1103/PhysRevLett.110.241305} {\bibfield  {journal} {\bibinfo
  {journal} {\prl}\ }\textbf {\bibinfo {volume} {110}},\ \bibinfo {eid}
  {241305} (\bibinfo {year} {2013})},\ \Eprint {http://arxiv.org/abs/1303.3121}
  {arXiv:1303.3121 [astro-ph.CO]} \BibitemShut {NoStop}%
\bibitem [{\citenamefont {{Wojtak}}\ \emph {et~al.}(2014)\citenamefont
  {{Wojtak}}, \citenamefont {{Knebe}}, \citenamefont {{Watson}}, \citenamefont
  {{Iliev}}, \citenamefont {{He{\ss}}}, \citenamefont {{Rapetti}},
  \citenamefont {{Yepes}},\ and\ \citenamefont {{Gottl{\"o}ber}}}]{Wojtak2014}%
  \BibitemOpen
  \bibfield  {author} {\bibinfo {author} {\bibfnamefont {R.}~\bibnamefont
  {{Wojtak}}}, \bibinfo {author} {\bibfnamefont {A.}~\bibnamefont {{Knebe}}},
  \bibinfo {author} {\bibfnamefont {W.~A.}\ \bibnamefont {{Watson}}}, \bibinfo
  {author} {\bibfnamefont {I.~T.}\ \bibnamefont {{Iliev}}}, \bibinfo {author}
  {\bibfnamefont {S.}~\bibnamefont {{He{\ss}}}}, \bibinfo {author}
  {\bibfnamefont {D.}~\bibnamefont {{Rapetti}}}, \bibinfo {author}
  {\bibfnamefont {G.}~\bibnamefont {{Yepes}}}, \ and\ \bibinfo {author}
  {\bibfnamefont {S.}~\bibnamefont {{Gottl{\"o}ber}}},\ }\href {\doibase
  10.1093/mnras/stt2321} {\bibfield  {journal} {\bibinfo  {journal} {\mnras}\
  }\textbf {\bibinfo {volume} {438}},\ \bibinfo {pages} {1805} (\bibinfo {year}
  {2014})},\ \Eprint {http://arxiv.org/abs/1312.0276} {arXiv:1312.0276
  [astro-ph.CO]} \BibitemShut {NoStop}%
\bibitem [{\citenamefont {{Enea Romano}}(2016)}]{Romano2016}%
  \BibitemOpen
  \bibfield  {author} {\bibinfo {author} {\bibfnamefont {A.}~\bibnamefont
  {{Enea Romano}}},\ }\href@noop {} {\bibfield  {journal} {\bibinfo  {journal}
  {arXiv e-prints}\ ,\ \bibinfo {eid} {arXiv:1609.04081}} (\bibinfo {year}
  {2016})},\ \Eprint {http://arxiv.org/abs/1609.04081} {arXiv:1609.04081
  [astro-ph.CO]} \BibitemShut {NoStop}%
\bibitem [{\citenamefont {{Wu}}\ and\ \citenamefont
  {{Huterer}}(2017)}]{Wu2017}%
  \BibitemOpen
  \bibfield  {author} {\bibinfo {author} {\bibfnamefont {H.-Y.}\ \bibnamefont
  {{Wu}}}\ and\ \bibinfo {author} {\bibfnamefont {D.}~\bibnamefont
  {{Huterer}}},\ }\href {\doibase 10.1093/mnras/stx1967} {\bibfield  {journal}
  {\bibinfo  {journal} {\mnras}\ }\textbf {\bibinfo {volume} {471}},\ \bibinfo
  {pages} {4946} (\bibinfo {year} {2017})},\ \Eprint
  {http://arxiv.org/abs/1706.09723} {arXiv:1706.09723 [astro-ph.CO]}
  \BibitemShut {NoStop}%
\bibitem [{\citenamefont {{Sasaki}}(1987)}]{Sasaki1987}%
  \BibitemOpen
  \bibfield  {author} {\bibinfo {author} {\bibfnamefont {M.}~\bibnamefont
  {{Sasaki}}},\ }\href {\doibase 10.1093/mnras/228.3.653} {\bibfield  {journal}
  {\bibinfo  {journal} {\mnras}\ }\textbf {\bibinfo {volume} {228}},\ \bibinfo
  {pages} {653} (\bibinfo {year} {1987})}\BibitemShut {NoStop}%
\bibitem [{\citenamefont {{Sirko}}(2005)}]{Sirko2005}%
  \BibitemOpen
  \bibfield  {author} {\bibinfo {author} {\bibfnamefont {E.}~\bibnamefont
  {{Sirko}}},\ }\href {\doibase 10.1086/497090} {\bibfield  {journal} {\bibinfo
   {journal} {\apj}\ }\textbf {\bibinfo {volume} {634}},\ \bibinfo {pages}
  {728} (\bibinfo {year} {2005})},\ \Eprint
  {http://arxiv.org/abs/astro-ph/0503106} {arXiv:astro-ph/0503106 [astro-ph]}
  \BibitemShut {NoStop}%
\bibitem [{\citenamefont {{Percival}}(2005)}]{Percival2005}%
  \BibitemOpen
  \bibfield  {author} {\bibinfo {author} {\bibfnamefont {W.~J.}\ \bibnamefont
  {{Percival}}},\ }\href {\doibase 10.1051/0004-6361:20053637} {\bibfield
  {journal} {\bibinfo  {journal} {\aap}\ }\textbf {\bibinfo {volume} {443}},\
  \bibinfo {pages} {819} (\bibinfo {year} {2005})},\ \Eprint
  {http://arxiv.org/abs/astro-ph/0508156} {arXiv:astro-ph/0508156 [astro-ph]}
  \BibitemShut {NoStop}%
\bibitem [{\citenamefont {{Scolnic}}\ \emph {et~al.}(2018)\citenamefont
  {{Scolnic}}, \citenamefont {{Jones}}, \citenamefont {{Rest}}, \citenamefont
  {{Pan}}, \citenamefont {{Chornock}}, \citenamefont {{Foley}}, \citenamefont
  {{Huber}}, \citenamefont {{Kessler}}, \citenamefont {{Narayan}},
  \citenamefont {{Riess}}, \citenamefont {{Rodney}}, \citenamefont {{Berger}},
  \citenamefont {{Brout}}, \citenamefont {{Challis}}, \citenamefont {{Drout}},
  \citenamefont {{Finkbeiner}}, \citenamefont {{Lunnan}}, \citenamefont
  {{Kirshner}}, \citenamefont {{Sanders}}, \citenamefont {{Schlafly}},
  \citenamefont {{Smartt}}, \citenamefont {{Stubbs}}, \citenamefont {{Tonry}},
  \citenamefont {{Wood-Vasey}}, \citenamefont {{Foley}}, \citenamefont
  {{Hand}}, \citenamefont {{Johnson}}, \citenamefont {{Burgett}}, \citenamefont
  {{Chambers}}, \citenamefont {{Draper}}, \citenamefont {{Hodapp}},
  \citenamefont {{Kaiser}}, \citenamefont {{Kudritzki}}, \citenamefont
  {{Magnier}}, \citenamefont {{Metcalfe}}, \citenamefont {{Bresolin}},
  \citenamefont {{Gall}}, \citenamefont {{Kotak}}, \citenamefont {{McCrum}},\
  and\ \citenamefont {{Smith}}}]{Pantheon}%
  \BibitemOpen
  \bibfield  {author} {\bibinfo {author} {\bibfnamefont {D.~M.}\ \bibnamefont
  {{Scolnic}}}, \bibinfo {author} {\bibfnamefont {D.~O.}\ \bibnamefont
  {{Jones}}}, \bibinfo {author} {\bibfnamefont {A.}~\bibnamefont {{Rest}}},
  \bibinfo {author} {\bibfnamefont {Y.~C.}\ \bibnamefont {{Pan}}}, \bibinfo
  {author} {\bibfnamefont {R.}~\bibnamefont {{Chornock}}}, \bibinfo {author}
  {\bibfnamefont {R.~J.}\ \bibnamefont {{Foley}}}, \bibinfo {author}
  {\bibfnamefont {M.~E.}\ \bibnamefont {{Huber}}}, \bibinfo {author}
  {\bibfnamefont {R.}~\bibnamefont {{Kessler}}}, \bibinfo {author}
  {\bibfnamefont {G.}~\bibnamefont {{Narayan}}}, \bibinfo {author}
  {\bibfnamefont {A.~G.}\ \bibnamefont {{Riess}}}, \bibinfo {author}
  {\bibfnamefont {S.}~\bibnamefont {{Rodney}}}, \bibinfo {author}
  {\bibfnamefont {E.}~\bibnamefont {{Berger}}}, \bibinfo {author}
  {\bibfnamefont {D.~J.}\ \bibnamefont {{Brout}}}, \bibinfo {author}
  {\bibfnamefont {P.~J.}\ \bibnamefont {{Challis}}}, \bibinfo {author}
  {\bibfnamefont {M.}~\bibnamefont {{Drout}}}, \bibinfo {author} {\bibfnamefont
  {D.}~\bibnamefont {{Finkbeiner}}}, \bibinfo {author} {\bibfnamefont
  {R.}~\bibnamefont {{Lunnan}}}, \bibinfo {author} {\bibfnamefont {R.~P.}\
  \bibnamefont {{Kirshner}}}, \bibinfo {author} {\bibfnamefont {N.~E.}\
  \bibnamefont {{Sanders}}}, \bibinfo {author} {\bibfnamefont {E.}~\bibnamefont
  {{Schlafly}}}, \bibinfo {author} {\bibfnamefont {S.}~\bibnamefont
  {{Smartt}}}, \bibinfo {author} {\bibfnamefont {C.~W.}\ \bibnamefont
  {{Stubbs}}}, \bibinfo {author} {\bibfnamefont {J.}~\bibnamefont {{Tonry}}},
  \bibinfo {author} {\bibfnamefont {W.~M.}\ \bibnamefont {{Wood-Vasey}}},
  \bibinfo {author} {\bibfnamefont {M.}~\bibnamefont {{Foley}}}, \bibinfo
  {author} {\bibfnamefont {J.}~\bibnamefont {{Hand}}}, \bibinfo {author}
  {\bibfnamefont {E.}~\bibnamefont {{Johnson}}}, \bibinfo {author}
  {\bibfnamefont {W.~S.}\ \bibnamefont {{Burgett}}}, \bibinfo {author}
  {\bibfnamefont {K.~C.}\ \bibnamefont {{Chambers}}}, \bibinfo {author}
  {\bibfnamefont {P.~W.}\ \bibnamefont {{Draper}}}, \bibinfo {author}
  {\bibfnamefont {K.~W.}\ \bibnamefont {{Hodapp}}}, \bibinfo {author}
  {\bibfnamefont {N.}~\bibnamefont {{Kaiser}}}, \bibinfo {author}
  {\bibfnamefont {R.~P.}\ \bibnamefont {{Kudritzki}}}, \bibinfo {author}
  {\bibfnamefont {E.~A.}\ \bibnamefont {{Magnier}}}, \bibinfo {author}
  {\bibfnamefont {N.}~\bibnamefont {{Metcalfe}}}, \bibinfo {author}
  {\bibfnamefont {F.}~\bibnamefont {{Bresolin}}}, \bibinfo {author}
  {\bibfnamefont {E.}~\bibnamefont {{Gall}}}, \bibinfo {author} {\bibfnamefont
  {R.}~\bibnamefont {{Kotak}}}, \bibinfo {author} {\bibfnamefont
  {M.}~\bibnamefont {{McCrum}}}, \ and\ \bibinfo {author} {\bibfnamefont
  {K.~W.}\ \bibnamefont {{Smith}}},\ }\href {\doibase 10.3847/1538-4357/aab9bb}
  {\bibfield  {journal} {\bibinfo  {journal} {\apj}\ }\textbf {\bibinfo
  {volume} {859}},\ \bibinfo {eid} {101} (\bibinfo {year} {2018})},\ \Eprint
  {http://arxiv.org/abs/1710.00845} {arXiv:1710.00845 [astro-ph.CO]}
  \BibitemShut {NoStop}%
\bibitem [{\citenamefont {{Barausse}}\ \emph {et~al.}(2005)\citenamefont
  {{Barausse}}, \citenamefont {{Matarrese}},\ and\ \citenamefont
  {{Riotto}}}]{Barausse2005}%
  \BibitemOpen
  \bibfield  {author} {\bibinfo {author} {\bibfnamefont {E.}~\bibnamefont
  {{Barausse}}}, \bibinfo {author} {\bibfnamefont {S.}~\bibnamefont
  {{Matarrese}}}, \ and\ \bibinfo {author} {\bibfnamefont {A.}~\bibnamefont
  {{Riotto}}},\ }\href {\doibase 10.1103/PhysRevD.71.063537} {\bibfield
  {journal} {\bibinfo  {journal} {\prd}\ }\textbf {\bibinfo {volume} {71}},\
  \bibinfo {eid} {063537} (\bibinfo {year} {2005})},\ \Eprint
  {http://arxiv.org/abs/astro-ph/0501152} {arXiv:astro-ph/0501152 [astro-ph]}
  \BibitemShut {NoStop}%
\bibitem [{\citenamefont {{Bonvin}}\ \emph {et~al.}(2006)\citenamefont
  {{Bonvin}}, \citenamefont {{Durrer}},\ and\ \citenamefont
  {{Gasparini}}}]{Bonvin2006}%
  \BibitemOpen
  \bibfield  {author} {\bibinfo {author} {\bibfnamefont {C.}~\bibnamefont
  {{Bonvin}}}, \bibinfo {author} {\bibfnamefont {R.}~\bibnamefont {{Durrer}}},
  \ and\ \bibinfo {author} {\bibfnamefont {M.~A.}\ \bibnamefont
  {{Gasparini}}},\ }\href {\doibase 10.1103/PhysRevD.73.023523} {\bibfield
  {journal} {\bibinfo  {journal} {\prd}\ }\textbf {\bibinfo {volume} {73}},\
  \bibinfo {eid} {023523} (\bibinfo {year} {2006})},\ \Eprint
  {http://arxiv.org/abs/astro-ph/0511183} {arXiv:astro-ph/0511183 [astro-ph]}
  \BibitemShut {NoStop}%
\bibitem [{\citenamefont {{Hui}}\ and\ \citenamefont
  {{Greene}}(2006)}]{Hui2006}%
  \BibitemOpen
  \bibfield  {author} {\bibinfo {author} {\bibfnamefont {L.}~\bibnamefont
  {{Hui}}}\ and\ \bibinfo {author} {\bibfnamefont {P.~B.}\ \bibnamefont
  {{Greene}}},\ }\href {\doibase 10.1103/PhysRevD.73.123526} {\bibfield
  {journal} {\bibinfo  {journal} {\prd}\ }\textbf {\bibinfo {volume} {73}},\
  \bibinfo {eid} {123526} (\bibinfo {year} {2006})},\ \Eprint
  {http://arxiv.org/abs/astro-ph/0512159} {arXiv:astro-ph/0512159 [astro-ph]}
  \BibitemShut {NoStop}%
\bibitem [{\citenamefont {{Takada}}\ and\ \citenamefont
  {{Hu}}(2013)}]{Takada2013}%
  \BibitemOpen
  \bibfield  {author} {\bibinfo {author} {\bibfnamefont {M.}~\bibnamefont
  {{Takada}}}\ and\ \bibinfo {author} {\bibfnamefont {W.}~\bibnamefont
  {{Hu}}},\ }\href {\doibase 10.1103/PhysRevD.87.123504} {\bibfield  {journal}
  {\bibinfo  {journal} {\prd}\ }\textbf {\bibinfo {volume} {87}},\ \bibinfo
  {eid} {123504} (\bibinfo {year} {2013})},\ \Eprint
  {http://arxiv.org/abs/1302.6994} {arXiv:1302.6994 [astro-ph.CO]} \BibitemShut
  {NoStop}%
\bibitem [{\citenamefont {{Frenk}}\ \emph {et~al.}(1988)\citenamefont
  {{Frenk}}, \citenamefont {{White}}, \citenamefont {{Davis}},\ and\
  \citenamefont {{Efstathiou}}}]{Frenk88}%
  \BibitemOpen
  \bibfield  {author} {\bibinfo {author} {\bibfnamefont {C.~S.}\ \bibnamefont
  {{Frenk}}}, \bibinfo {author} {\bibfnamefont {S.~D.~M.}\ \bibnamefont
  {{White}}}, \bibinfo {author} {\bibfnamefont {M.}~\bibnamefont {{Davis}}}, \
  and\ \bibinfo {author} {\bibfnamefont {G.}~\bibnamefont {{Efstathiou}}},\
  }\href {\doibase 10.1086/166213} {\bibfield  {journal} {\bibinfo  {journal}
  {\apj}\ }\textbf {\bibinfo {volume} {327}},\ \bibinfo {pages} {507} (\bibinfo
  {year} {1988})}\BibitemShut {NoStop}%
\bibitem [{\citenamefont {{Howlett}}\ and\ \citenamefont
  {{Percival}}(2017)}]{Howlett2017}%
  \BibitemOpen
  \bibfield  {author} {\bibinfo {author} {\bibfnamefont {C.}~\bibnamefont
  {{Howlett}}}\ and\ \bibinfo {author} {\bibfnamefont {W.~J.}\ \bibnamefont
  {{Percival}}},\ }\href {\doibase 10.1093/mnras/stx2342} {\bibfield  {journal}
  {\bibinfo  {journal} {\mnras}\ }\textbf {\bibinfo {volume} {472}},\ \bibinfo
  {pages} {4935} (\bibinfo {year} {2017})},\ \Eprint
  {http://arxiv.org/abs/1709.03057} {arXiv:1709.03057 [astro-ph.CO]}
  \BibitemShut {NoStop}%
\bibitem [{\citenamefont {{Gunn}}\ and\ \citenamefont {{Gott}}(1972)}]{gunn72}%
  \BibitemOpen
  \bibfield  {author} {\bibinfo {author} {\bibfnamefont {J.~E.}\ \bibnamefont
  {{Gunn}}}\ and\ \bibinfo {author} {\bibfnamefont {I.}~\bibnamefont {{Gott}},
  \bibfnamefont {J.~Richard}},\ }\href {\doibase 10.1086/151605} {\bibfield
  {journal} {\bibinfo  {journal} {\apj}\ }\textbf {\bibinfo {volume} {176}},\
  \bibinfo {pages} {1} (\bibinfo {year} {1972})}\BibitemShut {NoStop}%
\bibitem [{\citenamefont {{Lacey}}\ and\ \citenamefont
  {{Cole}}(1993)}]{lacey93}%
  \BibitemOpen
  \bibfield  {author} {\bibinfo {author} {\bibfnamefont {C.}~\bibnamefont
  {{Lacey}}}\ and\ \bibinfo {author} {\bibfnamefont {S.}~\bibnamefont
  {{Cole}}},\ }\href {\doibase 10.1093/mnras/262.3.627} {\bibfield  {journal}
  {\bibinfo  {journal} {\mnras}\ }\textbf {\bibinfo {volume} {262}},\ \bibinfo
  {pages} {627} (\bibinfo {year} {1993})}\BibitemShut {NoStop}%
\bibitem [{\citenamefont {{Eke}}\ \emph {et~al.}(1996)\citenamefont {{Eke}},
  \citenamefont {{Cole}},\ and\ \citenamefont {{Frenk}}}]{eke96}%
  \BibitemOpen
  \bibfield  {author} {\bibinfo {author} {\bibfnamefont {V.~R.}\ \bibnamefont
  {{Eke}}}, \bibinfo {author} {\bibfnamefont {S.}~\bibnamefont {{Cole}}}, \
  and\ \bibinfo {author} {\bibfnamefont {C.~S.}\ \bibnamefont {{Frenk}}},\
  }\href {\doibase 10.1093/mnras/282.1.263} {\bibfield  {journal} {\bibinfo
  {journal} {\mnras}\ }\textbf {\bibinfo {volume} {282}},\ \bibinfo {pages}
  {263} (\bibinfo {year} {1996})},\ \Eprint
  {http://arxiv.org/abs/astro-ph/9601088} {arXiv:astro-ph/9601088 [astro-ph]}
  \BibitemShut {NoStop}%
\bibitem [{\citenamefont {{Odderskov}}\ \emph {et~al.}(2014)\citenamefont
  {{Odderskov}}, \citenamefont {{Hannestad}},\ and\ \citenamefont
  {{Haugb{\o}lle}}}]{Odderskov2014}%
  \BibitemOpen
  \bibfield  {author} {\bibinfo {author} {\bibfnamefont {I.}~\bibnamefont
  {{Odderskov}}}, \bibinfo {author} {\bibfnamefont {S.}~\bibnamefont
  {{Hannestad}}}, \ and\ \bibinfo {author} {\bibfnamefont {T.}~\bibnamefont
  {{Haugb{\o}lle}}},\ }\href {\doibase 10.1088/1475-7516/2014/10/028}
  {\bibfield  {journal} {\bibinfo  {journal} {\jcap}\ }\textbf {\bibinfo
  {volume} {2014}},\ \bibinfo {eid} {028} (\bibinfo {year} {2014})},\ \Eprint
  {http://arxiv.org/abs/1407.7364} {arXiv:1407.7364 [astro-ph.CO]} \BibitemShut
  {NoStop}%
\bibitem [{\citenamefont {{Odderskov}}\ \emph {et~al.}(2017)\citenamefont
  {{Odderskov}}, \citenamefont {{Hannestad}},\ and\ \citenamefont
  {{Brandbyge}}}]{Odderskov2017}%
  \BibitemOpen
  \bibfield  {author} {\bibinfo {author} {\bibfnamefont {I.}~\bibnamefont
  {{Odderskov}}}, \bibinfo {author} {\bibfnamefont {S.}~\bibnamefont
  {{Hannestad}}}, \ and\ \bibinfo {author} {\bibfnamefont {J.}~\bibnamefont
  {{Brandbyge}}},\ }\href {\doibase 10.1088/1475-7516/2017/03/022} {\bibfield
  {journal} {\bibinfo  {journal} {\jcap}\ }\textbf {\bibinfo {volume} {2017}},\
  \bibinfo {eid} {022} (\bibinfo {year} {2017})},\ \Eprint
  {http://arxiv.org/abs/1701.05391} {arXiv:1701.05391 [astro-ph.CO]}
  \BibitemShut {NoStop}%
\bibitem [{\citenamefont {{Chuang}}\ \emph {et~al.}(2019)\citenamefont
  {{Chuang}}, \citenamefont {{Yepes}}, \citenamefont {{Kitaura}}, \citenamefont
  {{Pellejero-Ibanez}}, \citenamefont {{Rodr{\'\i}guez-Torres}}, \citenamefont
  {{Feng}}, \citenamefont {{Metcalf}}, \citenamefont {{Wechsler}},
  \citenamefont {{Zhao}}, \citenamefont {{To}}, \citenamefont {{Alam}},
  \citenamefont {{Banerjee}}, \citenamefont {{DeRose}}, \citenamefont
  {{Giocoli}}, \citenamefont {{Knebe}},\ and\ \citenamefont
  {{Reyes}}}]{Chuang2019}%
  \BibitemOpen
  \bibfield  {author} {\bibinfo {author} {\bibfnamefont {C.-H.}\ \bibnamefont
  {{Chuang}}}, \bibinfo {author} {\bibfnamefont {G.}~\bibnamefont {{Yepes}}},
  \bibinfo {author} {\bibfnamefont {F.-S.}\ \bibnamefont {{Kitaura}}}, \bibinfo
  {author} {\bibfnamefont {M.}~\bibnamefont {{Pellejero-Ibanez}}}, \bibinfo
  {author} {\bibfnamefont {S.}~\bibnamefont {{Rodr{\'\i}guez-Torres}}},
  \bibinfo {author} {\bibfnamefont {Y.}~\bibnamefont {{Feng}}}, \bibinfo
  {author} {\bibfnamefont {R.~B.}\ \bibnamefont {{Metcalf}}}, \bibinfo {author}
  {\bibfnamefont {R.~H.}\ \bibnamefont {{Wechsler}}}, \bibinfo {author}
  {\bibfnamefont {C.}~\bibnamefont {{Zhao}}}, \bibinfo {author} {\bibfnamefont
  {C.-H.}\ \bibnamefont {{To}}}, \bibinfo {author} {\bibfnamefont
  {S.}~\bibnamefont {{Alam}}}, \bibinfo {author} {\bibfnamefont
  {A.}~\bibnamefont {{Banerjee}}}, \bibinfo {author} {\bibfnamefont
  {J.}~\bibnamefont {{DeRose}}}, \bibinfo {author} {\bibfnamefont
  {C.}~\bibnamefont {{Giocoli}}}, \bibinfo {author} {\bibfnamefont
  {A.}~\bibnamefont {{Knebe}}}, \ and\ \bibinfo {author} {\bibfnamefont
  {G.}~\bibnamefont {{Reyes}}},\ }\href {\doibase 10.1093/mnras/stz1233}
  {\bibfield  {journal} {\bibinfo  {journal} {\mnras}\ }\textbf {\bibinfo
  {volume} {487}},\ \bibinfo {pages} {48} (\bibinfo {year} {2019})},\ \Eprint
  {http://arxiv.org/abs/1811.02111} {arXiv:1811.02111 [astro-ph.CO]}
  \BibitemShut {NoStop}%
\bibitem [{\citenamefont {{Planck Collaboration}}\ \emph
  {et~al.}(2016)\citenamefont {{Planck Collaboration}}, \citenamefont {{Ade}},
  \citenamefont {{Aghanim}}, \citenamefont {{Arnaud}}, \citenamefont
  {{Ashdown}}, \citenamefont {{Aumont}}, \citenamefont {{Baccigalupi}},
  \citenamefont {{Banday}}, \citenamefont {{Barreiro}}, \citenamefont
  {{Bartlett}}, \citenamefont {{Bartolo}}, \citenamefont {{Battaner}},
  \citenamefont {{Battye}}, \citenamefont {{Benabed}}, \citenamefont
  {{Beno{\^\i}t}}, \citenamefont {{Benoit-L{\'e}vy}}, \citenamefont
  {{Bernard}}, \citenamefont {{Bersanelli}}, \citenamefont {{Bielewicz}},
  \citenamefont {{Bock}}, \citenamefont {{Bonaldi}}, \citenamefont
  {{Bonavera}}, \citenamefont {{Bond}}, \citenamefont {{Borrill}},
  \citenamefont {{Bouchet}}, \citenamefont {{Boulanger}}, \citenamefont
  {{Bucher}}, \citenamefont {{Burigana}}, \citenamefont {{Butler}},
  \citenamefont {{Calabrese}}, \citenamefont {{Cardoso}}, \citenamefont
  {{Catalano}}, \citenamefont {{Challinor}}, \citenamefont {{Chamballu}},
  \citenamefont {{Chary}}, \citenamefont {{Chiang}}, \citenamefont {{Chluba}},
  \citenamefont {{Christensen}}, \citenamefont {{Church}}, \citenamefont
  {{Clements}}, \citenamefont {{Colombi}}, \citenamefont {{Colombo}},
  \citenamefont {{Combet}}, \citenamefont {{Coulais}}, \citenamefont {{Crill}},
  \citenamefont {{Curto}}, \citenamefont {{Cuttaia}}, \citenamefont {{Danese}},
  \citenamefont {{Davies}}, \citenamefont {{Davis}}, \citenamefont {{de
  Bernardis}}, \citenamefont {{de Rosa}}, \citenamefont {{de Zotti}},
  \citenamefont {{Delabrouille}}, \citenamefont {{D{\'e}sert}}, \citenamefont
  {{Di Valentino}}, \citenamefont {{Dickinson}}, \citenamefont {{Diego}},
  \citenamefont {{Dolag}}, \citenamefont {{Dole}}, \citenamefont {{Donzelli}},
  \citenamefont {{Dor{\'e}}}, \citenamefont {{Douspis}}, \citenamefont
  {{Ducout}}, \citenamefont {{Dunkley}}, \citenamefont {{Dupac}}, \citenamefont
  {{Efstathiou}}, \citenamefont {{Elsner}}, \citenamefont {{En{\ss}lin}},
  \citenamefont {{Eriksen}}, \citenamefont {{Farhang}}, \citenamefont
  {{Fergusson}}, \citenamefont {{Finelli}}, \citenamefont {{Forni}},
  \citenamefont {{Frailis}}, \citenamefont {{Fraisse}}, \citenamefont
  {{Franceschi}}, \citenamefont {{Frejsel}}, \citenamefont {{Galeotta}},
  \citenamefont {{Galli}}, \citenamefont {{Ganga}}, \citenamefont {{Gauthier}},
  \citenamefont {{Gerbino}}, \citenamefont {{Ghosh}}, \citenamefont {{Giard}},
  \citenamefont {{Giraud-H{\'e}raud}}, \citenamefont {{Giusarma}},
  \citenamefont {{Gjerl{\o}w}}, \citenamefont {{Gonz{\'a}lez-Nuevo}},
  \citenamefont {{G{\'o}rski}}, \citenamefont {{Gratton}}, \citenamefont
  {{Gregorio}}, \citenamefont {{Gruppuso}}, \citenamefont {{Gudmundsson}},
  \citenamefont {{Hamann}}, \citenamefont {{Hansen}}, \citenamefont {{Hanson}},
  \citenamefont {{Harrison}}, \citenamefont {{Helou}}, \citenamefont
  {{Henrot-Versill{\'e}}}, \citenamefont {{Hern{\'a}ndez-Monteagudo}},
  \citenamefont {{Herranz}}, \citenamefont {{Hildebrandt}}, \citenamefont
  {{Hivon}}, \citenamefont {{Hobson}}, \citenamefont {{Holmes}}, \citenamefont
  {{Hornstrup}}, \citenamefont {{Hovest}}, \citenamefont {{Huang}},
  \citenamefont {{Huffenberger}}, \citenamefont {{Hurier}}, \citenamefont
  {{Jaffe}}, \citenamefont {{Jaffe}}, \citenamefont {{Jones}}, \citenamefont
  {{Juvela}}, \citenamefont {{Keih{\"a}nen}}, \citenamefont {{Keskitalo}},
  \citenamefont {{Kisner}}, \citenamefont {{Kneissl}}, \citenamefont
  {{Knoche}}, \citenamefont {{Knox}}, \citenamefont {{Kunz}}, \citenamefont
  {{Kurki-Suonio}}, \citenamefont {{Lagache}}, \citenamefont
  {{L{\"a}hteenm{\"a}ki}}, \citenamefont {{Lamarre}}, \citenamefont
  {{Lasenby}}, \citenamefont {{Lattanzi}}, \citenamefont {{Lawrence}},
  \citenamefont {{Leahy}}, \citenamefont {{Leonardi}}, \citenamefont
  {{Lesgourgues}}, \citenamefont {{Levrier}}, \citenamefont {{Lewis}},
  \citenamefont {{Liguori}}, \citenamefont {{Lilje}}, \citenamefont
  {{Linden-V{\o}rnle}}, \citenamefont {{L{\'o}pez-Caniego}}, \citenamefont
  {{Lubin}}, \citenamefont {{Mac{\'\i}as-P{\'e}rez}}, \citenamefont {{Maggio}},
  \citenamefont {{Maino}}, \citenamefont {{Mandolesi}}, \citenamefont
  {{Mangilli}}, \citenamefont {{Marchini}}, \citenamefont {{Maris}},
  \citenamefont {{Martin}}, \citenamefont {{Martinelli}}, \citenamefont
  {{Mart{\'\i}nez-Gonz{\'a}lez}}, \citenamefont {{Masi}}, \citenamefont
  {{Matarrese}}, \citenamefont {{McGehee}}, \citenamefont {{Meinhold}},
  \citenamefont {{Melchiorri}}, \citenamefont {{Melin}}, \citenamefont
  {{Mendes}}, \citenamefont {{Mennella}}, \citenamefont {{Migliaccio}},
  \citenamefont {{Millea}}, \citenamefont {{Mitra}}, \citenamefont
  {{Miville-Desch{\^e}nes}}, \citenamefont {{Moneti}}, \citenamefont
  {{Montier}}, \citenamefont {{Morgante}}, \citenamefont {{Mortlock}},
  \citenamefont {{Moss}}, \citenamefont {{Munshi}}, \citenamefont {{Murphy}},
  \citenamefont {{Naselsky}}, \citenamefont {{Nati}}, \citenamefont {{Natoli}},
  \citenamefont {{Netterfield}}, \citenamefont {{N{\o}rgaard-Nielsen}},
  \citenamefont {{Noviello}}, \citenamefont {{Novikov}}, \citenamefont
  {{Novikov}}, \citenamefont {{Oxborrow}}, \citenamefont {{Paci}},
  \citenamefont {{Pagano}}, \citenamefont {{Pajot}}, \citenamefont
  {{Paladini}}, \citenamefont {{Paoletti}}, \citenamefont {{Partridge}},
  \citenamefont {{Pasian}}, \citenamefont {{Patanchon}}, \citenamefont
  {{Pearson}}, \citenamefont {{Perdereau}}, \citenamefont {{Perotto}},
  \citenamefont {{Perrotta}}, \citenamefont {{Pettorino}}, \citenamefont
  {{Piacentini}}, \citenamefont {{Piat}}, \citenamefont {{Pierpaoli}},
  \citenamefont {{Pietrobon}}, \citenamefont {{Plaszczynski}}, \citenamefont
  {{Pointecouteau}}, \citenamefont {{Polenta}}, \citenamefont {{Popa}},
  \citenamefont {{Pratt}}, \citenamefont {{Pr{\'e}zeau}}, \citenamefont
  {{Prunet}}, \citenamefont {{Puget}}, \citenamefont {{Rachen}}, \citenamefont
  {{Reach}}, \citenamefont {{Rebolo}}, \citenamefont {{Reinecke}},
  \citenamefont {{Remazeilles}}, \citenamefont {{Renault}}, \citenamefont
  {{Renzi}}, \citenamefont {{Ristorcelli}}, \citenamefont {{Rocha}},
  \citenamefont {{Rosset}}, \citenamefont {{Rossetti}}, \citenamefont
  {{Roudier}}, \citenamefont {{Rouill{\'e} d'Orfeuil}}, \citenamefont
  {{Rowan-Robinson}}, \citenamefont {{Rubi{\~n}o-Mart{\'\i}n}}, \citenamefont
  {{Rusholme}}, \citenamefont {{Said}}, \citenamefont {{Salvatelli}},
  \citenamefont {{Salvati}}, \citenamefont {{Sandri}}, \citenamefont
  {{Santos}}, \citenamefont {{Savelainen}}, \citenamefont {{Savini}},
  \citenamefont {{Scott}}, \citenamefont {{Seiffert}}, \citenamefont {{Serra}},
  \citenamefont {{Shellard}}, \citenamefont {{Spencer}}, \citenamefont
  {{Spinelli}}, \citenamefont {{Stolyarov}}, \citenamefont {{Stompor}},
  \citenamefont {{Sudiwala}}, \citenamefont {{Sunyaev}}, \citenamefont
  {{Sutton}}, \citenamefont {{Suur-Uski}}, \citenamefont {{Sygnet}},
  \citenamefont {{Tauber}}, \citenamefont {{Terenzi}}, \citenamefont
  {{Toffolatti}}, \citenamefont {{Tomasi}}, \citenamefont {{Tristram}},
  \citenamefont {{Trombetti}}, \citenamefont {{Tucci}}, \citenamefont
  {{Tuovinen}}, \citenamefont {{T{\"u}rler}}, \citenamefont {{Umana}},
  \citenamefont {{Valenziano}}, \citenamefont {{Valiviita}}, \citenamefont
  {{Van Tent}}, \citenamefont {{Vielva}}, \citenamefont {{Villa}},
  \citenamefont {{Wade}}, \citenamefont {{Wandelt}}, \citenamefont {{Wehus}},
  \citenamefont {{White}}, \citenamefont {{White}}, \citenamefont
  {{Wilkinson}}, \citenamefont {{Yvon}}, \citenamefont {{Zacchei}},\ and\
  \citenamefont {{Zonca}}}]{Planck_2016}%
  \BibitemOpen
  \bibfield  {author} {\bibinfo {author} {\bibnamefont {{Planck
  Collaboration}}}, \bibinfo {author} {\bibfnamefont {P.~A.~R.}\ \bibnamefont
  {{Ade}}}, \bibinfo {author} {\bibfnamefont {N.}~\bibnamefont {{Aghanim}}},
  \bibinfo {author} {\bibfnamefont {M.}~\bibnamefont {{Arnaud}}}, \bibinfo
  {author} {\bibfnamefont {M.}~\bibnamefont {{Ashdown}}}, \bibinfo {author}
  {\bibfnamefont {J.}~\bibnamefont {{Aumont}}}, \bibinfo {author}
  {\bibfnamefont {C.}~\bibnamefont {{Baccigalupi}}}, \bibinfo {author}
  {\bibfnamefont {A.~J.}\ \bibnamefont {{Banday}}}, \bibinfo {author}
  {\bibfnamefont {R.~B.}\ \bibnamefont {{Barreiro}}}, \bibinfo {author}
  {\bibfnamefont {J.~G.}\ \bibnamefont {{Bartlett}}}, \bibinfo {author}
  {\bibfnamefont {N.}~\bibnamefont {{Bartolo}}}, \bibinfo {author}
  {\bibfnamefont {E.}~\bibnamefont {{Battaner}}}, \bibinfo {author}
  {\bibfnamefont {R.}~\bibnamefont {{Battye}}}, \bibinfo {author}
  {\bibfnamefont {K.}~\bibnamefont {{Benabed}}}, \bibinfo {author}
  {\bibfnamefont {A.}~\bibnamefont {{Beno{\^\i}t}}}, \bibinfo {author}
  {\bibfnamefont {A.}~\bibnamefont {{Benoit-L{\'e}vy}}}, \bibinfo {author}
  {\bibfnamefont {J.~P.}\ \bibnamefont {{Bernard}}}, \bibinfo {author}
  {\bibfnamefont {M.}~\bibnamefont {{Bersanelli}}}, \bibinfo {author}
  {\bibfnamefont {P.}~\bibnamefont {{Bielewicz}}}, \bibinfo {author}
  {\bibfnamefont {J.~J.}\ \bibnamefont {{Bock}}}, \bibinfo {author}
  {\bibfnamefont {A.}~\bibnamefont {{Bonaldi}}}, \bibinfo {author}
  {\bibfnamefont {L.}~\bibnamefont {{Bonavera}}}, \bibinfo {author}
  {\bibfnamefont {J.~R.}\ \bibnamefont {{Bond}}}, \bibinfo {author}
  {\bibfnamefont {J.}~\bibnamefont {{Borrill}}}, \bibinfo {author}
  {\bibfnamefont {F.~R.}\ \bibnamefont {{Bouchet}}}, \bibinfo {author}
  {\bibfnamefont {F.}~\bibnamefont {{Boulanger}}}, \bibinfo {author}
  {\bibfnamefont {M.}~\bibnamefont {{Bucher}}}, \bibinfo {author}
  {\bibfnamefont {C.}~\bibnamefont {{Burigana}}}, \bibinfo {author}
  {\bibfnamefont {R.~C.}\ \bibnamefont {{Butler}}}, \bibinfo {author}
  {\bibfnamefont {E.}~\bibnamefont {{Calabrese}}}, \bibinfo {author}
  {\bibfnamefont {J.~F.}\ \bibnamefont {{Cardoso}}}, \bibinfo {author}
  {\bibfnamefont {A.}~\bibnamefont {{Catalano}}}, \bibinfo {author}
  {\bibfnamefont {A.}~\bibnamefont {{Challinor}}}, \bibinfo {author}
  {\bibfnamefont {A.}~\bibnamefont {{Chamballu}}}, \bibinfo {author}
  {\bibfnamefont {R.~R.}\ \bibnamefont {{Chary}}}, \bibinfo {author}
  {\bibfnamefont {H.~C.}\ \bibnamefont {{Chiang}}}, \bibinfo {author}
  {\bibfnamefont {J.}~\bibnamefont {{Chluba}}}, \bibinfo {author}
  {\bibfnamefont {P.~R.}\ \bibnamefont {{Christensen}}}, \bibinfo {author}
  {\bibfnamefont {S.}~\bibnamefont {{Church}}}, \bibinfo {author}
  {\bibfnamefont {D.~L.}\ \bibnamefont {{Clements}}}, \bibinfo {author}
  {\bibfnamefont {S.}~\bibnamefont {{Colombi}}}, \bibinfo {author}
  {\bibfnamefont {L.~P.~L.}\ \bibnamefont {{Colombo}}}, \bibinfo {author}
  {\bibfnamefont {C.}~\bibnamefont {{Combet}}}, \bibinfo {author}
  {\bibfnamefont {A.}~\bibnamefont {{Coulais}}}, \bibinfo {author}
  {\bibfnamefont {B.~P.}\ \bibnamefont {{Crill}}}, \bibinfo {author}
  {\bibfnamefont {A.}~\bibnamefont {{Curto}}}, \bibinfo {author} {\bibfnamefont
  {F.}~\bibnamefont {{Cuttaia}}}, \bibinfo {author} {\bibfnamefont
  {L.}~\bibnamefont {{Danese}}}, \bibinfo {author} {\bibfnamefont {R.~D.}\
  \bibnamefont {{Davies}}}, \bibinfo {author} {\bibfnamefont {R.~J.}\
  \bibnamefont {{Davis}}}, \bibinfo {author} {\bibfnamefont {P.}~\bibnamefont
  {{de Bernardis}}}, \bibinfo {author} {\bibfnamefont {A.}~\bibnamefont {{de
  Rosa}}}, \bibinfo {author} {\bibfnamefont {G.}~\bibnamefont {{de Zotti}}},
  \bibinfo {author} {\bibfnamefont {J.}~\bibnamefont {{Delabrouille}}},
  \bibinfo {author} {\bibfnamefont {F.~X.}\ \bibnamefont {{D{\'e}sert}}},
  \bibinfo {author} {\bibfnamefont {E.}~\bibnamefont {{Di Valentino}}},
  \bibinfo {author} {\bibfnamefont {C.}~\bibnamefont {{Dickinson}}}, \bibinfo
  {author} {\bibfnamefont {J.~M.}\ \bibnamefont {{Diego}}}, \bibinfo {author}
  {\bibfnamefont {K.}~\bibnamefont {{Dolag}}}, \bibinfo {author} {\bibfnamefont
  {H.}~\bibnamefont {{Dole}}}, \bibinfo {author} {\bibfnamefont
  {S.}~\bibnamefont {{Donzelli}}}, \bibinfo {author} {\bibfnamefont
  {O.}~\bibnamefont {{Dor{\'e}}}}, \bibinfo {author} {\bibfnamefont
  {M.}~\bibnamefont {{Douspis}}}, \bibinfo {author} {\bibfnamefont
  {A.}~\bibnamefont {{Ducout}}}, \bibinfo {author} {\bibfnamefont
  {J.}~\bibnamefont {{Dunkley}}}, \bibinfo {author} {\bibfnamefont
  {X.}~\bibnamefont {{Dupac}}}, \bibinfo {author} {\bibfnamefont
  {G.}~\bibnamefont {{Efstathiou}}}, \bibinfo {author} {\bibfnamefont
  {F.}~\bibnamefont {{Elsner}}}, \bibinfo {author} {\bibfnamefont {T.~A.}\
  \bibnamefont {{En{\ss}lin}}}, \bibinfo {author} {\bibfnamefont {H.~K.}\
  \bibnamefont {{Eriksen}}}, \bibinfo {author} {\bibfnamefont {M.}~\bibnamefont
  {{Farhang}}}, \bibinfo {author} {\bibfnamefont {J.}~\bibnamefont
  {{Fergusson}}}, \bibinfo {author} {\bibfnamefont {F.}~\bibnamefont
  {{Finelli}}}, \bibinfo {author} {\bibfnamefont {O.}~\bibnamefont {{Forni}}},
  \bibinfo {author} {\bibfnamefont {M.}~\bibnamefont {{Frailis}}}, \bibinfo
  {author} {\bibfnamefont {A.~A.}\ \bibnamefont {{Fraisse}}}, \bibinfo {author}
  {\bibfnamefont {E.}~\bibnamefont {{Franceschi}}}, \bibinfo {author}
  {\bibfnamefont {A.}~\bibnamefont {{Frejsel}}}, \bibinfo {author}
  {\bibfnamefont {S.}~\bibnamefont {{Galeotta}}}, \bibinfo {author}
  {\bibfnamefont {S.}~\bibnamefont {{Galli}}}, \bibinfo {author} {\bibfnamefont
  {K.}~\bibnamefont {{Ganga}}}, \bibinfo {author} {\bibfnamefont
  {C.}~\bibnamefont {{Gauthier}}}, \bibinfo {author} {\bibfnamefont
  {M.}~\bibnamefont {{Gerbino}}}, \bibinfo {author} {\bibfnamefont
  {T.}~\bibnamefont {{Ghosh}}}, \bibinfo {author} {\bibfnamefont
  {M.}~\bibnamefont {{Giard}}}, \bibinfo {author} {\bibfnamefont
  {Y.}~\bibnamefont {{Giraud-H{\'e}raud}}}, \bibinfo {author} {\bibfnamefont
  {E.}~\bibnamefont {{Giusarma}}}, \bibinfo {author} {\bibfnamefont
  {E.}~\bibnamefont {{Gjerl{\o}w}}}, \bibinfo {author} {\bibfnamefont
  {J.}~\bibnamefont {{Gonz{\'a}lez-Nuevo}}}, \bibinfo {author} {\bibfnamefont
  {K.~M.}\ \bibnamefont {{G{\'o}rski}}}, \bibinfo {author} {\bibfnamefont
  {S.}~\bibnamefont {{Gratton}}}, \bibinfo {author} {\bibfnamefont
  {A.}~\bibnamefont {{Gregorio}}}, \bibinfo {author} {\bibfnamefont
  {A.}~\bibnamefont {{Gruppuso}}}, \bibinfo {author} {\bibfnamefont {J.~E.}\
  \bibnamefont {{Gudmundsson}}}, \bibinfo {author} {\bibfnamefont
  {J.}~\bibnamefont {{Hamann}}}, \bibinfo {author} {\bibfnamefont {F.~K.}\
  \bibnamefont {{Hansen}}}, \bibinfo {author} {\bibfnamefont {D.}~\bibnamefont
  {{Hanson}}}, \bibinfo {author} {\bibfnamefont {D.~L.}\ \bibnamefont
  {{Harrison}}}, \bibinfo {author} {\bibfnamefont {G.}~\bibnamefont {{Helou}}},
  \bibinfo {author} {\bibfnamefont {S.}~\bibnamefont {{Henrot-Versill{\'e}}}},
  \bibinfo {author} {\bibfnamefont {C.}~\bibnamefont
  {{Hern{\'a}ndez-Monteagudo}}}, \bibinfo {author} {\bibfnamefont
  {D.}~\bibnamefont {{Herranz}}}, \bibinfo {author} {\bibfnamefont {S.~R.}\
  \bibnamefont {{Hildebrandt}}}, \bibinfo {author} {\bibfnamefont
  {E.}~\bibnamefont {{Hivon}}}, \bibinfo {author} {\bibfnamefont
  {M.}~\bibnamefont {{Hobson}}}, \bibinfo {author} {\bibfnamefont {W.~A.}\
  \bibnamefont {{Holmes}}}, \bibinfo {author} {\bibfnamefont {A.}~\bibnamefont
  {{Hornstrup}}}, \bibinfo {author} {\bibfnamefont {W.}~\bibnamefont
  {{Hovest}}}, \bibinfo {author} {\bibfnamefont {Z.}~\bibnamefont {{Huang}}},
  \bibinfo {author} {\bibfnamefont {K.~M.}\ \bibnamefont {{Huffenberger}}},
  \bibinfo {author} {\bibfnamefont {G.}~\bibnamefont {{Hurier}}}, \bibinfo
  {author} {\bibfnamefont {A.~H.}\ \bibnamefont {{Jaffe}}}, \bibinfo {author}
  {\bibfnamefont {T.~R.}\ \bibnamefont {{Jaffe}}}, \bibinfo {author}
  {\bibfnamefont {W.~C.}\ \bibnamefont {{Jones}}}, \bibinfo {author}
  {\bibfnamefont {M.}~\bibnamefont {{Juvela}}}, \bibinfo {author}
  {\bibfnamefont {E.}~\bibnamefont {{Keih{\"a}nen}}}, \bibinfo {author}
  {\bibfnamefont {R.}~\bibnamefont {{Keskitalo}}}, \bibinfo {author}
  {\bibfnamefont {T.~S.}\ \bibnamefont {{Kisner}}}, \bibinfo {author}
  {\bibfnamefont {R.}~\bibnamefont {{Kneissl}}}, \bibinfo {author}
  {\bibfnamefont {J.}~\bibnamefont {{Knoche}}}, \bibinfo {author}
  {\bibfnamefont {L.}~\bibnamefont {{Knox}}}, \bibinfo {author} {\bibfnamefont
  {M.}~\bibnamefont {{Kunz}}}, \bibinfo {author} {\bibfnamefont
  {H.}~\bibnamefont {{Kurki-Suonio}}}, \bibinfo {author} {\bibfnamefont
  {G.}~\bibnamefont {{Lagache}}}, \bibinfo {author} {\bibfnamefont
  {A.}~\bibnamefont {{L{\"a}hteenm{\"a}ki}}}, \bibinfo {author} {\bibfnamefont
  {J.~M.}\ \bibnamefont {{Lamarre}}}, \bibinfo {author} {\bibfnamefont
  {A.}~\bibnamefont {{Lasenby}}}, \bibinfo {author} {\bibfnamefont
  {M.}~\bibnamefont {{Lattanzi}}}, \bibinfo {author} {\bibfnamefont {C.~R.}\
  \bibnamefont {{Lawrence}}}, \bibinfo {author} {\bibfnamefont {J.~P.}\
  \bibnamefont {{Leahy}}}, \bibinfo {author} {\bibfnamefont {R.}~\bibnamefont
  {{Leonardi}}}, \bibinfo {author} {\bibfnamefont {J.}~\bibnamefont
  {{Lesgourgues}}}, \bibinfo {author} {\bibfnamefont {F.}~\bibnamefont
  {{Levrier}}}, \bibinfo {author} {\bibfnamefont {A.}~\bibnamefont {{Lewis}}},
  \bibinfo {author} {\bibfnamefont {M.}~\bibnamefont {{Liguori}}}, \bibinfo
  {author} {\bibfnamefont {P.~B.}\ \bibnamefont {{Lilje}}}, \bibinfo {author}
  {\bibfnamefont {M.}~\bibnamefont {{Linden-V{\o}rnle}}}, \bibinfo {author}
  {\bibfnamefont {M.}~\bibnamefont {{L{\'o}pez-Caniego}}}, \bibinfo {author}
  {\bibfnamefont {P.~M.}\ \bibnamefont {{Lubin}}}, \bibinfo {author}
  {\bibfnamefont {J.~F.}\ \bibnamefont {{Mac{\'\i}as-P{\'e}rez}}}, \bibinfo
  {author} {\bibfnamefont {G.}~\bibnamefont {{Maggio}}}, \bibinfo {author}
  {\bibfnamefont {D.}~\bibnamefont {{Maino}}}, \bibinfo {author} {\bibfnamefont
  {N.}~\bibnamefont {{Mandolesi}}}, \bibinfo {author} {\bibfnamefont
  {A.}~\bibnamefont {{Mangilli}}}, \bibinfo {author} {\bibfnamefont
  {A.}~\bibnamefont {{Marchini}}}, \bibinfo {author} {\bibfnamefont
  {M.}~\bibnamefont {{Maris}}}, \bibinfo {author} {\bibfnamefont {P.~G.}\
  \bibnamefont {{Martin}}}, \bibinfo {author} {\bibfnamefont {M.}~\bibnamefont
  {{Martinelli}}}, \bibinfo {author} {\bibfnamefont {E.}~\bibnamefont
  {{Mart{\'\i}nez-Gonz{\'a}lez}}}, \bibinfo {author} {\bibfnamefont
  {S.}~\bibnamefont {{Masi}}}, \bibinfo {author} {\bibfnamefont
  {S.}~\bibnamefont {{Matarrese}}}, \bibinfo {author} {\bibfnamefont
  {P.}~\bibnamefont {{McGehee}}}, \bibinfo {author} {\bibfnamefont {P.~R.}\
  \bibnamefont {{Meinhold}}}, \bibinfo {author} {\bibfnamefont
  {A.}~\bibnamefont {{Melchiorri}}}, \bibinfo {author} {\bibfnamefont {J.~B.}\
  \bibnamefont {{Melin}}}, \bibinfo {author} {\bibfnamefont {L.}~\bibnamefont
  {{Mendes}}}, \bibinfo {author} {\bibfnamefont {A.}~\bibnamefont
  {{Mennella}}}, \bibinfo {author} {\bibfnamefont {M.}~\bibnamefont
  {{Migliaccio}}}, \bibinfo {author} {\bibfnamefont {M.}~\bibnamefont
  {{Millea}}}, \bibinfo {author} {\bibfnamefont {S.}~\bibnamefont {{Mitra}}},
  \bibinfo {author} {\bibfnamefont {M.~A.}\ \bibnamefont
  {{Miville-Desch{\^e}nes}}}, \bibinfo {author} {\bibfnamefont
  {A.}~\bibnamefont {{Moneti}}}, \bibinfo {author} {\bibfnamefont
  {L.}~\bibnamefont {{Montier}}}, \bibinfo {author} {\bibfnamefont
  {G.}~\bibnamefont {{Morgante}}}, \bibinfo {author} {\bibfnamefont
  {D.}~\bibnamefont {{Mortlock}}}, \bibinfo {author} {\bibfnamefont
  {A.}~\bibnamefont {{Moss}}}, \bibinfo {author} {\bibfnamefont
  {D.}~\bibnamefont {{Munshi}}}, \bibinfo {author} {\bibfnamefont {J.~A.}\
  \bibnamefont {{Murphy}}}, \bibinfo {author} {\bibfnamefont {P.}~\bibnamefont
  {{Naselsky}}}, \bibinfo {author} {\bibfnamefont {F.}~\bibnamefont {{Nati}}},
  \bibinfo {author} {\bibfnamefont {P.}~\bibnamefont {{Natoli}}}, \bibinfo
  {author} {\bibfnamefont {C.~B.}\ \bibnamefont {{Netterfield}}}, \bibinfo
  {author} {\bibfnamefont {H.~U.}\ \bibnamefont {{N{\o}rgaard-Nielsen}}},
  \bibinfo {author} {\bibfnamefont {F.}~\bibnamefont {{Noviello}}}, \bibinfo
  {author} {\bibfnamefont {D.}~\bibnamefont {{Novikov}}}, \bibinfo {author}
  {\bibfnamefont {I.}~\bibnamefont {{Novikov}}}, \bibinfo {author}
  {\bibfnamefont {C.~A.}\ \bibnamefont {{Oxborrow}}}, \bibinfo {author}
  {\bibfnamefont {F.}~\bibnamefont {{Paci}}}, \bibinfo {author} {\bibfnamefont
  {L.}~\bibnamefont {{Pagano}}}, \bibinfo {author} {\bibfnamefont
  {F.}~\bibnamefont {{Pajot}}}, \bibinfo {author} {\bibfnamefont
  {R.}~\bibnamefont {{Paladini}}}, \bibinfo {author} {\bibfnamefont
  {D.}~\bibnamefont {{Paoletti}}}, \bibinfo {author} {\bibfnamefont
  {B.}~\bibnamefont {{Partridge}}}, \bibinfo {author} {\bibfnamefont
  {F.}~\bibnamefont {{Pasian}}}, \bibinfo {author} {\bibfnamefont
  {G.}~\bibnamefont {{Patanchon}}}, \bibinfo {author} {\bibfnamefont {T.~J.}\
  \bibnamefont {{Pearson}}}, \bibinfo {author} {\bibfnamefont {O.}~\bibnamefont
  {{Perdereau}}}, \bibinfo {author} {\bibfnamefont {L.}~\bibnamefont
  {{Perotto}}}, \bibinfo {author} {\bibfnamefont {F.}~\bibnamefont
  {{Perrotta}}}, \bibinfo {author} {\bibfnamefont {V.}~\bibnamefont
  {{Pettorino}}}, \bibinfo {author} {\bibfnamefont {F.}~\bibnamefont
  {{Piacentini}}}, \bibinfo {author} {\bibfnamefont {M.}~\bibnamefont
  {{Piat}}}, \bibinfo {author} {\bibfnamefont {E.}~\bibnamefont {{Pierpaoli}}},
  \bibinfo {author} {\bibfnamefont {D.}~\bibnamefont {{Pietrobon}}}, \bibinfo
  {author} {\bibfnamefont {S.}~\bibnamefont {{Plaszczynski}}}, \bibinfo
  {author} {\bibfnamefont {E.}~\bibnamefont {{Pointecouteau}}}, \bibinfo
  {author} {\bibfnamefont {G.}~\bibnamefont {{Polenta}}}, \bibinfo {author}
  {\bibfnamefont {L.}~\bibnamefont {{Popa}}}, \bibinfo {author} {\bibfnamefont
  {G.~W.}\ \bibnamefont {{Pratt}}}, \bibinfo {author} {\bibfnamefont
  {G.}~\bibnamefont {{Pr{\'e}zeau}}}, \bibinfo {author} {\bibfnamefont
  {S.}~\bibnamefont {{Prunet}}}, \bibinfo {author} {\bibfnamefont {J.~L.}\
  \bibnamefont {{Puget}}}, \bibinfo {author} {\bibfnamefont {J.~P.}\
  \bibnamefont {{Rachen}}}, \bibinfo {author} {\bibfnamefont {W.~T.}\
  \bibnamefont {{Reach}}}, \bibinfo {author} {\bibfnamefont {R.}~\bibnamefont
  {{Rebolo}}}, \bibinfo {author} {\bibfnamefont {M.}~\bibnamefont
  {{Reinecke}}}, \bibinfo {author} {\bibfnamefont {M.}~\bibnamefont
  {{Remazeilles}}}, \bibinfo {author} {\bibfnamefont {C.}~\bibnamefont
  {{Renault}}}, \bibinfo {author} {\bibfnamefont {A.}~\bibnamefont {{Renzi}}},
  \bibinfo {author} {\bibfnamefont {I.}~\bibnamefont {{Ristorcelli}}}, \bibinfo
  {author} {\bibfnamefont {G.}~\bibnamefont {{Rocha}}}, \bibinfo {author}
  {\bibfnamefont {C.}~\bibnamefont {{Rosset}}}, \bibinfo {author}
  {\bibfnamefont {M.}~\bibnamefont {{Rossetti}}}, \bibinfo {author}
  {\bibfnamefont {G.}~\bibnamefont {{Roudier}}}, \bibinfo {author}
  {\bibfnamefont {B.}~\bibnamefont {{Rouill{\'e} d'Orfeuil}}}, \bibinfo
  {author} {\bibfnamefont {M.}~\bibnamefont {{Rowan-Robinson}}}, \bibinfo
  {author} {\bibfnamefont {J.~A.}\ \bibnamefont {{Rubi{\~n}o-Mart{\'\i}n}}},
  \bibinfo {author} {\bibfnamefont {B.}~\bibnamefont {{Rusholme}}}, \bibinfo
  {author} {\bibfnamefont {N.}~\bibnamefont {{Said}}}, \bibinfo {author}
  {\bibfnamefont {V.}~\bibnamefont {{Salvatelli}}}, \bibinfo {author}
  {\bibfnamefont {L.}~\bibnamefont {{Salvati}}}, \bibinfo {author}
  {\bibfnamefont {M.}~\bibnamefont {{Sandri}}}, \bibinfo {author}
  {\bibfnamefont {D.}~\bibnamefont {{Santos}}}, \bibinfo {author}
  {\bibfnamefont {M.}~\bibnamefont {{Savelainen}}}, \bibinfo {author}
  {\bibfnamefont {G.}~\bibnamefont {{Savini}}}, \bibinfo {author}
  {\bibfnamefont {D.}~\bibnamefont {{Scott}}}, \bibinfo {author} {\bibfnamefont
  {M.~D.}\ \bibnamefont {{Seiffert}}}, \bibinfo {author} {\bibfnamefont
  {P.}~\bibnamefont {{Serra}}}, \bibinfo {author} {\bibfnamefont {E.~P.~S.}\
  \bibnamefont {{Shellard}}}, \bibinfo {author} {\bibfnamefont {L.~D.}\
  \bibnamefont {{Spencer}}}, \bibinfo {author} {\bibfnamefont {M.}~\bibnamefont
  {{Spinelli}}}, \bibinfo {author} {\bibfnamefont {V.}~\bibnamefont
  {{Stolyarov}}}, \bibinfo {author} {\bibfnamefont {R.}~\bibnamefont
  {{Stompor}}}, \bibinfo {author} {\bibfnamefont {R.}~\bibnamefont
  {{Sudiwala}}}, \bibinfo {author} {\bibfnamefont {R.}~\bibnamefont
  {{Sunyaev}}}, \bibinfo {author} {\bibfnamefont {D.}~\bibnamefont {{Sutton}}},
  \bibinfo {author} {\bibfnamefont {A.~S.}\ \bibnamefont {{Suur-Uski}}},
  \bibinfo {author} {\bibfnamefont {J.~F.}\ \bibnamefont {{Sygnet}}}, \bibinfo
  {author} {\bibfnamefont {J.~A.}\ \bibnamefont {{Tauber}}}, \bibinfo {author}
  {\bibfnamefont {L.}~\bibnamefont {{Terenzi}}}, \bibinfo {author}
  {\bibfnamefont {L.}~\bibnamefont {{Toffolatti}}}, \bibinfo {author}
  {\bibfnamefont {M.}~\bibnamefont {{Tomasi}}}, \bibinfo {author}
  {\bibfnamefont {M.}~\bibnamefont {{Tristram}}}, \bibinfo {author}
  {\bibfnamefont {T.}~\bibnamefont {{Trombetti}}}, \bibinfo {author}
  {\bibfnamefont {M.}~\bibnamefont {{Tucci}}}, \bibinfo {author} {\bibfnamefont
  {J.}~\bibnamefont {{Tuovinen}}}, \bibinfo {author} {\bibfnamefont
  {M.}~\bibnamefont {{T{\"u}rler}}}, \bibinfo {author} {\bibfnamefont
  {G.}~\bibnamefont {{Umana}}}, \bibinfo {author} {\bibfnamefont
  {L.}~\bibnamefont {{Valenziano}}}, \bibinfo {author} {\bibfnamefont
  {J.}~\bibnamefont {{Valiviita}}}, \bibinfo {author} {\bibfnamefont
  {F.}~\bibnamefont {{Van Tent}}}, \bibinfo {author} {\bibfnamefont
  {P.}~\bibnamefont {{Vielva}}}, \bibinfo {author} {\bibfnamefont
  {F.}~\bibnamefont {{Villa}}}, \bibinfo {author} {\bibfnamefont {L.~A.}\
  \bibnamefont {{Wade}}}, \bibinfo {author} {\bibfnamefont {B.~D.}\
  \bibnamefont {{Wandelt}}}, \bibinfo {author} {\bibfnamefont {I.~K.}\
  \bibnamefont {{Wehus}}}, \bibinfo {author} {\bibfnamefont {M.}~\bibnamefont
  {{White}}}, \bibinfo {author} {\bibfnamefont {S.~D.~M.}\ \bibnamefont
  {{White}}}, \bibinfo {author} {\bibfnamefont {A.}~\bibnamefont
  {{Wilkinson}}}, \bibinfo {author} {\bibfnamefont {D.}~\bibnamefont {{Yvon}}},
  \bibinfo {author} {\bibfnamefont {A.}~\bibnamefont {{Zacchei}}}, \ and\
  \bibinfo {author} {\bibfnamefont {A.}~\bibnamefont {{Zonca}}},\ }\href
  {\doibase 10.1051/0004-6361/201525830} {\bibfield  {journal} {\bibinfo
  {journal} {\aap}\ }\textbf {\bibinfo {volume} {594}},\ \bibinfo {eid} {A13}
  (\bibinfo {year} {2016})},\ \Eprint {http://arxiv.org/abs/1502.01589}
  {arXiv:1502.01589 [astro-ph.CO]} \BibitemShut {NoStop}%
\bibitem [{\citenamefont {{Howlett}}\ \emph {et~al.}(2015)\citenamefont
  {{Howlett}}, \citenamefont {{Manera}},\ and\ \citenamefont
  {{Percival}}}]{L-PICOLA}%
  \BibitemOpen
  \bibfield  {author} {\bibinfo {author} {\bibfnamefont {C.}~\bibnamefont
  {{Howlett}}}, \bibinfo {author} {\bibfnamefont {M.}~\bibnamefont {{Manera}}},
  \ and\ \bibinfo {author} {\bibfnamefont {W.~J.}\ \bibnamefont {{Percival}}},\
  }\href {\doibase 10.1016/j.ascom.2015.07.003} {\bibfield  {journal} {\bibinfo
   {journal} {Astronomy and Computing}\ }\textbf {\bibinfo {volume} {12}},\
  \bibinfo {pages} {109} (\bibinfo {year} {2015})},\ \Eprint
  {http://arxiv.org/abs/1506.03737} {arXiv:1506.03737 [astro-ph.CO]}
  \BibitemShut {NoStop}%
\bibitem [{\citenamefont {{Merson}}\ \emph {et~al.}(2013)\citenamefont
  {{Merson}}, \citenamefont {{Baugh}}, \citenamefont {{Helly}}, \citenamefont
  {{Gonzalez-Perez}}, \citenamefont {{Cole}}, \citenamefont {{Bielby}},
  \citenamefont {{Norberg}}, \citenamefont {{Frenk}}, \citenamefont {{Benson}},
  \citenamefont {{Bower}}, \citenamefont {{Lacey}},\ and\ \citenamefont
  {{Lagos}}}]{Merson_2013}%
  \BibitemOpen
  \bibfield  {author} {\bibinfo {author} {\bibfnamefont {A.~I.}\ \bibnamefont
  {{Merson}}}, \bibinfo {author} {\bibfnamefont {C.~M.}\ \bibnamefont
  {{Baugh}}}, \bibinfo {author} {\bibfnamefont {J.~C.}\ \bibnamefont
  {{Helly}}}, \bibinfo {author} {\bibfnamefont {V.}~\bibnamefont
  {{Gonzalez-Perez}}}, \bibinfo {author} {\bibfnamefont {S.}~\bibnamefont
  {{Cole}}}, \bibinfo {author} {\bibfnamefont {R.}~\bibnamefont {{Bielby}}},
  \bibinfo {author} {\bibfnamefont {P.}~\bibnamefont {{Norberg}}}, \bibinfo
  {author} {\bibfnamefont {C.~S.}\ \bibnamefont {{Frenk}}}, \bibinfo {author}
  {\bibfnamefont {A.~J.}\ \bibnamefont {{Benson}}}, \bibinfo {author}
  {\bibfnamefont {R.~G.}\ \bibnamefont {{Bower}}}, \bibinfo {author}
  {\bibfnamefont {C.~G.}\ \bibnamefont {{Lacey}}}, \ and\ \bibinfo {author}
  {\bibfnamefont {C.~d.~P.}\ \bibnamefont {{Lagos}}},\ }\href {\doibase
  10.1093/mnras/sts355} {\bibfield  {journal} {\bibinfo  {journal} {\mnras}\
  }\textbf {\bibinfo {volume} {429}},\ \bibinfo {pages} {556} (\bibinfo {year}
  {2013})},\ \Eprint {http://arxiv.org/abs/1206.4049} {arXiv:1206.4049
  [astro-ph.CO]} \BibitemShut {NoStop}%
\bibitem [{\citenamefont {{Riess}}\ \emph {et~al.}(2009)\citenamefont
  {{Riess}}, \citenamefont {{Macri}}, \citenamefont {{Casertano}},
  \citenamefont {{Sosey}}, \citenamefont {{Lampeitl}}, \citenamefont
  {{Ferguson}}, \citenamefont {{Filippenko}}, \citenamefont {{Jha}},
  \citenamefont {{Li}}, \citenamefont {{Chornock}},\ and\ \citenamefont
  {{Sarkar}}}]{localH0-2009}%
  \BibitemOpen
  \bibfield  {author} {\bibinfo {author} {\bibfnamefont {A.~G.}\ \bibnamefont
  {{Riess}}}, \bibinfo {author} {\bibfnamefont {L.}~\bibnamefont {{Macri}}},
  \bibinfo {author} {\bibfnamefont {S.}~\bibnamefont {{Casertano}}}, \bibinfo
  {author} {\bibfnamefont {M.}~\bibnamefont {{Sosey}}}, \bibinfo {author}
  {\bibfnamefont {H.}~\bibnamefont {{Lampeitl}}}, \bibinfo {author}
  {\bibfnamefont {H.~C.}\ \bibnamefont {{Ferguson}}}, \bibinfo {author}
  {\bibfnamefont {A.~V.}\ \bibnamefont {{Filippenko}}}, \bibinfo {author}
  {\bibfnamefont {S.~W.}\ \bibnamefont {{Jha}}}, \bibinfo {author}
  {\bibfnamefont {W.}~\bibnamefont {{Li}}}, \bibinfo {author} {\bibfnamefont
  {R.}~\bibnamefont {{Chornock}}}, \ and\ \bibinfo {author} {\bibfnamefont
  {D.}~\bibnamefont {{Sarkar}}},\ }\href {\doibase 10.1088/0004-637X/699/1/539}
  {\bibfield  {journal} {\bibinfo  {journal} {\apj}\ }\textbf {\bibinfo
  {volume} {699}},\ \bibinfo {pages} {539} (\bibinfo {year} {2009})},\ \Eprint
  {http://arxiv.org/abs/0905.0695} {arXiv:0905.0695 [astro-ph.CO]} \BibitemShut
  {NoStop}%
\bibitem [{\citenamefont {{Scolnic}}\ \emph {et~al.}(2021)\citenamefont
  {{Scolnic}}, \citenamefont {{Brout}}, \citenamefont {{Carr}}, \citenamefont
  {{Riess}}, \citenamefont {{Davis}}, \citenamefont {{Dwomoh}}, \citenamefont
  {{Jones}}, \citenamefont {{Ali}}, \citenamefont {{Charvu}}, \citenamefont
  {{Chen}}, \citenamefont {{Peterson}}, \citenamefont {{Popovic}},
  \citenamefont {{Rose}}, \citenamefont {{Wood}}, \citenamefont {{Brown}},
  \citenamefont {{Chambers}}, \citenamefont {{Coulter}}, \citenamefont
  {{Dettman}}, \citenamefont {{Dimitriadis}}, \citenamefont {{Filippenko}},
  \citenamefont {{Foley}}, \citenamefont {{Jha}}, \citenamefont {{Kilpatrick}},
  \citenamefont {{Kirshner}}, \citenamefont {{Pan}}, \citenamefont {{Rest}},
  \citenamefont {{Rojas-Bravo}}, \citenamefont {{Siebert}}, \citenamefont
  {{Stahl}},\ and\ \citenamefont {{Zheng}}}]{PantheonPlus}%
  \BibitemOpen
  \bibfield  {author} {\bibinfo {author} {\bibfnamefont {D.}~\bibnamefont
  {{Scolnic}}}, \bibinfo {author} {\bibfnamefont {D.}~\bibnamefont {{Brout}}},
  \bibinfo {author} {\bibfnamefont {A.}~\bibnamefont {{Carr}}}, \bibinfo
  {author} {\bibfnamefont {A.~G.}\ \bibnamefont {{Riess}}}, \bibinfo {author}
  {\bibfnamefont {T.~M.}\ \bibnamefont {{Davis}}}, \bibinfo {author}
  {\bibfnamefont {A.}~\bibnamefont {{Dwomoh}}}, \bibinfo {author}
  {\bibfnamefont {D.~O.}\ \bibnamefont {{Jones}}}, \bibinfo {author}
  {\bibfnamefont {N.}~\bibnamefont {{Ali}}}, \bibinfo {author} {\bibfnamefont
  {P.}~\bibnamefont {{Charvu}}}, \bibinfo {author} {\bibfnamefont
  {R.}~\bibnamefont {{Chen}}}, \bibinfo {author} {\bibfnamefont {E.~R.}\
  \bibnamefont {{Peterson}}}, \bibinfo {author} {\bibfnamefont
  {B.}~\bibnamefont {{Popovic}}}, \bibinfo {author} {\bibfnamefont {B.~M.}\
  \bibnamefont {{Rose}}}, \bibinfo {author} {\bibfnamefont {C.}~\bibnamefont
  {{Wood}}}, \bibinfo {author} {\bibfnamefont {P.~J.}\ \bibnamefont {{Brown}}},
  \bibinfo {author} {\bibfnamefont {K.}~\bibnamefont {{Chambers}}}, \bibinfo
  {author} {\bibfnamefont {D.~A.}\ \bibnamefont {{Coulter}}}, \bibinfo {author}
  {\bibfnamefont {K.~G.}\ \bibnamefont {{Dettman}}}, \bibinfo {author}
  {\bibfnamefont {G.}~\bibnamefont {{Dimitriadis}}}, \bibinfo {author}
  {\bibfnamefont {A.~V.}\ \bibnamefont {{Filippenko}}}, \bibinfo {author}
  {\bibfnamefont {R.~J.}\ \bibnamefont {{Foley}}}, \bibinfo {author}
  {\bibfnamefont {S.~W.}\ \bibnamefont {{Jha}}}, \bibinfo {author}
  {\bibfnamefont {C.~D.}\ \bibnamefont {{Kilpatrick}}}, \bibinfo {author}
  {\bibfnamefont {R.~P.}\ \bibnamefont {{Kirshner}}}, \bibinfo {author}
  {\bibfnamefont {Y.-C.}\ \bibnamefont {{Pan}}}, \bibinfo {author}
  {\bibfnamefont {A.}~\bibnamefont {{Rest}}}, \bibinfo {author} {\bibfnamefont
  {C.}~\bibnamefont {{Rojas-Bravo}}}, \bibinfo {author} {\bibfnamefont {M.~R.}\
  \bibnamefont {{Siebert}}}, \bibinfo {author} {\bibfnamefont {B.~E.}\
  \bibnamefont {{Stahl}}}, \ and\ \bibinfo {author} {\bibfnamefont
  {W.}~\bibnamefont {{Zheng}}},\ }\href@noop {} {\bibfield  {journal} {\bibinfo
   {journal} {arXiv e-prints}\ ,\ \bibinfo {eid} {arXiv:2112.03863}} (\bibinfo
  {year} {2021})},\ \Eprint {http://arxiv.org/abs/2112.03863} {arXiv:2112.03863
  [astro-ph.CO]} \BibitemShut {NoStop}%
\bibitem [{\citenamefont {{G{\'o}rski}}\ \emph {et~al.}(2005)\citenamefont
  {{G{\'o}rski}}, \citenamefont {{Hivon}}, \citenamefont {{Banday}},
  \citenamefont {{Wandelt}}, \citenamefont {{Hansen}}, \citenamefont
  {{Reinecke}},\ and\ \citenamefont {{Bartelmann}}}]{healpix}%
  \BibitemOpen
  \bibfield  {author} {\bibinfo {author} {\bibfnamefont {K.~M.}\ \bibnamefont
  {{G{\'o}rski}}}, \bibinfo {author} {\bibfnamefont {E.}~\bibnamefont
  {{Hivon}}}, \bibinfo {author} {\bibfnamefont {A.~J.}\ \bibnamefont
  {{Banday}}}, \bibinfo {author} {\bibfnamefont {B.~D.}\ \bibnamefont
  {{Wandelt}}}, \bibinfo {author} {\bibfnamefont {F.~K.}\ \bibnamefont
  {{Hansen}}}, \bibinfo {author} {\bibfnamefont {M.}~\bibnamefont
  {{Reinecke}}}, \ and\ \bibinfo {author} {\bibfnamefont {M.}~\bibnamefont
  {{Bartelmann}}},\ }\href {\doibase 10.1086/427976} {\bibfield  {journal}
  {\bibinfo  {journal} {\apj}\ }\textbf {\bibinfo {volume} {622}},\ \bibinfo
  {pages} {759} (\bibinfo {year} {2005})},\ \Eprint
  {http://arxiv.org/abs/arXiv:astro-ph/0409513} {arXiv:astro-ph/0409513}
  \BibitemShut {NoStop}%
\bibitem [{\citenamefont {Zonca}\ \emph {et~al.}(2019)\citenamefont {Zonca},
  \citenamefont {Singer}, \citenamefont {Lenz}, \citenamefont {Reinecke},
  \citenamefont {Rosset}, \citenamefont {Hivon},\ and\ \citenamefont
  {Gorski}}]{Zonca2019}%
  \BibitemOpen
  \bibfield  {author} {\bibinfo {author} {\bibfnamefont {A.}~\bibnamefont
  {Zonca}}, \bibinfo {author} {\bibfnamefont {L.}~\bibnamefont {Singer}},
  \bibinfo {author} {\bibfnamefont {D.}~\bibnamefont {Lenz}}, \bibinfo {author}
  {\bibfnamefont {M.}~\bibnamefont {Reinecke}}, \bibinfo {author}
  {\bibfnamefont {C.}~\bibnamefont {Rosset}}, \bibinfo {author} {\bibfnamefont
  {E.}~\bibnamefont {Hivon}}, \ and\ \bibinfo {author} {\bibfnamefont
  {K.}~\bibnamefont {Gorski}},\ }\href {\doibase 10.21105/joss.01298}
  {\bibfield  {journal} {\bibinfo  {journal} {Journal of Open Source Software}\
  }\textbf {\bibinfo {volume} {4}},\ \bibinfo {pages} {1298} (\bibinfo {year}
  {2019})}\BibitemShut {NoStop}%
\bibitem [{\citenamefont {{Planck Collaboration}}\ \emph
  {et~al.}(2020{\natexlab{b}})\citenamefont {{Planck Collaboration}},
  \citenamefont {{Akrami}}, \citenamefont {{Ashdown}}, \citenamefont
  {{Aumont}}, \citenamefont {{Baccigalupi}}, \citenamefont {{Ballardini}},
  \citenamefont {{Banday}}, \citenamefont {{Barreiro}}, \citenamefont
  {{Bartolo}}, \citenamefont {{Basak}}, \citenamefont {{Benabed}},
  \citenamefont {{Bersanelli}}, \citenamefont {{Bielewicz}}, \citenamefont
  {{Bock}}, \citenamefont {{Bond}}, \citenamefont {{Borrill}}, \citenamefont
  {{Bouchet}}, \citenamefont {{Boulanger}}, \citenamefont {{Bucher}},
  \citenamefont {{Burigana}}, \citenamefont {{Butler}}, \citenamefont
  {{Calabrese}}, \citenamefont {{Cardoso}}, \citenamefont {{Casaponsa}},
  \citenamefont {{Chiang}}, \citenamefont {{Colombo}}, \citenamefont
  {{Combet}}, \citenamefont {{Contreras}}, \citenamefont {{Crill}},
  \citenamefont {{de Bernardis}}, \citenamefont {{de Zotti}}, \citenamefont
  {{Delabrouille}}, \citenamefont {{Delouis}}, \citenamefont {{Di Valentino}},
  \citenamefont {{Diego}}, \citenamefont {{Dor{\'e}}}, \citenamefont
  {{Douspis}}, \citenamefont {{Ducout}}, \citenamefont {{Dupac}}, \citenamefont
  {{Efstathiou}}, \citenamefont {{Elsner}}, \citenamefont {{En{\ss}lin}},
  \citenamefont {{Eriksen}}, \citenamefont {{Fantaye}}, \citenamefont
  {{Fernandez-Cobos}}, \citenamefont {{Finelli}}, \citenamefont {{Frailis}},
  \citenamefont {{Fraisse}}, \citenamefont {{Franceschi}}, \citenamefont
  {{Frolov}}, \citenamefont {{Galeotta}}, \citenamefont {{Galli}},
  \citenamefont {{Ganga}}, \citenamefont {{G{\'e}nova-Santos}}, \citenamefont
  {{Gerbino}}, \citenamefont {{Ghosh}}, \citenamefont {{Gonz{\'a}lez-Nuevo}},
  \citenamefont {{G{\'o}rski}}, \citenamefont {{Gruppuso}}, \citenamefont
  {{Gudmundsson}}, \citenamefont {{Hamann}}, \citenamefont {{Handley}},
  \citenamefont {{Hansen}}, \citenamefont {{Herranz}}, \citenamefont {{Hivon}},
  \citenamefont {{Huang}}, \citenamefont {{Jaffe}}, \citenamefont {{Jones}},
  \citenamefont {{Keih{\"a}nen}}, \citenamefont {{Keskitalo}}, \citenamefont
  {{Kiiveri}}, \citenamefont {{Kim}}, \citenamefont {{Krachmalnicoff}},
  \citenamefont {{Kunz}}, \citenamefont {{Kurki-Suonio}}, \citenamefont
  {{Lagache}}, \citenamefont {{Lamarre}}, \citenamefont {{Lasenby}},
  \citenamefont {{Lattanzi}}, \citenamefont {{Lawrence}}, \citenamefont {{Le
  Jeune}}, \citenamefont {{Levrier}}, \citenamefont {{Liguori}}, \citenamefont
  {{Lilje}}, \citenamefont {{Lindholm}}, \citenamefont {{L{\'o}pez-Caniego}},
  \citenamefont {{Ma}}, \citenamefont {{Mac{\'\i}as-P{\'e}rez}}, \citenamefont
  {{Maggio}}, \citenamefont {{Maino}}, \citenamefont {{Mandolesi}},
  \citenamefont {{Mangilli}}, \citenamefont {{Marcos-Caballero}}, \citenamefont
  {{Maris}}, \citenamefont {{Martin}}, \citenamefont
  {{Mart{\'\i}nez-Gonz{\'a}lez}}, \citenamefont {{Matarrese}}, \citenamefont
  {{Mauri}}, \citenamefont {{McEwen}}, \citenamefont {{Meinhold}},
  \citenamefont {{Mennella}}, \citenamefont {{Migliaccio}}, \citenamefont
  {{Miville-Desch{\^e}nes}}, \citenamefont {{Molinari}}, \citenamefont
  {{Moneti}}, \citenamefont {{Montier}}, \citenamefont {{Morgante}},
  \citenamefont {{Moss}}, \citenamefont {{Natoli}}, \citenamefont {{Pagano}},
  \citenamefont {{Paoletti}}, \citenamefont {{Partridge}}, \citenamefont
  {{Perrotta}}, \citenamefont {{Pettorino}}, \citenamefont {{Piacentini}},
  \citenamefont {{Polenta}}, \citenamefont {{Puget}}, \citenamefont {{Rachen}},
  \citenamefont {{Reinecke}}, \citenamefont {{Remazeilles}}, \citenamefont
  {{Renzi}}, \citenamefont {{Rocha}}, \citenamefont {{Rosset}}, \citenamefont
  {{Roudier}}, \citenamefont {{Rubi{\~n}o-Mart{\'\i}n}}, \citenamefont
  {{Ruiz-Granados}}, \citenamefont {{Salvati}}, \citenamefont {{Savelainen}},
  \citenamefont {{Scott}}, \citenamefont {{Shellard}}, \citenamefont
  {{Sirignano}}, \citenamefont {{Sunyaev}}, \citenamefont {{Suur-Uski}},
  \citenamefont {{Tauber}}, \citenamefont {{Tavagnacco}}, \citenamefont
  {{Tenti}}, \citenamefont {{Toffolatti}}, \citenamefont {{Tomasi}},
  \citenamefont {{Trombetti}}, \citenamefont {{Valenziano}}, \citenamefont
  {{Valiviita}}, \citenamefont {{Van Tent}}, \citenamefont {{Vielva}},
  \citenamefont {{Villa}}, \citenamefont {{Vittorio}}, \citenamefont
  {{Wandelt}}, \citenamefont {{Wehus}}, \citenamefont {{Zacchei}},
  \citenamefont {{Zibin}},\ and\ \citenamefont {{Zonca}}}]{Planck_CMB_dipole}%
  \BibitemOpen
  \bibfield  {author} {\bibinfo {author} {\bibnamefont {{Planck
  Collaboration}}}, \bibinfo {author} {\bibfnamefont {Y.}~\bibnamefont
  {{Akrami}}}, \bibinfo {author} {\bibfnamefont {M.}~\bibnamefont {{Ashdown}}},
  \bibinfo {author} {\bibfnamefont {J.}~\bibnamefont {{Aumont}}}, \bibinfo
  {author} {\bibfnamefont {C.}~\bibnamefont {{Baccigalupi}}}, \bibinfo {author}
  {\bibfnamefont {M.}~\bibnamefont {{Ballardini}}}, \bibinfo {author}
  {\bibfnamefont {A.~J.}\ \bibnamefont {{Banday}}}, \bibinfo {author}
  {\bibfnamefont {R.~B.}\ \bibnamefont {{Barreiro}}}, \bibinfo {author}
  {\bibfnamefont {N.}~\bibnamefont {{Bartolo}}}, \bibinfo {author}
  {\bibfnamefont {S.}~\bibnamefont {{Basak}}}, \bibinfo {author} {\bibfnamefont
  {K.}~\bibnamefont {{Benabed}}}, \bibinfo {author} {\bibfnamefont
  {M.}~\bibnamefont {{Bersanelli}}}, \bibinfo {author} {\bibfnamefont
  {P.}~\bibnamefont {{Bielewicz}}}, \bibinfo {author} {\bibfnamefont {J.~J.}\
  \bibnamefont {{Bock}}}, \bibinfo {author} {\bibfnamefont {J.~R.}\
  \bibnamefont {{Bond}}}, \bibinfo {author} {\bibfnamefont {J.}~\bibnamefont
  {{Borrill}}}, \bibinfo {author} {\bibfnamefont {F.~R.}\ \bibnamefont
  {{Bouchet}}}, \bibinfo {author} {\bibfnamefont {F.}~\bibnamefont
  {{Boulanger}}}, \bibinfo {author} {\bibfnamefont {M.}~\bibnamefont
  {{Bucher}}}, \bibinfo {author} {\bibfnamefont {C.}~\bibnamefont
  {{Burigana}}}, \bibinfo {author} {\bibfnamefont {R.~C.}\ \bibnamefont
  {{Butler}}}, \bibinfo {author} {\bibfnamefont {E.}~\bibnamefont
  {{Calabrese}}}, \bibinfo {author} {\bibfnamefont {J.~F.}\ \bibnamefont
  {{Cardoso}}}, \bibinfo {author} {\bibfnamefont {B.}~\bibnamefont
  {{Casaponsa}}}, \bibinfo {author} {\bibfnamefont {H.~C.}\ \bibnamefont
  {{Chiang}}}, \bibinfo {author} {\bibfnamefont {L.~P.~L.}\ \bibnamefont
  {{Colombo}}}, \bibinfo {author} {\bibfnamefont {C.}~\bibnamefont {{Combet}}},
  \bibinfo {author} {\bibfnamefont {D.}~\bibnamefont {{Contreras}}}, \bibinfo
  {author} {\bibfnamefont {B.~P.}\ \bibnamefont {{Crill}}}, \bibinfo {author}
  {\bibfnamefont {P.}~\bibnamefont {{de Bernardis}}}, \bibinfo {author}
  {\bibfnamefont {G.}~\bibnamefont {{de Zotti}}}, \bibinfo {author}
  {\bibfnamefont {J.}~\bibnamefont {{Delabrouille}}}, \bibinfo {author}
  {\bibfnamefont {J.~M.}\ \bibnamefont {{Delouis}}}, \bibinfo {author}
  {\bibfnamefont {E.}~\bibnamefont {{Di Valentino}}}, \bibinfo {author}
  {\bibfnamefont {J.~M.}\ \bibnamefont {{Diego}}}, \bibinfo {author}
  {\bibfnamefont {O.}~\bibnamefont {{Dor{\'e}}}}, \bibinfo {author}
  {\bibfnamefont {M.}~\bibnamefont {{Douspis}}}, \bibinfo {author}
  {\bibfnamefont {A.}~\bibnamefont {{Ducout}}}, \bibinfo {author}
  {\bibfnamefont {X.}~\bibnamefont {{Dupac}}}, \bibinfo {author} {\bibfnamefont
  {G.}~\bibnamefont {{Efstathiou}}}, \bibinfo {author} {\bibfnamefont
  {F.}~\bibnamefont {{Elsner}}}, \bibinfo {author} {\bibfnamefont {T.~A.}\
  \bibnamefont {{En{\ss}lin}}}, \bibinfo {author} {\bibfnamefont {H.~K.}\
  \bibnamefont {{Eriksen}}}, \bibinfo {author} {\bibfnamefont {Y.}~\bibnamefont
  {{Fantaye}}}, \bibinfo {author} {\bibfnamefont {R.}~\bibnamefont
  {{Fernandez-Cobos}}}, \bibinfo {author} {\bibfnamefont {F.}~\bibnamefont
  {{Finelli}}}, \bibinfo {author} {\bibfnamefont {M.}~\bibnamefont
  {{Frailis}}}, \bibinfo {author} {\bibfnamefont {A.~A.}\ \bibnamefont
  {{Fraisse}}}, \bibinfo {author} {\bibfnamefont {E.}~\bibnamefont
  {{Franceschi}}}, \bibinfo {author} {\bibfnamefont {A.}~\bibnamefont
  {{Frolov}}}, \bibinfo {author} {\bibfnamefont {S.}~\bibnamefont
  {{Galeotta}}}, \bibinfo {author} {\bibfnamefont {S.}~\bibnamefont {{Galli}}},
  \bibinfo {author} {\bibfnamefont {K.}~\bibnamefont {{Ganga}}}, \bibinfo
  {author} {\bibfnamefont {R.~T.}\ \bibnamefont {{G{\'e}nova-Santos}}},
  \bibinfo {author} {\bibfnamefont {M.}~\bibnamefont {{Gerbino}}}, \bibinfo
  {author} {\bibfnamefont {T.}~\bibnamefont {{Ghosh}}}, \bibinfo {author}
  {\bibfnamefont {J.}~\bibnamefont {{Gonz{\'a}lez-Nuevo}}}, \bibinfo {author}
  {\bibfnamefont {K.~M.}\ \bibnamefont {{G{\'o}rski}}}, \bibinfo {author}
  {\bibfnamefont {A.}~\bibnamefont {{Gruppuso}}}, \bibinfo {author}
  {\bibfnamefont {J.~E.}\ \bibnamefont {{Gudmundsson}}}, \bibinfo {author}
  {\bibfnamefont {J.}~\bibnamefont {{Hamann}}}, \bibinfo {author}
  {\bibfnamefont {W.}~\bibnamefont {{Handley}}}, \bibinfo {author}
  {\bibfnamefont {F.~K.}\ \bibnamefont {{Hansen}}}, \bibinfo {author}
  {\bibfnamefont {D.}~\bibnamefont {{Herranz}}}, \bibinfo {author}
  {\bibfnamefont {E.}~\bibnamefont {{Hivon}}}, \bibinfo {author} {\bibfnamefont
  {Z.}~\bibnamefont {{Huang}}}, \bibinfo {author} {\bibfnamefont {A.~H.}\
  \bibnamefont {{Jaffe}}}, \bibinfo {author} {\bibfnamefont {W.~C.}\
  \bibnamefont {{Jones}}}, \bibinfo {author} {\bibfnamefont {E.}~\bibnamefont
  {{Keih{\"a}nen}}}, \bibinfo {author} {\bibfnamefont {R.}~\bibnamefont
  {{Keskitalo}}}, \bibinfo {author} {\bibfnamefont {K.}~\bibnamefont
  {{Kiiveri}}}, \bibinfo {author} {\bibfnamefont {J.}~\bibnamefont {{Kim}}},
  \bibinfo {author} {\bibfnamefont {N.}~\bibnamefont {{Krachmalnicoff}}},
  \bibinfo {author} {\bibfnamefont {M.}~\bibnamefont {{Kunz}}}, \bibinfo
  {author} {\bibfnamefont {H.}~\bibnamefont {{Kurki-Suonio}}}, \bibinfo
  {author} {\bibfnamefont {G.}~\bibnamefont {{Lagache}}}, \bibinfo {author}
  {\bibfnamefont {J.~M.}\ \bibnamefont {{Lamarre}}}, \bibinfo {author}
  {\bibfnamefont {A.}~\bibnamefont {{Lasenby}}}, \bibinfo {author}
  {\bibfnamefont {M.}~\bibnamefont {{Lattanzi}}}, \bibinfo {author}
  {\bibfnamefont {C.~R.}\ \bibnamefont {{Lawrence}}}, \bibinfo {author}
  {\bibfnamefont {M.}~\bibnamefont {{Le Jeune}}}, \bibinfo {author}
  {\bibfnamefont {F.}~\bibnamefont {{Levrier}}}, \bibinfo {author}
  {\bibfnamefont {M.}~\bibnamefont {{Liguori}}}, \bibinfo {author}
  {\bibfnamefont {P.~B.}\ \bibnamefont {{Lilje}}}, \bibinfo {author}
  {\bibfnamefont {V.}~\bibnamefont {{Lindholm}}}, \bibinfo {author}
  {\bibfnamefont {M.}~\bibnamefont {{L{\'o}pez-Caniego}}}, \bibinfo {author}
  {\bibfnamefont {Y.~Z.}\ \bibnamefont {{Ma}}}, \bibinfo {author}
  {\bibfnamefont {J.~F.}\ \bibnamefont {{Mac{\'\i}as-P{\'e}rez}}}, \bibinfo
  {author} {\bibfnamefont {G.}~\bibnamefont {{Maggio}}}, \bibinfo {author}
  {\bibfnamefont {D.}~\bibnamefont {{Maino}}}, \bibinfo {author} {\bibfnamefont
  {N.}~\bibnamefont {{Mandolesi}}}, \bibinfo {author} {\bibfnamefont
  {A.}~\bibnamefont {{Mangilli}}}, \bibinfo {author} {\bibfnamefont
  {A.}~\bibnamefont {{Marcos-Caballero}}}, \bibinfo {author} {\bibfnamefont
  {M.}~\bibnamefont {{Maris}}}, \bibinfo {author} {\bibfnamefont {P.~G.}\
  \bibnamefont {{Martin}}}, \bibinfo {author} {\bibfnamefont {E.}~\bibnamefont
  {{Mart{\'\i}nez-Gonz{\'a}lez}}}, \bibinfo {author} {\bibfnamefont
  {S.}~\bibnamefont {{Matarrese}}}, \bibinfo {author} {\bibfnamefont
  {N.}~\bibnamefont {{Mauri}}}, \bibinfo {author} {\bibfnamefont {J.~D.}\
  \bibnamefont {{McEwen}}}, \bibinfo {author} {\bibfnamefont {P.~R.}\
  \bibnamefont {{Meinhold}}}, \bibinfo {author} {\bibfnamefont
  {A.}~\bibnamefont {{Mennella}}}, \bibinfo {author} {\bibfnamefont
  {M.}~\bibnamefont {{Migliaccio}}}, \bibinfo {author} {\bibfnamefont {M.~A.}\
  \bibnamefont {{Miville-Desch{\^e}nes}}}, \bibinfo {author} {\bibfnamefont
  {D.}~\bibnamefont {{Molinari}}}, \bibinfo {author} {\bibfnamefont
  {A.}~\bibnamefont {{Moneti}}}, \bibinfo {author} {\bibfnamefont
  {L.}~\bibnamefont {{Montier}}}, \bibinfo {author} {\bibfnamefont
  {G.}~\bibnamefont {{Morgante}}}, \bibinfo {author} {\bibfnamefont
  {A.}~\bibnamefont {{Moss}}}, \bibinfo {author} {\bibfnamefont
  {P.}~\bibnamefont {{Natoli}}}, \bibinfo {author} {\bibfnamefont
  {L.}~\bibnamefont {{Pagano}}}, \bibinfo {author} {\bibfnamefont
  {D.}~\bibnamefont {{Paoletti}}}, \bibinfo {author} {\bibfnamefont
  {B.}~\bibnamefont {{Partridge}}}, \bibinfo {author} {\bibfnamefont
  {F.}~\bibnamefont {{Perrotta}}}, \bibinfo {author} {\bibfnamefont
  {V.}~\bibnamefont {{Pettorino}}}, \bibinfo {author} {\bibfnamefont
  {F.}~\bibnamefont {{Piacentini}}}, \bibinfo {author} {\bibfnamefont
  {G.}~\bibnamefont {{Polenta}}}, \bibinfo {author} {\bibfnamefont {J.~L.}\
  \bibnamefont {{Puget}}}, \bibinfo {author} {\bibfnamefont {J.~P.}\
  \bibnamefont {{Rachen}}}, \bibinfo {author} {\bibfnamefont {M.}~\bibnamefont
  {{Reinecke}}}, \bibinfo {author} {\bibfnamefont {M.}~\bibnamefont
  {{Remazeilles}}}, \bibinfo {author} {\bibfnamefont {A.}~\bibnamefont
  {{Renzi}}}, \bibinfo {author} {\bibfnamefont {G.}~\bibnamefont {{Rocha}}},
  \bibinfo {author} {\bibfnamefont {C.}~\bibnamefont {{Rosset}}}, \bibinfo
  {author} {\bibfnamefont {G.}~\bibnamefont {{Roudier}}}, \bibinfo {author}
  {\bibfnamefont {J.~A.}\ \bibnamefont {{Rubi{\~n}o-Mart{\'\i}n}}}, \bibinfo
  {author} {\bibfnamefont {B.}~\bibnamefont {{Ruiz-Granados}}}, \bibinfo
  {author} {\bibfnamefont {L.}~\bibnamefont {{Salvati}}}, \bibinfo {author}
  {\bibfnamefont {M.}~\bibnamefont {{Savelainen}}}, \bibinfo {author}
  {\bibfnamefont {D.}~\bibnamefont {{Scott}}}, \bibinfo {author} {\bibfnamefont
  {E.~P.~S.}\ \bibnamefont {{Shellard}}}, \bibinfo {author} {\bibfnamefont
  {C.}~\bibnamefont {{Sirignano}}}, \bibinfo {author} {\bibfnamefont
  {R.}~\bibnamefont {{Sunyaev}}}, \bibinfo {author} {\bibfnamefont {A.~S.}\
  \bibnamefont {{Suur-Uski}}}, \bibinfo {author} {\bibfnamefont {J.~A.}\
  \bibnamefont {{Tauber}}}, \bibinfo {author} {\bibfnamefont {D.}~\bibnamefont
  {{Tavagnacco}}}, \bibinfo {author} {\bibfnamefont {M.}~\bibnamefont
  {{Tenti}}}, \bibinfo {author} {\bibfnamefont {L.}~\bibnamefont
  {{Toffolatti}}}, \bibinfo {author} {\bibfnamefont {M.}~\bibnamefont
  {{Tomasi}}}, \bibinfo {author} {\bibfnamefont {T.}~\bibnamefont
  {{Trombetti}}}, \bibinfo {author} {\bibfnamefont {L.}~\bibnamefont
  {{Valenziano}}}, \bibinfo {author} {\bibfnamefont {J.}~\bibnamefont
  {{Valiviita}}}, \bibinfo {author} {\bibfnamefont {B.}~\bibnamefont {{Van
  Tent}}}, \bibinfo {author} {\bibfnamefont {P.}~\bibnamefont {{Vielva}}},
  \bibinfo {author} {\bibfnamefont {F.}~\bibnamefont {{Villa}}}, \bibinfo
  {author} {\bibfnamefont {N.}~\bibnamefont {{Vittorio}}}, \bibinfo {author}
  {\bibfnamefont {B.~D.}\ \bibnamefont {{Wandelt}}}, \bibinfo {author}
  {\bibfnamefont {I.~K.}\ \bibnamefont {{Wehus}}}, \bibinfo {author}
  {\bibfnamefont {A.}~\bibnamefont {{Zacchei}}}, \bibinfo {author}
  {\bibfnamefont {J.~P.}\ \bibnamefont {{Zibin}}}, \ and\ \bibinfo {author}
  {\bibfnamefont {A.}~\bibnamefont {{Zonca}}},\ }\href {\doibase
  10.1051/0004-6361/201935201} {\bibfield  {journal} {\bibinfo  {journal}
  {\aap}\ }\textbf {\bibinfo {volume} {641}},\ \bibinfo {eid} {A7} (\bibinfo
  {year} {2020}{\natexlab{b}})},\ \Eprint {http://arxiv.org/abs/1906.02552}
  {arXiv:1906.02552 [astro-ph.CO]} \BibitemShut {NoStop}%
\bibitem [{\citenamefont {{Kenworthy}}\ \emph {et~al.}(2022)\citenamefont
  {{Kenworthy}}, \citenamefont {{Riess}}, \citenamefont {{Scolnic}},
  \citenamefont {{Yuan}}, \citenamefont {{Bernal}}, \citenamefont {{Brout}},
  \citenamefont {{Cassertano}}, \citenamefont {{Jones}}, \citenamefont
  {{Macri}},\ and\ \citenamefont {{Peterson}}}]{Kenworthy_2022}%
  \BibitemOpen
  \bibfield  {author} {\bibinfo {author} {\bibfnamefont {W.~D.}\ \bibnamefont
  {{Kenworthy}}}, \bibinfo {author} {\bibfnamefont {A.~G.}\ \bibnamefont
  {{Riess}}}, \bibinfo {author} {\bibfnamefont {D.}~\bibnamefont {{Scolnic}}},
  \bibinfo {author} {\bibfnamefont {W.}~\bibnamefont {{Yuan}}}, \bibinfo
  {author} {\bibfnamefont {J.~L.}\ \bibnamefont {{Bernal}}}, \bibinfo {author}
  {\bibfnamefont {D.}~\bibnamefont {{Brout}}}, \bibinfo {author} {\bibfnamefont
  {S.}~\bibnamefont {{Cassertano}}}, \bibinfo {author} {\bibfnamefont {D.~O.}\
  \bibnamefont {{Jones}}}, \bibinfo {author} {\bibfnamefont {L.}~\bibnamefont
  {{Macri}}}, \ and\ \bibinfo {author} {\bibfnamefont {E.}~\bibnamefont
  {{Peterson}}},\ }\href@noop {} {\bibfield  {journal} {\bibinfo  {journal}
  {arXiv e-prints}\ ,\ \bibinfo {eid} {arXiv:2204.10866}} (\bibinfo {year}
  {2022})},\ \Eprint {http://arxiv.org/abs/2204.10866} {arXiv:2204.10866
  [astro-ph.CO]} \BibitemShut {NoStop}%
\bibitem [{\citenamefont {{Krishnan}}\ \emph {et~al.}(2021)\citenamefont
  {{Krishnan}}, \citenamefont {{Mohayaee}}, \citenamefont {{Colg{\'a}in}},
  \citenamefont {{Sheikh-Jabbari}},\ and\ \citenamefont
  {{Yin}}}]{Krishnan_2021}%
  \BibitemOpen
  \bibfield  {author} {\bibinfo {author} {\bibfnamefont {C.}~\bibnamefont
  {{Krishnan}}}, \bibinfo {author} {\bibfnamefont {R.}~\bibnamefont
  {{Mohayaee}}}, \bibinfo {author} {\bibfnamefont {E.~{\'O}.}\ \bibnamefont
  {{Colg{\'a}in}}}, \bibinfo {author} {\bibfnamefont {M.~M.}\ \bibnamefont
  {{Sheikh-Jabbari}}}, \ and\ \bibinfo {author} {\bibfnamefont
  {L.}~\bibnamefont {{Yin}}},\ }\href {\doibase 10.1088/1361-6382/ac1a81}
  {\bibfield  {journal} {\bibinfo  {journal} {Classical and Quantum Gravity}\
  }\textbf {\bibinfo {volume} {38}},\ \bibinfo {eid} {184001} (\bibinfo {year}
  {2021})},\ \Eprint {http://arxiv.org/abs/2105.09790} {arXiv:2105.09790
  [astro-ph.CO]} \BibitemShut {NoStop}%
\bibitem [{\citenamefont {{Krishnan}}\ \emph {et~al.}(2022)\citenamefont
  {{Krishnan}}, \citenamefont {{Mohayaee}}, \citenamefont {{Colg{\'a}in}},
  \citenamefont {{Sheikh-Jabbari}},\ and\ \citenamefont
  {{Yin}}}]{Krishnan_2022}%
  \BibitemOpen
  \bibfield  {author} {\bibinfo {author} {\bibfnamefont {C.}~\bibnamefont
  {{Krishnan}}}, \bibinfo {author} {\bibfnamefont {R.}~\bibnamefont
  {{Mohayaee}}}, \bibinfo {author} {\bibfnamefont {E.~{\~A}.~.}\ \bibnamefont
  {{Colg{\'a}in}}}, \bibinfo {author} {\bibfnamefont {M.~M.}\ \bibnamefont
  {{Sheikh-Jabbari}}}, \ and\ \bibinfo {author} {\bibfnamefont
  {L.}~\bibnamefont {{Yin}}},\ }\href {\doibase 10.1103/PhysRevD.105.063514}
  {\bibfield  {journal} {\bibinfo  {journal} {\prd}\ }\textbf {\bibinfo
  {volume} {105}},\ \bibinfo {eid} {063514} (\bibinfo {year} {2022})},\ \Eprint
  {http://arxiv.org/abs/2106.02532} {arXiv:2106.02532 [astro-ph.CO]}
  \BibitemShut {NoStop}%
\bibitem [{\citenamefont {{Luongo}}\ \emph {et~al.}(2022)\citenamefont
  {{Luongo}}, \citenamefont {{Muccino}}, \citenamefont {{Colg{\'a}in}},
  \citenamefont {{Sheikh-Jabbari}},\ and\ \citenamefont {{Yin}}}]{Luongo_2022}%
  \BibitemOpen
  \bibfield  {author} {\bibinfo {author} {\bibfnamefont {O.}~\bibnamefont
  {{Luongo}}}, \bibinfo {author} {\bibfnamefont {M.}~\bibnamefont {{Muccino}}},
  \bibinfo {author} {\bibfnamefont {E.~{\'O}.}\ \bibnamefont {{Colg{\'a}in}}},
  \bibinfo {author} {\bibfnamefont {M.~M.}\ \bibnamefont {{Sheikh-Jabbari}}}, \
  and\ \bibinfo {author} {\bibfnamefont {L.}~\bibnamefont {{Yin}}},\ }\href
  {\doibase 10.1103/PhysRevD.105.103510} {\bibfield  {journal} {\bibinfo
  {journal} {\prd}\ }\textbf {\bibinfo {volume} {105}},\ \bibinfo {eid}
  {103510} (\bibinfo {year} {2022})},\ \Eprint
  {http://arxiv.org/abs/2108.13228} {arXiv:2108.13228 [astro-ph.CO]}
  \BibitemShut {NoStop}%
\bibitem [{\citenamefont {{Garc{\'\i}a-Bellido}}\ and\ \citenamefont
  {{Haugb{\o}lle}}(2008)}]{LTB-DE1}%
  \BibitemOpen
  \bibfield  {author} {\bibinfo {author} {\bibfnamefont {J.}~\bibnamefont
  {{Garc{\'\i}a-Bellido}}}\ and\ \bibinfo {author} {\bibfnamefont
  {T.}~\bibnamefont {{Haugb{\o}lle}}},\ }\href {\doibase
  10.1088/1475-7516/2008/09/016} {\bibfield  {journal} {\bibinfo  {journal}
  {\jcap}\ }\textbf {\bibinfo {volume} {2008}},\ \bibinfo {eid} {016} (\bibinfo
  {year} {2008})},\ \Eprint {http://arxiv.org/abs/0807.1326} {arXiv:0807.1326
  [astro-ph]} \BibitemShut {NoStop}%
\bibitem [{\citenamefont {{Moss}}\ \emph {et~al.}(2011)\citenamefont {{Moss}},
  \citenamefont {{Zibin}},\ and\ \citenamefont {{Scott}}}]{LTB-DE2}%
  \BibitemOpen
  \bibfield  {author} {\bibinfo {author} {\bibfnamefont {A.}~\bibnamefont
  {{Moss}}}, \bibinfo {author} {\bibfnamefont {J.~P.}\ \bibnamefont {{Zibin}}},
  \ and\ \bibinfo {author} {\bibfnamefont {D.}~\bibnamefont {{Scott}}},\ }\href
  {\doibase 10.1103/PhysRevD.83.103515} {\bibfield  {journal} {\bibinfo
  {journal} {\prd}\ }\textbf {\bibinfo {volume} {83}},\ \bibinfo {eid} {103515}
  (\bibinfo {year} {2011})},\ \Eprint {http://arxiv.org/abs/1007.3725}
  {arXiv:1007.3725 [astro-ph.CO]} \BibitemShut {NoStop}%
\bibitem [{\citenamefont {{Camarena}}\ \emph {et~al.}(2022)\citenamefont
  {{Camarena}}, \citenamefont {{Marra}}, \citenamefont {{Sakr}},\ and\
  \citenamefont {{Clarkson}}}]{LTB-H0}%
  \BibitemOpen
  \bibfield  {author} {\bibinfo {author} {\bibfnamefont {D.}~\bibnamefont
  {{Camarena}}}, \bibinfo {author} {\bibfnamefont {V.}~\bibnamefont {{Marra}}},
  \bibinfo {author} {\bibfnamefont {Z.}~\bibnamefont {{Sakr}}}, \ and\ \bibinfo
  {author} {\bibfnamefont {C.}~\bibnamefont {{Clarkson}}},\ }\href@noop {}
  {\bibfield  {journal} {\bibinfo  {journal} {arXiv e-prints}\ ,\ \bibinfo
  {eid} {arXiv:2205.05422}} (\bibinfo {year} {2022})},\ \Eprint
  {http://arxiv.org/abs/2205.05422} {arXiv:2205.05422 [astro-ph.CO]}
  \BibitemShut {NoStop}%
\bibitem [{\citenamefont {{Kenworthy}}\ \emph {et~al.}(2019)\citenamefont
  {{Kenworthy}}, \citenamefont {{Scolnic}},\ and\ \citenamefont
  {{Riess}}}]{LTB-H02}%
  \BibitemOpen
  \bibfield  {author} {\bibinfo {author} {\bibfnamefont {W.~D.}\ \bibnamefont
  {{Kenworthy}}}, \bibinfo {author} {\bibfnamefont {D.}~\bibnamefont
  {{Scolnic}}}, \ and\ \bibinfo {author} {\bibfnamefont {A.}~\bibnamefont
  {{Riess}}},\ }\href {\doibase 10.3847/1538-4357/ab0ebf} {\bibfield  {journal}
  {\bibinfo  {journal} {\apj}\ }\textbf {\bibinfo {volume} {875}},\ \bibinfo
  {eid} {145} (\bibinfo {year} {2019})},\ \Eprint
  {http://arxiv.org/abs/1901.08681} {arXiv:1901.08681 [astro-ph.CO]}
  \BibitemShut {NoStop}%
\bibitem [{\citenamefont {{Wojtak}}\ \emph {et~al.}(2015)\citenamefont
  {{Wojtak}}, \citenamefont {{Davis}},\ and\ \citenamefont
  {{Wiis}}}]{Wojtak2015}%
  \BibitemOpen
  \bibfield  {author} {\bibinfo {author} {\bibfnamefont {R.}~\bibnamefont
  {{Wojtak}}}, \bibinfo {author} {\bibfnamefont {T.~M.}\ \bibnamefont
  {{Davis}}}, \ and\ \bibinfo {author} {\bibfnamefont {J.}~\bibnamefont
  {{Wiis}}},\ }\href {\doibase 10.1088/1475-7516/2015/07/025} {\bibfield
  {journal} {\bibinfo  {journal} {\jcap}\ }\textbf {\bibinfo {volume} {2015}},\
  \bibinfo {eid} {025} (\bibinfo {year} {2015})},\ \Eprint
  {http://arxiv.org/abs/1504.00718} {arXiv:1504.00718 [astro-ph.CO]}
  \BibitemShut {NoStop}%
\bibitem [{\citenamefont {{Bolton}}\ \emph {et~al.}(2012)\citenamefont
  {{Bolton}}, \citenamefont {{Schlegel}}, \citenamefont {{Aubourg}},
  \citenamefont {{Bailey}}, \citenamefont {{Bhardwaj}}, \citenamefont
  {{Brownstein}}, \citenamefont {{Burles}}, \citenamefont {{Chen}},
  \citenamefont {{Dawson}}, \citenamefont {{Eisenstein}}, \citenamefont
  {{Gunn}}, \citenamefont {{Knapp}}, \citenamefont {{Loomis}}, \citenamefont
  {{Lupton}}, \citenamefont {{Maraston}}, \citenamefont {{Muna}}, \citenamefont
  {{Myers}}, \citenamefont {{Olmstead}}, \citenamefont {{Padmanabhan}},
  \citenamefont {{P{\^a}ris}}, \citenamefont {{Percival}}, \citenamefont
  {{Petitjean}}, \citenamefont {{Rockosi}}, \citenamefont {{Ross}},
  \citenamefont {{Schneider}}, \citenamefont {{Shu}}, \citenamefont
  {{Strauss}}, \citenamefont {{Thomas}}, \citenamefont {{Tremonti}},
  \citenamefont {{Wake}}, \citenamefont {{Weaver}},\ and\ \citenamefont
  {{Wood-Vasey}}}]{Bolton2012}%
  \BibitemOpen
  \bibfield  {author} {\bibinfo {author} {\bibfnamefont {A.~S.}\ \bibnamefont
  {{Bolton}}}, \bibinfo {author} {\bibfnamefont {D.~J.}\ \bibnamefont
  {{Schlegel}}}, \bibinfo {author} {\bibfnamefont {{\'E}.}~\bibnamefont
  {{Aubourg}}}, \bibinfo {author} {\bibfnamefont {S.}~\bibnamefont {{Bailey}}},
  \bibinfo {author} {\bibfnamefont {V.}~\bibnamefont {{Bhardwaj}}}, \bibinfo
  {author} {\bibfnamefont {J.~R.}\ \bibnamefont {{Brownstein}}}, \bibinfo
  {author} {\bibfnamefont {S.}~\bibnamefont {{Burles}}}, \bibinfo {author}
  {\bibfnamefont {Y.-M.}\ \bibnamefont {{Chen}}}, \bibinfo {author}
  {\bibfnamefont {K.}~\bibnamefont {{Dawson}}}, \bibinfo {author}
  {\bibfnamefont {D.~J.}\ \bibnamefont {{Eisenstein}}}, \bibinfo {author}
  {\bibfnamefont {J.~E.}\ \bibnamefont {{Gunn}}}, \bibinfo {author}
  {\bibfnamefont {G.~R.}\ \bibnamefont {{Knapp}}}, \bibinfo {author}
  {\bibfnamefont {C.~P.}\ \bibnamefont {{Loomis}}}, \bibinfo {author}
  {\bibfnamefont {R.~H.}\ \bibnamefont {{Lupton}}}, \bibinfo {author}
  {\bibfnamefont {C.}~\bibnamefont {{Maraston}}}, \bibinfo {author}
  {\bibfnamefont {D.}~\bibnamefont {{Muna}}}, \bibinfo {author} {\bibfnamefont
  {A.~D.}\ \bibnamefont {{Myers}}}, \bibinfo {author} {\bibfnamefont {M.~D.}\
  \bibnamefont {{Olmstead}}}, \bibinfo {author} {\bibfnamefont
  {N.}~\bibnamefont {{Padmanabhan}}}, \bibinfo {author} {\bibfnamefont
  {I.}~\bibnamefont {{P{\^a}ris}}}, \bibinfo {author} {\bibfnamefont {W.~J.}\
  \bibnamefont {{Percival}}}, \bibinfo {author} {\bibfnamefont
  {P.}~\bibnamefont {{Petitjean}}}, \bibinfo {author} {\bibfnamefont {C.~M.}\
  \bibnamefont {{Rockosi}}}, \bibinfo {author} {\bibfnamefont {N.~P.}\
  \bibnamefont {{Ross}}}, \bibinfo {author} {\bibfnamefont {D.~P.}\
  \bibnamefont {{Schneider}}}, \bibinfo {author} {\bibfnamefont
  {Y.}~\bibnamefont {{Shu}}}, \bibinfo {author} {\bibfnamefont {M.~A.}\
  \bibnamefont {{Strauss}}}, \bibinfo {author} {\bibfnamefont {D.}~\bibnamefont
  {{Thomas}}}, \bibinfo {author} {\bibfnamefont {C.~A.}\ \bibnamefont
  {{Tremonti}}}, \bibinfo {author} {\bibfnamefont {D.~A.}\ \bibnamefont
  {{Wake}}}, \bibinfo {author} {\bibfnamefont {B.~A.}\ \bibnamefont
  {{Weaver}}}, \ and\ \bibinfo {author} {\bibfnamefont {W.~M.}\ \bibnamefont
  {{Wood-Vasey}}},\ }\href {\doibase 10.1088/0004-6256/144/5/144} {\bibfield
  {journal} {\bibinfo  {journal} {\aj}\ }\textbf {\bibinfo {volume} {144}},\
  \bibinfo {eid} {144} (\bibinfo {year} {2012})},\ \Eprint
  {http://arxiv.org/abs/1207.7326} {arXiv:1207.7326 [astro-ph.CO]} \BibitemShut
  {NoStop}%
\end{thebibliography}%

\end{document}